\documentclass[11pt,a4paper]{article}
\usepackage{jheppub}
\allowdisplaybreaks[1]

\usepackage{bm}
\usepackage{bbold}
\usepackage[compat=1.1.0]{tikz-feynman}
\tikzfeynmanset{ warn luatex = false }
\usepackage{contour}

\usepackage{cancel}

\usepackage{comment}
\usepackage{multirow}
\DeclareUnicodeCharacter{2212}{-}

\usepackage{slashed}
\usepackage{amsfonts}

\usepackage[normalem]{ulem} %crossing out text

\begin{document}

%\preprint{APS/123-QED}

\title{On the On-Shell Renormalization of\\ Fermion Masses, Fields, and Mixing Matrices at 1-loop}

\author{Simonas Draukšas}
\emailAdd{simonas.drauksas@ff.stud.vu.lt}
\affiliation{%
Institute of Theoretical Physics and Astronomy, Faculty of Physics,\\Vilnius University, 9 Saulėtekio, LT-10222 Vilnius, Lithuania}%

\date{\today}% It is always \today, today,
             %  but any date may be explicitly specified

\abstract{
We propose a new and simple On-Shell definition of off-diagonal fermion field and mass counterterms at 1-loop in terms of self-energy scalar functions. Further, we show that the anti-hermitian part of the field renormalization is always finite and that mass counterterms can be chosen gauge-independent. It is noteworthy that our definition relies on mass structures, which is a novel approach, and does not require to drop absorptive parts as is usually the case. Definitions of the off-diagonal mass and field counterterms allow us to comment on the renormalization of mixing matrices with focus on the quark mixing matrix in the Standard Model. As an example of our scheme, we provide computations in the Two Higgs Doublet Model with an additional Heavy Majorana neutrino --- the Grimus-Neufeld model. This allows for the comparison with other schemes known in literature and also provides examples for the case of massless fermions. The examples also serve as genuine results in this particular model as well as this scheme. In the appendix the scheme is extended to arbitrary orders, although without example computations in the Grimus-Neufeld model.}

%Use showkeys class option if keyword display desired
\maketitle

%\tableofcontents

\section{Introduction}
While renormalization of models without mixing is more or less straightforward, particle mixing introduces a few complications. For example, the renormalization constants develop off-diagonal terms which have to be fixed, the appearance of mixing matrices poses the question whether they also should be renormalized. In terms of the Standard Model (SM) probably the first attempt to account for the quark mixing is attempted in~\cite{Denner1990_CKMreno} in renormalizing the CKM matrix~\cite{Cabibbo1963, Kobayashi1973}, however, the proposed scheme later showed its shortcomings. There the authors chose to absorb the whole anti-hermitian part of field renormalization into the CKM matrix counterterm, which resulted in a gauge-dependent counterterm, a feature which breaks Ward-Takahashi identities~\cite{Gambino1999}. An analogous approach for neutrinos was introduced in~\cite{Kniehl1996}.

Another approach in~\cite{Kniehl2006} relies on defining off-diagonal mass counterterms and performing a 1-loop rotation. The definition is done using the self-energy decomposition used by Feynman in QED \cite{Feynman1949} and noting how different terms contribute to external leg corrections. The same authors later also provided mass counterterms in the SM in terms of the conventional self-energy decomposition~\cite{Kniehl2009a}. While the authors do compute field renormalization contributions, they remain undefined in terms of the conventional self-energy decomposition. 

We also note that the approach and shortcomings of~\cite{Denner1990_CKMreno} sparked a series of papers trying to remedy the gauge-dependence of the CKM counterterm. One such approach in~\cite{Gambino1999} considers a zero-momentum subtraction scheme. Other approaches propose various methods to separate the gauge-dependent part of field renormalization such that only the gauge-independent part goes into the definition of the CKM counterterm: computing field renormalization in the 't Hooft-Feynman gauge~\cite{Yamada2001}, using a reference theory in which there is no mixing~\cite{Diener2001}, obtaining a renormalization condition via modified minimal subtraction ($\overline{\mathrm{MS}}$)~\cite{Pilaftsis2002}, proposing physical renormalization conditions based on BRS symmetry and $W\rightarrow u_i d_j$ and $t\rightarrow W d_j$ decay amplitudes~\cite{Denner2004} (and without BRS in~\cite{Zhou2003}), using the difference between "leptonic" and quark transition amplitudes~\cite{Barroso2000}. However, our scheme is closest to that of~\cite{Kniehl2006,Almasy2009}, hence the schemes in this paragraph are rather removed from our approach and will not be discussed any further.

In this paper we propose a simple on-shell scheme that defines off-diagonal fermion mass and field counterterms in terms of scalar self-energy functions. The off-diagonal mass counterterms are related to the anti-hermitian part of the field renormalization. At 1-loop, at least for Dirac fermions, the definition avoids the usage of the $\widetilde{Re}$ operator~\cite{Denner2007} which drops absorptive parts. Clear definitions allow us to discuss the renormalization of mixing matrices and to comment on rotating away the off-diagonal mass counterterms as done in~\cite{Kniehl2006,Kniehl2009a}. We call this the 1-loop rotation as is customary. We also apply the scheme in the Grimus-Neufeld model~\cite{Grimus1989} which is a Two Higgs Doublet Model with an additional heavy Majorana neutrino. This allows for the comparison of divergent parts with~\cite{Denner1990_CKMreno, Kniehl2006} for the quarks in the SM, for the comparison with~\cite{Almasy2009} in case of Majorana neutrinos and the corresponding mixing matrix while also providing new results.

The paper is structured as follows: in Section~\ref{sec:big_ct} we write up all the needed renormalization constants, the fermion self-energy decomposition, and specify the choice of renormalization conditions. In Section~\ref{sec:big_income} we find that at 1-loop there are mass structures related to field and mass counterterms. After considering UV and gauge properties related with these structures we make a definition of off-diagonal mass and anti-hermitian part of field renormalization counterterms.~These definitions are model independent and are written in terms of self-energies.~The section is concluded by a discussion on the 1-loop rotation and mixing matrix counterterm. In Section~\ref{sec:big_examples} we introduce the Grimus-Neufeld model and provide computations of quark, charged lepton, and neutrino counterterms as well as comparisons to existing literature. In Section~\ref{sec:big_conclusions} we give our conclusions. In Appendix~\ref{sec:beyond_1loop} we extend the scheme beyond 1-loop.

\section{Counterterms and self-energy}\label{sec:big_ct}
\subsection{Field renormalization}\label{sec:field_reno}
Let us begin by introducing the necessary renormalization constants. We choose to renormalize fermion fields as
\begin{align}\label{eq:reno_conditions}
        \nonumber
        \psi_0& \rightarrow Z^{1/2}\psi,\\
        \nonumber
        \bar{\psi}_0& \rightarrow \bar{\psi}\gamma^0 Z^{1/2\dagger}\gamma^0,\\
        \nonumber
        Z&=Z_L P_L + Z_R P_R,\\
        Z_{L,R}&=1+\delta Z_{L,R},
\end{align}
where $P_{L,R}=\frac{1}{2}\left(1\mp\gamma^5\right)$. This field renormalization is in line with keeping the hermiticity of the Lagrangian, which is a feature not to be taken for granted~\cite{Espriu2002}. We want to apply the off-diagonal renormalization conditions that prohibit particle mixing on external legs as in~\cite{Aoki1982}
\begin{subequations}
\begin{equation}\label{eq:incoming}
    \Sigma_{ji}(\cancel{p})u_i=0,
\end{equation}
\begin{equation}
    \bar{u}_j\Sigma_{ji}(\cancel{p})=0,
\end{equation}
\end{subequations}
where $\Sigma_{ij}(\cancel{p})$ is the self-energy, $i\neq j$, incoming particles are denoted by $i$ and outgoing ones by $j$. It has been noted in~\cite{Espriu2000,Espriu2002} that these conditions suffer from over specification. Upon application of the above conditions one arrives at expressions for $Z^{}_{L,R}$ and $Z^{\dagger}_{L,R}$ in terms of self-energy scalar functions, however, one finds that due to absorptive parts one cannot get $Z^{\dagger}_{L,R}$ from $Z^{}_{L,R}$ (or vice versa) by naive hermitian conjugation and the resulting expressions are different. The authors of~\cite{Espriu2002} propose to relax the no-mixing conditions or to use two sets of field renormalization constants $Z$ (associated with $\psi$) and $\bar{Z}$ (associated with $\bar{\psi}$), for which the hermiticity condition $\bar{Z}=\gamma^0 Z^\dagger \gamma^0$ does not hold. The latter apparently makes the Lagrangian written in terms of renormalized fields non-hermitian, but is, however, fine in terms of external leg renormalization. While neither of the proposed solutions is satisfactory, for our scheme we choose to keep the Lagrangian hermitian as already chosen in eq.~(\ref{eq:reno_conditions}) and to drop one of the renormalization conditions. We choose to keep the no-mixing requirement for incoming particles in eq.~(\ref{eq:incoming}), which is equivalent to no-mixing requirement for outgoing anti-particles. This is a possible choice as noted in~\cite{Espriu2000}, the other choice is to keep the no-mixing condition only for outgoing particles. Finally, it is worth noting, that there are no absorptive parts if a particle's mass is below particle production thresholds, in that case the no-mixing condition automatically holds for outgoing particles, too.

\subsubsection{A note on Majorana fermions}
For Majorana fermions the story is a little different due to the Majorana condition imposing an additional requirement on field renormalization. The Majorana condition on bare fields~\cite{Pal2010}
\begin{equation}
    \nu_0=\nu_0^c=\gamma^0 \mathcal{C} \nu_0^\star,
\end{equation}
where $\mathcal{C}$ is the charge conjugation matrix, implies
\begin{equation}
 \Longrightarrow Z=\gamma^0 \mathcal{C} Z^\star \mathcal{C}^{-1} \gamma^0.
\end{equation}
The implication is there if the renormalized fields are also Majorana. More simply, this means $Z^{}_L=Z^\star_R$, which is an additional relation for field renormalization constants. The relation involves complex conjugation, which leads to the same problem of over-specification and does not hold due to absorptive parts even if one of the no-mixing conditions is dropped. In other words, the bare Majorana condition is incompatible with the no-mixing conditions above particle production thresholds. There is now a number of ways to proceed.

First, we may consider relaxing the no-mixing condition on incoming legs. Since eq.~(\ref{eq:incoming}) has left and right projections, we may require only one of those to hold. This approach, however, does not seem to provide us with convenient mass structures, which will appear in upcoming sections. Another way to relax the condition would be to fully use eq.~(\ref{eq:incoming}) and solve for field renormalization, but then only use one solution (e.g., use $Z_L$ and get $Z_R$ by complex conjugation of $Z_L$) -  this will keep all of the nice properties we are about to find, but the expressions of mass counterterms are then fairly cumbersome (not a limit of the Dirac case). We will not relax the no-mixing condition in this way.

Second and maybe the easiest approach is to use the $\widetilde{Re}$ operator in case of Majorana particles, such that the no-mixing condition is
\begin{equation}
    \widetilde{Re}\left[\Sigma_{ji}(\cancel{p})\right]u_i=0.
\end{equation}
This, of course, relaxes the no-mixing condition, but keeps the Majorana condition in full for both bare and renormalized fields, the Lagrangian is also hermitian.

Currently, this seems to be the best approach when considering Majorana fermions, hence, in further expressions the counterterms of Majorana particles will come with the $\widetilde{Re}$ operator, except for expressions containing only divergent parts or that do not contain absorptive parts.

The takeaway is that the no-mixing conditions, hermiticity of the Lagrangian, and the Majorana condition are incompatible if considered together. There is simply not enough freedom to impose everything.

\subsection{Mass Renormalization}

The standard for mass renormalization is to only introduce the diagonal mass counterterm and then account for the mixing solely with field renormalization (e.g., see Section III in~\cite{Denner2020} for a recent review of radiative corrections in the Standard Model). This approach seems to introduce difficulties when discussing the renormalization of mixing matrices, hence, we follow~\cite{Kniehl2006, Kniehl2009a} and also introduce off-diagonal mass counterterms
\begin{equation}
    m_{0ji} \rightarrow m_{i}\delta_{ji}+\delta m^{L}_{ji}P_L + \delta m^{R}_{ji} P_R .
\end{equation}
Here the renormalized mass $m$ is real and diagonal, while $\delta m^{R,L}_{ji}$ contain off-diagonal contributions. One could imagine starting out in a non-diagonal-mass basis, introducing non-diagonal counterterms, and diagonalizing the renormalized mass. Diagonalization of the renormalized mass does not automatically diagonlize the counterterms. Hermiticity of the Lagrangian imposes the following relation on mass counterterms (in matrix notation)
\begin{equation}\label{eq:hermiticity}
    (\delta m^{L})^\dagger=\delta m^{R}.
\end{equation}

\subsection{Self-Energy Decomposition}\label{sec:SE_decomposition}
In order to write down the self-energy decomposition let us also write down the kinetic and mass terms of the renormalized fermion Lagrangian for clarity
\begin{equation}
\begin{split}
    \mathcal{L}_\mathrm{kin.+mass}=&\bar{\psi}^{}_{L}Z^{1/2\dagger}_{L}i\cancel{\partial}Z^{1/2}_{L}\psi^{}_{L}
    -\bar{\psi}^{}_{R}Z^{1/2\dagger}_{R}(m+\delta m^{L})Z^{1/2}_{L}\psi^{}_{L}
    +(L \leftrightarrow R).
\end{split}
\end{equation}
We can now decompose the renormalized self-energy at 1-loop\footnote{The indices work as follows: $(A^\dagger)_{ji}\equiv A^\dagger_{ji}=A^\star_{ij}$.}
\begin{equation}\label{eq:SE_deco}
    \begin{split}
        \Sigma^{}_{ji}(\cancel{p})=&\Sigma^{L}_{ji}(p^2)\cancel{p} P_{L}+\Sigma^{R}_{ji}(p^2) \cancel{p}P_{R}
        +\Sigma^{sL}_{ji}(p^2) P_{L}+\Sigma^{sR}_{ji}(p^2) P_{R}\\
        &+\frac{1}{2}\left(\delta Z^\dagger_{Lji}+\delta Z^{}_{Lji}\right)\cancel{p}P_L
        +\frac{1}{2}\left(\delta Z^\dagger_{Rji}+\delta Z^{}_{Rji}\right)\cancel{p}P_R\\
        &-\left(\delta m^L_{ji}+\frac{1}{2}\delta Z^\dagger_{Rji}m_i+\frac{1}{2}m_j\delta Z_{Lji}\right)P_L\\
        &-\left(\delta m^R_{ji}+\frac{1}{2}\delta Z^\dagger_{Lji}m_i+\frac{1}{2}m_j\delta Z_{Rji}\right)P_R.
    \end{split}
\end{equation}
Here everything is completely general and model independent, for example, we do not introduce relations for $\Sigma^{sL,sR}(p^2)$ as done in the SM. We also introduced family indices $i$ and $j$. In addition, we note that we include tadpole diagrams in the self-energy computation, so in terms of diagrams the self-energy is
\begin{equation}
\begin{split}
\Sigma_{ji}\left(\cancel{p}\right)&=
\vcenter{\hbox{\begin{tikzpicture}
  \begin{feynman}
    \vertex (a) at (-1,0);
    \vertex (b) at ( 1,0);
    \diagram* {
    (a)--[fermion, edge label=\small $\delta_{ij}$](b)};
  \end{feynman}
\end{tikzpicture}}}
+
\vcenter{\vspace{0.45cm}\hbox{\begin{tikzpicture}
  \begin{feynman}
    \vertex[blob] (m) at (0,0) {\contour{gray}{1PI}};
    \vertex (a) at (-1.6,0) ;
    \vertex (b) at ( 1.6,0);
    \diagram* {
    (a)--[fermion, edge label=\small $i$](m)--[fermion, edge label=\small $j$](b)};
  \end{feynman}
\end{tikzpicture}}}
+
\vcenter{\hbox{\begin{tikzpicture}
  \begin{feynman}
    \vertex[blob] (m) at (0,1) {\contour{gray}{1PI}};
    \vertex (c) at (0,0);
    \vertex (a) at (-0.9,0) ;
    \vertex (b) at ( 0.9,0);
    \diagram* {
    (a)--[fermion, edge label=\small $i$](c)--[fermion, edge label=\small $j$](b),
    (c)--(m)};
  \end{feynman}
\end{tikzpicture}}}
+
\vcenter{\hbox{\begin{tikzpicture}
  \begin{feynman}
    \vertex (a) at (-1.1,0) ;
    \vertex (b) at (1.1,0);
    \vertex (x) at (0,0);
    \diagram* {
    (a)--[fermion, edge label=\small $i$](x)--[fermion,insertion={[size=5pt]0}, edge label=\small $j$](b)};
  \end{feynman}
\end{tikzpicture}}}
\end{split},
\end{equation}
where the propagator with the cross marks counterterm insertions and 1PI stands for 1-particle-irreducible diagrams. The choice of including tadpoles corresponds to a scheme where the tadpoles are not renormalized, which is also equivalent to the the FJ scheme~\cite{Fleischer1981,Denner2016}.
\section{Incoming renormalization}\label{sec:big_income}

We now apply the condition of no-mixing on incoming particles in eq.~(\ref{eq:incoming}) to get a relation between field and mass counterterms. We arrive at the standard field renormalization expression except for the shift due to mass counterterms
\begin{equation}\label{eq:dZL}
\begin{split}
    \delta Z^{}_{Lji}=-\frac{2}{m^2_i-m^2_j}&\left(m^2_i\Sigma^{L}_{ji}(m^2_i)+m^{}_i m^{}_j\Sigma^{R}_{ji}(m^2_i)+m^{}_j\Sigma^{sL}_{ji}(m^2_i)+m^{}_i\Sigma^{sR}_{ji}(m^2_i)\right)\\
    +\frac{2}{m^2_i-m^2_j}&\left(m^{}_j\delta m^L_{ji}+m^{}_i\delta m^R_{ji}\right), \qquad (i\neq j).
\end{split}
\end{equation}
For $\delta Z_R$ we get an analogous expression
\begin{equation}\label{eq:dZR}
\begin{split}
    \delta Z^{}_{Rji}=-\frac{2}{m^2_i-m^2_j}&\left(m^2_i\Sigma^{R}_{ji}(m^2_i)+m^{}_i m^{}_j\Sigma^{L}_{ji}(m^2_i)+m^{}_j\Sigma^{sR}_{ji}(m^2_i)+m^{}_i\Sigma^{sL}_{ji}(m^2_i)\right)\\
    +\frac{2}{m^2_i-m^2_j}&\left(m^{}_j\delta m^R_{ji}+m^{}_i\delta m^L_{ji}\right), \qquad (i\neq j).
\end{split}
\end{equation}
It may seem that this is the endpoint and any further definitions are completely arbitrary, however, that is not the case. There is a number of properties that we can use to pick out definitions out of these relations. First, let us notice that, due to the hermiticity relation of mass counterterms, mass counterterms are related only to the anti-hermitian part. Therefore, we are only interested in the relation between anti-hermitian parts of the field renormalization and mass counterterms. Let us write down the anti-hermitian part of the left-handed field renormalization $Z^{A}_L$ in a slightly unconventional way
\begin{equation}\label{eq:dZL_mass}
    \begin{split}
        (m^2_i-m^2_j)\delta Z^A_{Lji}-2m_j\delta m^L_{ji}-2m_i\delta m^R_{ji}=&-\left(m^2_i\Sigma^{L}_{ji}(m^2_i)+m^{}_i m^{}_j\Sigma^{R}_{ji}(m^2_i)\right.\\
    &\left.+m^{}_j\Sigma^{sL}_{ji}(m^2_i)+m^{}_i\Sigma^{sR}_{ji}(m^2_i)\right)+H.C.
    \end{split}
\end{equation}
Here $H.C.$ stands for hermitian conjugation and $i\neq j$. An analogous expression also exists for the right-handed part and can be easily derived. The above expression allows to notice an interesting thing --- the mass structures multiplying field and mass counterterms are different: $(m^2_i-m^2_j)$ for $\delta Z^A$ and $m_i$ or $m_j$ for $\delta m^{L,R}$. In addition, mass factors in front of mass counterterms swap places in eqs.~(\ref{eq:dZL}) and (\ref{eq:dZR}), which also propagates into corresponding anti-hermitian parts. We would like to use these mass structures to define corresponding counterterms, so let us further investigate these mass structures. 

\subsection{Gauge Dependence}\label{sec:nielsen}
Naturally, one of the first questions is that of gauge dependence. Since mass is a physical parameter, we would like the corresponding counterterms to be gauge-independent, meaning that all of the gauge dependence should go into field renormalization. To investigate the gauge dependence we use Nielsen identities~\cite{Nielsen1975} as is the usual approach. For convenience, we write down the derivative w.r.t. a gauge parameter of the self-energy in general as in~\cite{Gambino2000}
\begin{equation}\label{eq:NI}
    \partial_\xi \Sigma_{ji}(\cancel{p})=\Lambda_{jj'}\Sigma_{j'i}(\cancel{p})+\Sigma_{ji'}(\cancel{p})\bar{\Lambda}_{i'i}.
\end{equation}
Here $\Lambda$'s are also functions of $\cancel{p}$ and arise from 1-loop vertex functions: $\Lambda_{ji}=-\Gamma_{\chi \bar{\psi}_j \eta_{\psi_i}}$ and $\bar{\Lambda}_{ji}=-\Gamma_{\chi \bar{\eta}_{\psi_j} \psi_i}$. Also, $\chi$ is the BRST source of a gauge parameter and $\eta_\psi$ is the BRST source of a fermion $\psi$~\cite{Gambino2000}. Note that $\Lambda$'s also have Dirac structure. 

Since we're interested in the 1-loop case, we write down the above expression at 1-loop as in~\cite{Yamada2001}
\begin{equation}
    \partial_\xi \Sigma_{ji}(\cancel{p})=\Lambda_{ji}\left(\cancel{p}-m_i\right)+\left(\cancel{p}-m_j\right)\bar{\Lambda}_{ji}.
\end{equation}
Here we simply made the replacement
\begin{equation}
 \Sigma_{ji}(\cancel{p})\rightarrow\left(\cancel{p}-m_i\right)\delta_{ji},
\end{equation}
because $\Lambda$'s vanish at tree-level. At 1-loop it is also obvious that gauge dependence is described by $\Lambda$'s as $\cancel{p}$ and $m$ are gauge-independent.

Since the derivative of the self-energy $\partial_\xi \Sigma_{ji}(\cancel{p})$ and $\Lambda$'s have Dirac structure, we may decompose them just like the self-energy in eq.~(\ref{eq:SE_deco}). It is then a matter of trivial rearrangements and collection of terms to arrive at
\begin{equation}
    \begin{split}
        \partial_\xi \Sigma^{L}_{ji}(p^2)&=-m_i \Lambda^{L}_{ji}-m_j \bar{\Lambda}^{L}_{ji}+\Lambda^{sR}_{ji}+\bar{\Lambda}^{sL}_{ji},\\
        \partial_\xi \Sigma^{R}_{ji}(p^2)&=-m_i \Lambda^{R}_{ji}-m_j \bar{\Lambda}^{R}_{ji}+\Lambda^{sL}_{ji}+\bar{\Lambda}^{sR}_{ji},\\
        \partial_\xi \Sigma^{sL}_{ji}(p^2)&=p^2\Lambda^{R}_{ji}+p^2\bar{\Lambda}^{L}_{ji}-m_i\Lambda^{sL}_{ji}-m_j\bar{\Lambda}^{sL}_{ji},\\
        \partial_\xi \Sigma^{sR}_{ji}(p^2)&=p^2\Lambda^{L}_{ji}+p^2\bar{\Lambda}^{R}_{ji}-m_i\Lambda^{sR}_{ji}-m_j\bar{\Lambda}^{sR}_{ji}.
    \end{split}
\end{equation}
As the result is general, analogous equations hold for any gauge parameter.

Let us apply the above to our expression containing the left-handed anti-hermitian part of field renormalization and mass counterterms in eq.~(\ref{eq:dZL_mass}). We arrive at
\begin{equation}\label{eq:gauge_dep}
\begin{split}
   &(m_i^2-m_j^2)\partial_\xi(\delta Z^{A}_{Lji})-2m_j\partial_\xi\delta m^L_{ji}-2m_i\partial_\xi\delta m^R_{ji}=\\
   &(m_i^2-m_j^2)\left(-m^{}_i\bar{\Lambda}^R_{ji}(m^2_i)-\bar{\Lambda}^{sL}_{ji}(m^2_i)\right)+H.C.
\end{split}
\end{equation}
Here $i\neq j$. We can obviously see that all of the gauge dependence on the r.h.s. is always multiplied by ${m_i^2-m_j^2}$, we will use this when defining counterterms. 
The above equation also shows the cancellation of the ${m_i^2-m_j^2}$ factor in the gauge-dependent part of the field renormalization as noticed in~\cite{Yamada2001}, however, the authors there did not have off-diagonal mass counterterms and could not use this fact to lay foundations for a definition. 

While all of the gauge dependence has a factor of ${m_i^2-m_j^2}$, one should be careful as there might also be gauge-independent contributions with the same factor.

\subsection{UV Divergences}\label{sec:UV_div}
As we are working with a 1-loop definition, we might as well investigate the UV divergences at 1-loop. To do so, let us first notice the pseudo-hermiticity property of the self-energy
\begin{equation}
    \Sigma(\cancel{p})=\gamma^0(\Sigma(\cancel{p}))^\dagger\gamma^0.
\end{equation}
Here hermitian conjugation acts both on flavour and Dirac structures. Pseudo-hermiticity simply relates the scalar self-energy functions
\begin{equation}
\begin{split}
    (\Sigma^{L,R}(p^2))^\dagger&=\Sigma^{L,R}(p^2),\\
    (\Sigma^{sL}(p^2))^\dagger&=\Sigma^{sR}(p^2).
\end{split}
\end{equation}
However, this property fully holds only below particle production thresholds and is spoiled by the absorptive parts. Since in this section we are interested in UV divergences, which do not contain the absorptive parts, pseudo-hermiticity holds for our purposes and we can write down the divergent part of eq.~(\ref{eq:dZL_mass}) in the following way
\begin{equation}\label{eq:divergent_part}
    \begin{split}
        \left[(m^2_i-m^2_j)\delta Z^A_{Lji}-2m_j\delta m^L_{ji}-2m_i\delta m^R_{ji}\right]_\mathrm{div.}=&
        -\left[m^2_i\Sigma^{L}_{ji}(m^2_i)+m^{}_i m^{}_j\Sigma^{R}_{ji}(m^2_i)\right.\\
    &+m^{}_j\Sigma^{sL}_{ji}(m^2_i)+m^{}_i\Sigma^{sR}_{ji}(m^2_i)\\
    &+m^2_j\Sigma^{L}_{ji}(m^2_j)+m^{}_i m^{}_j\Sigma^{R}_{ji}(m^2_j)\\
    &\left.+m^{}_i\Sigma^{sR}_{ji}(m^2_j)+m^{}_j\Sigma^{sL}_{ji}(m^2_j)\right]_\mathrm{div.}
    \end{split}
\end{equation}
Here $i\neq j$ and we have used pseudo-hermiticity. 

Now, whether a scalar or a vector contributes to the fermion self-energy there is barely a difference as one still gets the same Passarino-Veltman function~\cite{Passarino1979} contributions to the self-energies. In that regard we shall consider some boson contributing to fermion self-energy at 1-loop, be it a scalar or a vector. The contributions of a boson are as follows~\cite{Staub2010}
\begin{equation}
    \begin{split}
        \Sigma^{L}(p^2)&=f_L B_1(p^2, m^2_{\psi\mathrm{ loop}}, m^2_\mathrm{bos.}),\\
        \Sigma^{R}(p^2)&=f_R B_1(p^2, m^2_{\psi\mathrm{ loop}}, m^2_\mathrm{bos.}),\\
        \Sigma^{sL}(p^2)&=m_{\psi\mathrm{ loop}}f_{s} B_0(p^2, m^2_{\psi\mathrm{ loop}}, m^2_\mathrm{bos.}),\\
        \Sigma^{sR}(p^2)&=m_{\psi\mathrm{ loop}}f^\dagger_{s} B_0(p^2, m^2_{\psi\mathrm{ loop}}, m^2_\mathrm{bos.}).
    \end{split}
\end{equation}
Here $m_\mathrm{bos.}$ stands for the mass the boson in the loop and $m_{\psi\mathrm{ loop}}$ for the mass of the fermion in the loop, which is not necessarily the same as that of external fermions (e.g. up-type quarks may have down-type quarks in the loop). $B_1$ and $B_0$ are Passarino-Veltman functions. The constants $f_{L,R,s}$ include all the necessary symmetry factors and couplings of the theory, $f_{L}$ and $f_{R}$ are also hermitian as needed by pseudo-hermiticity. 
The divergent parts of Passarino-Veltman functions are well-known
\begin{equation}\label{eq:B_div}
\begin{split}
    \left[B_1(p^2, m^2_1, m^2_2)\right]_\mathrm{div.}&=\frac{1}{D-4},\\
    \left[B_0(p^2, m^2_1, m^2_2)\right]_\mathrm{div.}&=-\frac{2}{D-4}.\\
\end{split}
\end{equation}
Here $D$ is the spacetime dimension be it $4-\epsilon$ or $4-2\epsilon$ depending on the preferred regularization.
Now we can put everything into eq.~(\ref{eq:divergent_part}) and also write down the right-handed part
\begin{equation}\label{eq:explicit_div1}
    \begin{split}
        \left[(m^2_i-m^2_j)\delta Z^A_{Lji}-2m_j\delta m^L_{ji}-2m_i\delta m^R_{ji}\right]_\mathrm{div.}=&
        \frac{1}{D-4}\Big(-f_L(m^2_i+m^2_j)-f_R 2 m_i m_j\\
        &\qquad+4f^\dagger_s m_i m_{\psi\mathrm{ loop}}+4f_s m_j m_{\psi\mathrm{ loop}}\Big),
    \end{split}
\end{equation}
\begin{equation}\label{eq:explicit_div2}
    \begin{split}
        \left[(m^2_i-m^2_j)\delta Z^A_{Rji}-2m_i\delta m^L_{ji}-2m_j\delta m^R_{ji}\right]_\mathrm{div.}=&
        \frac{1}{D-4}Big(-f_R(m^2_i+m^2_j)-f_L 2 m_i m_j\\
        &\qquad+4f^\dagger_s m_j m_{\psi\mathrm{ loop}}+4f_s m_i m_{\psi\mathrm{ loop}}\Big).
    \end{split}
\end{equation}
Here $i\neq j$. There is a number of features to be discussed. First, we notice that the right-hand sides of the above expressions do not contain the ${m^2_i-m^2_j}$ factor and there is no way to arrange the couplings such that ${m^2_i-m^2_j}$ appears. On the other hand, if we were to look at UV divergences of the hermitian part of field renormalization, we would find only ${m^2_i-m^2_j}$ mass structure.

Another interesting feature is that we see more than just $m_i$ and $m_j$, but also $2m_i m_j$ and ${m^2_i+m^2_j}$. One might argue that we are missing something and that there should be corresponding counterterms on the l.h.s. However, it is fairly simple to check that if the mass is renormalized as
\begin{equation}
    m_0\rightarrow \delta m^{L}P_L + \delta m^{R} P_R+
    \left(1+\delta m^+ P_L+\delta m^- P_R\right)m\left(1+\delta m^+ P_R+\delta m^- P_L\right),
\end{equation}
the $2m_i m_j$ and ${m^2_i+m^2_j}$ mass structures appear along the corresponding mass counterterms $\delta m^\pm$. This kind of mass renormalization might be convenient for MS schemes as it is then easy to associate the mass structures with counterterms. However, in the On-Shell scheme, finite parts also matter and it is fairly complicated or impractical to try and associate all the mass counterterms with corresponding mass structures, therefore we only keep $\delta m^{L,R}$ for the mass counterterms. 

It is not hard to check that identical results hold if tadpole diagrams are included as well. Tadpoles contribute to $\Sigma^{sL,sR}$, which are multiplied by $m_i$ or $m_j$ in eq.~(\ref{eq:divergent_part}). Now, scalars and vectors in the loop can only contribute the corresponding scalar and vector masses, which does not change the above results. Tadpoles with fermion loops are also possible, so let us look at their contributions, in general~\cite{Staub2010}
\begin{equation}
\begin{split}
    \Sigma^{sL}&=f_T m_{\psi\mathrm{loop}} A_0(m^2_{\psi\mathrm{loop}}),\\
    \Sigma^{sR}&=f^\dagger_T m_{\psi\mathrm{loop}} A_0(m^2_{\psi\mathrm{loop}}).
\end{split}
\end{equation}
Here $f_T$ is some non-chiral coupling and $A_0$ is a Passarino-Veltman function with the divergence
\begin{equation}
    [A_0(m^2)]_\mathrm{div.}=-\frac{2m^2}{D-4}. 
\end{equation}
As a result, fermion tadpoles contribute mass of the fermion in the loop to the third power. We can insert our results into eq.~(\ref{eq:divergent_part})
\begin{equation}
    \left[(m^2_i-m^2_j)\delta Z^A_{Lji}-2m_j\delta m^L_{ji}-2m_i\delta m^R_{ji}\right]_\mathrm{div.}=
    \frac{4m^3_{\psi\mathrm{loop}}}{D-4}\left(m_j f_T+m_i f^\dagger_T\right).
\end{equation}
Here we see that it is also impossible to arrive at the mass factor ${m^2_i-m^2_j}$ (in the case of anti-hermitian $f_T$ the squares are still missing) and the above divergences can be neatly accounted for by the mass counterterms.

Combining the results of this and the previous section, it is clear that there are no UV divergent terms with the mass structure ${m_i^2-m_j^2}$ and, in addition, gauge-dependence always comes with a factor of ${m_i^2-m_j^2}$ in the expressions we discussed. However, gauge-dependence is a feature of the model and ${m_i^2-m_j^2}$ is only seen after the contributions from all particles have been considered. On the other hand, UV divergences with corresponding mass factors arise per particle species.

\subsection{The Definition of Mass and Field Counterterms}\label{sec:mas_field_def}
In the previous sections we have shown a number of properties regarding the mass structures, let us then use them to define the off-diagonal field and mass counterterms. For the field counterterms we simply take the coefficient of ${m^2_i-m^2_j}$ in eq.~(\ref{eq:dZL_mass}) and in an analogous equation for the right-handed part\footnote{Here $((m_i^2-m_j^2)A+B)\bigg|_{m_i^2-m_j^2}=A$}
    \begin{equation}\label{eq:dZL_def}
        \begin{split}
            \delta Z^A_{Lji}&\equiv\left[-\left(m^2_i\Sigma^{L}_{ji}(m^2_i)+m^{}_i m^{}_j\Sigma^{R}_{ji}(m^2_i)+m^{}_j\Sigma^{sL}_{ji}(m^2_i)+m^{}_i\Sigma^{sR}_{ji}(m^2_i)\right)+H.C.\right]\Big\vert_{(m^2_i-m^2_j)},\\
            \delta Z^A_{Rji}&\equiv\left[-\left(m^2_i\Sigma^{R}_{ji}(m^2_i)+m^{}_im^{}_j\Sigma^{L}_{ji}(m^2_i)+m^{}_j\Sigma^{sR}_{ji}(m^2_i)+m^{}_i\Sigma^{sL}_{ji}(m^2_i)\right)+H.C.\right]\Big\vert_{(m^2_i-m^2_j)},            
        \end{split}
    \end{equation}
in both expressions $i\neq j$. As this is model independent, it is really needed to evaluate the above expressions and then pick the needed terms, however, it is not too hard to pick out the difference of squared masses. 

For the mass counterterms we could try to evaluate the expressions and then also pick the coefficients of corresponding mass structures, however, once the field renormalization is defined and computed, we have two equations (eq.~(\ref{eq:dZL_mass}) and an equation with $\delta Z^A_R$) with two unknowns and can simply solve for the mass counterterms in terms of the self-energies and the field renormalization. The solutions are
\begin{equation}
    \begin{split}
        \delta m^L_{ji}&=\frac{1}{2}\left(m_i\Sigma^{R}_{ji}(m^2_i)+\Sigma^{sL}_{ji} (m^2_i)+m_j\Sigma^{L\dagger}_{ji}(m^2_j)+\Sigma^{sR\dagger}_{ji}(m^2_j)\right)+\frac{1}{2}\left(m_i\delta Z^{A}_{Rji}-m_j\delta Z^A_{Lji}\right),\\
        \delta m^R_{ji}&=\frac{1}{2}\left(m_i\Sigma^{L}_{ji}(m^2_i)+\Sigma^{sR}_{ji} (m^2_i)+m_j\Sigma^{R\dagger}_{ji}(m^2_j)+\Sigma^{sL\dagger}_{ji}(m^2_j)\right)+\frac{1}{2}\left(m_i \delta Z^{A}_{Lji}-m_j\delta Z^A_{Rji}\right),
    \end{split}
\end{equation}
for $i\neq j$. One can clearly see that the hermiticity constraint in eq.~(\ref{eq:hermiticity}) trivially holds. As we are keeping the hermitian conjugation of self-energies, these expressions also hold above thresholds and incorporate the absorptive parts. In addition, since gauge-dependence is contained in the anti-hermitian part of the field renormalization, the mass counterterms are gauge-independent by definition:
\begin{equation}
    \begin{split}
        \partial_\xi \delta m^L_{ji}=0 \qquad\text{and}\qquad\partial_\xi \delta m^R_{ji}=0.
    \end{split}
\end{equation}
As noted in~\cite{Gambino2000} it seems impossible to have gauge-independent quantities in terms of self-energies in general, hence, there is nothing wrong with  having the mass counterterms shifted by the anti-hermitian parts of the field renormalization. Although, in some specific models it is sometimes possible to find gauge-independent quantities in terms of the self-energies as found for the SM in~\cite{Espriu2002}.

Let us comment on the expressions we just derived. First, we chose to keep the hermiticity of the Lagrangian and use the no-mixing condition only on incoming particles. To get to the case where the Lagrangian is non-hermitian one only needs to drop all the $\dagger$'s in our expressions and change $\delta Z^A_{L,R}\rightarrow \frac{1}{2}\left(\delta Z_{L,R}-\bar{Z}_{L,R} \right)$. Here $\bar{Z}$ renormalizes $\bar{\psi}$ and is not related to $Z$ by pseudo-hermiticity, i.e. $Z\neq \gamma^0 \bar{Z}^\dagger \gamma ^0 $. $\bar{Z}$ may be found by additionally applying the no-mixing condition on outgoing particles.

Second, since diagonal components of the anti-hermitian part of field renormalization cannot be fixed and are usually set to zero, expressions for the mass counterterms $\delta m^{L,R}$ also have the same real part for $i=j$ and that real part also corresponds to the one found in~\cite{Espriu2002}. In our scheme the imaginary parts of diagonal components of $\delta m^{L,R}$ are complex conjugate to each other and are not the same as in~\cite{Espriu2002} due to absorptive parts and hermitian conjugation. In other words, we may cheat a little and use the off-diagonal mass counterterm expressions for diagonal terms as well, however, we will use this extension to all $i$ and $j$ only for the real part. 

Third, in the standard approach factors of ${1/(m_i^2-m_j^2)}$ are evident due to the fact that there are mass structures ${m_i^2+m_j^2},\ {2m_im_j},\  m_i,\  m_j$ that do not simplify with ${1/(m_i^2-m_j^2)}$. Our approach distributes these structures between field and mass counterterms by definition, such that the degenerate mass limit is non-singular.

Fourth, we may ask whether it is at all needed to have off-diagonal mass counterterms. Let us consider the case of the SM CKM matrix renormalization in~\cite{Denner1990_CKMreno}. There the authors do not introduce off-diagonal mass counterterms and field renormalization is used to account for all the UV divergences appearing in the mass term. In turn, this means that divergences are propagated by field renormalization from the mass term to other terms in the Lagrangian. That is why the authors in~\cite{Denner1990_CKMreno} \textit{needed} to renormalize the CKM matrix on the basis of a UV divergent $W$ vertex, and that is why the anti-hermitian part of the field renormalization appeared in their approach. In terms of divergences, by introducing the CKM counterterm one simply goes back to the case where all of the UV divergences are contained in the mass term. However, this also moves finite parts, which then makes the CKM counterterm and, at least in part, physical amplitudes gauge-dependent. On the other hand, in our approach the anti-hermitian part of field renormalization is always finite and the mass counterterms account for the UV divergences in the mass term without propagating them to other terms in the Lagrangian. In turn, the $W$ vertex is finite and there is no need to renormalize the CKM mixing matrix on the basis of divergences. In addition, our approach does not spoil gauge properties of the CKM matrix and its counterterm, as the counterterm is trivially 0. We may conclude that it is needed to have off-diagonal mass counterterms alongside field renormalization to avoid propagation of UV divergences --- mass counterterms in eq.~(\ref{eq:dZL}) and eq.~(\ref{eq:dZR}) are not due to a trivial shift of the field renormalization. 

Fifth, it is important to note that while our definitions of renormalization constants do solve the problem of gauge-dependence of the CKM counterterm, it does not completely remove the gauge-dependence for physical amplitudes. This is because our field renormalization constants incorporate the Lehmann-Symanzik-Zimmermann (LSZ)~\cite{Lehmann1955} factors for incoming particles, but not for the outgoing ones, all of this is due to absorptive parts. For example, Section 6 of~\cite{Espriu2002} explicitly shows the need to have different LSZ factors for incoming and outgoing particles to ensure that no gauge-dependence appears in $Wu_id_j$ vertices. While this is a downside, other schemes also suffer from the same problem since it is impossible to have both LSZ factors accounted for by field renormalization without losing Hermiticity of the Lagrangian~\cite{Espriu2002,Denner2004}. However, this is can be fixed by including the LSZ factor for outgoing particles (incoming antiparticles).

Finally, it has been brought to our attention\footnote{Thanks to H. Rzehak.} that the end result is rather similar to the approach considered for sfermions in MSSM in~\cite{Baro2009}. There the authors define the mass counterterm as the residue of the ${1/(m_i^2-m_j^2)}$ pole and effectively solve for field renormalization, our approach is in that sense opposite --- defining the field renormalization as the coefficient of ${m_i^2-m_j^2}$ and solving for mass counterterms. The two approaches arrive at (almost) analogous conclusions in different ways, which is only further emphasised by the fact that sfermions are scalars and here we are dealing with fermions. However, this only shows that the problems and their solutions both for fermions and scalars are rather universal.

Although similar, the schemes also have important differences. One of the important distinctions is that if the logic of~\cite{Baro2009} is applied to fermions one does not arrive at a UV-finite anti-hermitian part of the field renormalization for all masses. This is because the masses $m_i$ and $m_j$ are effectively expressed in terms of the sum and difference of their squares. When rewriting a UV-divergent term proportional to some arbitrary function of masses (e.g. $2m_im_j$) in terms of the new variables one in general generates UV-divergent terms proportional to powers of the difference of squares. Such terms are then attributed to the anti-hermitian part of field renormalization and vanish only in the degenerate mass limit. In contrast, our approach has a finite anti-hermitian part of field renormalization for all masses, which is very important for the renormalization of mixing matrices as discussed in the paragraphs above and below.

Another difference between the schemes is that the one presented in this work can also include absorptive parts, while $\widetilde{Re}$ must be used in~\cite{Baro2009}. Then the readers might also notice that the symmetric part of field renormalization in~\cite{Baro2009} is non-singular in the degenerate mass limit, however, in our case the absorptive parts in the hermitian part of field renormalization spoil this non-singular behaviour at first sight. This is remedied by including the remaining LSZ factor for outgoing particles such that these singular absorptive parts cancel in practice. Note that singularities for the anti-hermitian part could not be removed in this way as non-absorptive parts would also be singular. Since in our approach the remaining LSZ factor contains only absorptive parts we must make use of the mass counterterms in this case. The takeaway is that there is a lot of freedom when distributing contributions between LSZ factors and field renormalization, however, for consistency some contributions must also go into off-diagonal mass counterterms.

\subsection{1-loop rotation}\label{sec:1loop_rotation}

In this section we present the so-called 1-loop rotation which deals with diagonalization of the full mass matrix and in turn defines counterterms of mixing matrices. We follow the approach in~\cite{Kniehl2006, Almasy2009}. For Dirac fermions we also think of a biunitary rotation that diagonalizes the complete mass matrix ${m+\delta m}$, or rather a rotation that rotates away off-diagonal contributions of $\delta m$ as $m$ is taken to be already diagonal
\begin{equation}
    -\bar{\psi}^{}_R \mathcal{U}^\dagger_R\left(m+\delta m^L\right)\mathcal{U}_L\psi_L+ (L\leftrightarrow R)
\end{equation}
with
\begin{equation}
    \mathcal{U}_{L,R}=1+ih_{L,R},
\end{equation}
where $\mathcal{U}_{L,R}$ are unitary matrices and $h_{L,R}$ are hermitian matrices of 1-loop order. We can solve for the $ih_{L,R}$ matrices by noticing that
\begin{equation}
\begin{split}
    &\mathcal{U}^\dagger_L\left(m+\delta m^R\right)\left(m+\delta m^L\right)\mathcal{U}_L=\mathcal{D}^2,\\
    &\mathcal{U}^\dagger_R\left(m+\delta m^L\right)\left(m+\delta m^R\right)\mathcal{U}_R=\mathcal{D}^2,
\end{split}
\end{equation}
where $\mathcal{D}$ is a diagonal matrix. The solution method is the same for both matrices, so let us choose to solve for $ih_L$. Inserting the definitions and writing everything up to 1-loop order we have
\begin{equation}
    m^2+\delta m^R m+m\delta m^L+(-ih_L)m^2+m^2(ih_L)=\mathcal{D}^2.
\end{equation}
Splitting the above equation into diagonal and off-diagonal contributions we have
\begin{equation}
\begin{split}
    m^2_{i}+\delta m^R_{ii} m_{i}+m_{i}\delta m^L_{ii}&=\mathcal{D}^2_{ii},\\
    \delta m^R_{ji} m_i+m_j\delta m^L_{ji}+(-ih_{Lji})m^2_i+m^2_j(ih_{Lji})&=0.
\end{split}
\end{equation}
Here $i\neq j$ in the second line. For diagonal terms, the $h_L$ contributions cancel, hence, they are arbitrary and can be safely be set to $0$. For off-diagonal terms, we can easily solve for $ih_{Lji}$
\begin{equation}\label{eq:ihL}
    ih_{Lji}=\frac{m_j\delta m^L_{ji}+\delta m^R_{ji} m_i}{m^2_i-m^2_j}.
\end{equation}
The $ih_{Rji}$ solution is analogous
\begin{equation}\label{eq:ihR}
    ih_{Rji}=\frac{m_j\delta m^R_{ji}+\delta m^L_{ji} m_i}{m^2_i-m^2_j}.
\end{equation}
As a quick reminder, $j$ marks the outgoing particle and $i$ marks the incoming particle and the above expressions are for $i\neq j$.

As noted in~\cite{Almasy2009}, for Majorana fermions a unitary rotation is enough (as opposed to a biunitary one for Dirac fermions), such that only $ih^{}_L$ remains. Whether the fermions are Dirac or Majorana, after the 1-loop rotation that diagonalizes the mass, the $ih_{L,R}$ matrices appear in vertices that also contain the mixing-matrices. For example, let us consider fermions $f$ and $l$ that mix in a chiral interaction with some mixing matrix\footnote{By a mixing matrix we do not mean a matrix coupling, but a matrix that appears due to diagonalization of the mass term.} $V$, vector $A_\mu$, and coupling $g$ (as quarks do in the SM)
\begin{equation}
    g\bar{f}^0_L\cancel{A}V_{fl}l^0_L.
\end{equation}
Here $0$'s indicate bare quantities and we do not bother with renormalization of $g$ or $A_\mu$ as it does not change our results. Since we are also considering field renormalization, let us renormalize the fields before the 1-loop rotation
\begin{align}
    \nonumber
    g\bar{f}^0_L\cancel{A}V_{fl}l^0_L&=g_0\bar{f}_LZ^{f1/2\dagger}_{Lff'}\cancel{A}V_{f'l'}Z^{l1/2}_{Ll'l}l_L\\
    \nonumber
    &=g\left(V_{fl}+\frac{\delta Z^{f\dagger}_{Lff'}V_{f'l}}{2}+\frac{V_{fl'}\delta Z^l_{Ll'l}}{2}\right)\bar{f}_L\cancel{A}l_L\\
    &=g\left(V^{}_{fl}+\delta V^Z_{fl}\right)\bar{f}_L\cancel{A}l_L.
\end{align}
Here we defined $\delta V^Z_{fl}$, which is a simple shorthand notation for the terms containing $V$ and $\delta Z$'s in the second line. Also here and below summation over $f'$ and $l'$ is implied. After the additional 1-loop rotation (hence the hats over fermions) we have
\begin{align}
    &g\left(V_{fl}+\delta V^Z_{fl}+(-ih^f_{Lff'})V_{f'l}+V_{fl'}(ih^l_{Ll'l})\right)\bar{\hat{f}}_L\cancel{A}\hat{l}_L
    =g\left(V_{fl}+\delta V^Z_{fl}+\delta V^h_{fl}\right)\bar{\hat{f}}_L\cancel{A}\hat{l}_L.
\end{align}
 Here we also introduced the shorthand notation $\delta V^h_{fl}$ for terms containing $V$ and $ih$'s. This suggestive shorthand notation is because we would like to identify $\delta V^{Z}$ and/or $\delta V^{h}$ with the mixing matrix counterterm $\delta V$. The counterterm $\delta V$ should satisfy some properties required by unitarity~\cite{Kniehl2006}
\begin{equation}
    (V+\delta V)^\dagger (V+\delta V)=\delta \Rightarrow V^\dagger\delta V=-\delta V^\dagger V.
\end{equation}
Note that the relation holds up to the order of computation and we have dropped the $\delta V^\dagger\delta V$ term. One can check that $\delta V^h_{fl}$ satisfies these unitarity constraints by simply noting the anti-hermiticity of $ih$'s. In the case of $\delta V^Z_{fl}$ this requirement does not fully hold since both hermitian and anti-hermitian contributions of $\delta Z$'s are included in $\delta V^Z_{fl}$, hence, the part that includes hermitian parts of $\delta Z$'s cannot be identified with the mixing matrix counterterm $\delta V$ due to unitarity. We are then left with
\begin{equation}
    \delta V^{Z,A}_{fl}=-\delta Z^{A,f}_{Lff'}V_{f'l}+V_{fl'}\delta Z^{A,l}_{Ll'l}.
\end{equation}
and $\delta V^h_{fl}$. Even though $\delta V^{Z,A}$ is allowed by unitarity it is gauge-dependent and also arises differently than the mixing matrix $V$. If $\delta V^{Z,A}$ is included in the mixing matrix counterterm, it breaks Ward-Takahashi Identities as noted for the SM CKM matrix in~\cite{Gambino1999}. Therefore, we do not identify (or include) $\delta V^{Z,A}$ with the mixing matrix counterterm. We are then only left with the only option of identifying $\delta V^h_{fl}$ with the mixing matrix counterterm
 \begin{equation}
    \delta V^{}_{fl}=\delta V^h_{fl}=V_{fl'}ih^l_{Ll'l}-ih^f_{Lff'}V_{f'l}.
 \end{equation}
 
It is also important to comment on the meaning of the 1-loop rotation. The 1-loop rotation tells that the effect of mass counterterm insertions onto external legs can also be achieved by an appropriate insertion of mixing matrix counterterms (equivalence of these insertion is shown in~\cite{Kniehl2006}). On the other hand, in terms of the Lagrangian the 1-loop rotation leaves the mass term unrenormalized. This is easy to see since the 1-loop rotation only regards the counterterms, but not the renormalized parameters. In turn, if we again compute self-energies in the 1-loop rotated basis, we get the same self-energies as before the 1-loop rotation, but the needed counterterms are now missing. By a similar argument, the anti-hermitian part of the field renormalization is a rotation, so one could rotate it away from the Lagrangian \textit{entirely} and leave the fields (partly) unrenormalized.
%Not necessary
%\textbf{In addition, one could speak of the 1-loop rotation in a picture containing the mass counterterms and the LSZ factors. Essentially, the 1-loop rotation redefines the LSZ factors by including the mass counterterms into them and then the same mass contribution also goes into the definition of the mixing matrix counterterm. Being the same, these contributions cancel in amplitudes such as the $W$ decay, while the mass counterterms remain in the self-energy in the form of LSZ factors. Effectively, this the same as having only mass counterterms and no mixing matrix counterterms and the 1-loop rotation is redundant. Even more so, in intermediate steps, the 1-loop rotation introduces singular behaviour in the degenerate mass limit for the mixing matrix counterterms.}

Further, if we compare the expression for $ih_L$ in eq.~(\ref{eq:ihL}) with eq.~(\ref{eq:dZL_mass}) divided by $m^2_i-m^2_j$, it is not hard to see that in our formulation the l.h.s. of eq.~(\ref{eq:dZL_mass}) is $\delta Z^A_{Lji}-2ih_{ji}$. Importantly, the 1-loop rotation is also related only to the anti-hermitian part of field renormalization. It is then no surprise, that the authors of~\cite{Denner1990_CKMreno} and~\cite{Kniehl2006} got the same divergent parts when renormalizing the CKM matrix in their respective approaches. This also reiterates on our propagation of UV divergences argument from Section~\ref{sec:mas_field_def}.

Finally, since the 1-loop rotation means various inconsistencies, we discuss whether it is consistent to renormalize mixing matrices. There are two scenarios:
\begin{enumerate}
\item Mass is diagonalized and then the theory is renormalized. This is the usual approach, where masses are given diagonal counterterms and mixing matrices are renormalized. Schematically:
    \begin{equation}
    \begin{matrix}
      	m^0_{ji}  & \xrightarrow{\mathrm{diag.}} &  m^0_{i} & \xrightarrow{\mathrm{renorm.}}  &m_i+\delta \tilde{m}_i\\
       \cancel{V^0} & \xrightarrow{\mathrm{diag.}} &  V^0 &  \xrightarrow{\mathrm{renorm.}} & V+\delta V.
    \end{matrix}
    \end{equation}
\item The theory is first renormalized and only then the renormalized mass is diagonalized. This means non-diagonal mass counterterms and no counterterms for mixing matrices, since mixing matrices were not present during renormalization. Schematically:
    \begin{equation}
    \begin{matrix}
      	m^0_{ji} & \xrightarrow{\mathrm{renorm.}} & m_{ji}+\delta m_{ji} &  \xrightarrow{\mathrm{diag.}} & m_i+(V^\dagger \delta m V)_{ji} \\
       \cancel{V^0}& \xrightarrow{\mathrm{renorm.}}  & \cancel{V+\delta V} &  \xrightarrow{\mathrm{diag.}} & V\hspace{0.25cm}\cancel{+\delta V} .
    \end{matrix}
    \end{equation}
\end{enumerate}

Both scenarios lead to the same renormalized Lagrangian and only differ in their counterterms. One should be able to reach the second step of the second scenario from the final step of the first one via a finite basis rotation. However, such a rotation gives
\begin{equation}
    \begin{matrix}
      m_i+\delta \tilde{m}_i & \xrightarrow{\mathrm{rot.}} & m_{ji}+(V \delta \tilde{m} V^{\dagger})_{ji}&\neq & m_{ji}+\delta m_{ji}\\
      V+\delta V & \xrightarrow{\mathrm{rot.}} & \cancel{V+}\hspace{0.25cm}\delta V'&\neq & \cancel{V+\delta V}
    \end{matrix}
\end{equation}
with some rotated counterterm $\delta V'$ without an associated parameter. This connection is possible only if the mixing matrix counterterm is 0 and the mass counterterms are non-diagonal. In turn this means that mixing matrices cannot have counterterms and mass matrices must have off-diagonal contributions in all bases --- the bare mass matrix can be diagonal only up to its counterterms. Again, an infinite basis rotation is not suitable because it simply unrenormalizes the self-energy by removing the off-diagonal mass counterterms without changing the field renormalization in any way. The consistency condition is that renormalization and (finite) basis rotations must commute. In turn, this means that only the renormalized part of the mass matrix can be diagonalized and there can be no counterterms for mixing matrices.

It has also been brought to our attention\footnote{Again thanks to H. Rzehak.} that in models with extended scalar sectors authors sometimes also choose to not renormalize mixing angles (i.e. renormalize before diagonalization)~\cite{Fritzsche2002,Baro2009,Chatterjee2011,Fritzsche2012,Altenkamp2017}. However, such an approach is presented as an alternative to the one where the renormalization takes place after diagonalization, while the point made in the previous paragraph regards the usual renormalization-diagonalization approach as inconsistent. Although inconsistent in the above sense and with problems coming due to this inconsistency (e.g., singular behaviour), the usual approach can still be useful.

Nonetheless, mixing matrix counterterms for external legs from the 1-loop rotation allow us to compare the divergent parts of our scheme with other results in the literature and we will present them in upcoming sections. Apart for the reason of comparison, we consider the 1-loop rotation inconsistent in the above sense.

\section{Examples of the proposed scheme}\label{sec:big_examples}
Before proceeding to the introduction of the Grimus-Neufeld model and presentation of examples, we would like to mention the software used in the following computations. For the Grimus-Neufeld model we adapted the Two Higgs Doublet Model (THDM)~\cite{Lee1973} files~\cite{Degrande2015}\footnote{Version 1.2 of the THDM model files contains a bug in the ghost Lagrangian. Comparing that Lagrangian with eq.~(21.52) in Peskin\&Schroeder~\cite{Peskin1995} we see that the gauge parameter is missing.} for \texttt{FeynRules-2.3.41}~\cite{Alloul2014}. Using \texttt{FeynRules} we generated model files for \texttt{FeynArts-3.11}~\cite{Hahn2001} and performed the computations using \texttt{FeynCalc-9.3.1}~\cite{Mertig1991,Shtabovenko2016,Shtabovenko2020} in combination with \texttt{PackageX-2.1.1}~\cite{Patel2015}. Naturally, the convetions for Passarino-Veltman functions follow those used in the packages.

\subsection{Introduction to the Grimus-Neufeld model}

To present our scheme we use the Grimus-Neufeld model, which is the Standard Model extended with an additional Higgs doublet and a single heavy Majorana neutrino. Parts of this introduction can also be found in~\cite{Grimus1989,Dudenas2019,Dudenas2019a}.

In the Higgs sector we have the most general Higgs potential for the THDM but in the $CP$-conserving case where the parameters $\mu_{12}, \lambda_5, \lambda_6, \lambda_7$ are real~\cite{Ferreira2009,Branco2011}
\begin{align}   
    \nonumber
    -V_\mathrm{THDM}=&\mu_1 \left(H^\dagger_1 H_1\right)+\mu_2 \left(H^\dagger_2 H_2\right)
    +\left[\mu_{12}\left(H^\dagger_1 H_2\right)+H.C.\right]\\
    \nonumber
    &+\lambda_1\left(H^\dagger_1 H_1\right)\left(H^\dagger_1 H_1\right)+\lambda_2\left(H^\dagger_2 H_2\right)\left(H^\dagger_2 H_2\right)\\
    \nonumber
    &+\lambda_3\left(H^\dagger_1 H_1\right)\left(H^\dagger_2 H_2\right)+\lambda_4\left(H^\dagger_1 H_2\right)\left(H^\dagger_2 H_1\right)\\
    \nonumber
    &+\left[\lambda_5\left(H^\dagger_1 H_2\right)\left(H^\dagger_1 H_2\right)+\lambda_6\left(H^\dagger_1 H_1\right)\left(H^\dagger_1 H_2\right)\right.\\
    &\hspace{0.5cm}\left.+\lambda_7\left(H^\dagger_2 H_2\right)\left(H^\dagger_1 H_2\right)+H.C.\right].
\end{align}
We use the Higgs basis, meaning that the vacuum expectation value (VEV) $v$ as well as all of the Goldstone bosons $G^{0,\pm}$ are in the first Higgs doublet. In the upcoming computations we will be using the mass eigenstate basis and will express the couplings $\lambda_j$ in terms of masses where possible. The couplings $\lambda_3$ and $\lambda_7$ are among the ones undefined by going to the mass eigenstates and will appear in various expressions. Upon going to the mass eigenstates we get the usual Higgs $h$ with the mass $m_h$ and because of the second Higgs doublet there is also a scalar $H$, a pseudo scalar $A$, and a charged scalar $H^{\pm}$. The corresponding masses will be denoted as $m_H$, $m_A$, and $m_{H^\pm}$. 

The Higgs sector contains mixing. We chose to work with the CP-conserving case so that there is no mixing between CP-odd and CP-even scalars, however, the two CP-even scalars ($h$ and $H$) still mix. We denote this mixing angle as $\alpha$ and in upcoming examples we use the following shorthand notation:
\begin{equation}
    s_x\equiv \sin x \qquad\text{and}\qquad c_x\equiv \cos x,
\end{equation}
where $x$ is some argument of sines and cosines.

We also have the interaction between Higgs doublets and the fermions
\begin{equation}
    \begin{split}
        -\mathcal{L}_\mathrm{H-F}=&(Y^{}_{d})^{}_{ji}\bar{d}_{Rj}.(Q_i H^\dagger_1)+
        (Y^{}_{u})^{}_{ji}\bar{u}_{Rj}.(Q_i \tilde{H}_1)\\
        &+(Y^{}_{l})^{}_{ji}\bar{e}_{Rj}.(L_i H^\dagger_1)+(Y^{}_{\nu})^{}_{i}\bar{N}.(L_i \tilde{H}_1)\\
        &+(G^{}_{d})^{}_{ji}\bar{d}_{Rj}.(Q_i H^\dagger_2)+
        (G^{}_{u})^{}_{ji}\bar{u}_{Rj}.(Q_i \tilde{H}_2)\\
        &+(G^{}_{l})^{}_{ji}\bar{e}_{Rj}.(L_i H^\dagger_2)+(G^{}_{\nu})^{}_{i}\bar{N}.(L_i \tilde{H}_2)+H.C.
    \end{split}
\end{equation}
Here $N$ is the Majorana neutrino that couples as a right-handed particle, $L$ and $Q$ are the left-handed lepton and quark doublets, respectively, $u_R$, $d_R$ and $e_R$ are the right-handed up-quark, down-quark, and charged lepton singlet fields, $\tilde{H}_{1,2}$ correspond to $SU(2)$ conjugated Higgs doublets $\epsilon_{ji} (H_{1,2})_i$, $Y$'s and $G$'s are the Yukawa couplings to the first and second Higgs doublets. The dots emphasize the contraction of fermion indices, while $j,i$ are family indices and parentheses imply the contraction of $SU(2)$ indices.

We also add a Majorana mass term for the heavy Majorana singlet $N$
\begin{equation}
    \mathcal{L}_\mathrm{Maj.}=-\frac{1}{2}M_R\bar{N}.N+H.C.
\end{equation}

For convenience, let us take the quarks and charged leptons to already be in the mass eigenstate basis, i.e. $(Y_{u,d,l})_{ji}=\frac{\sqrt{2}}{v} m^{u,d,l}_j\delta_{ji}$. Then for the neutrinos at tree level we have a non-diagonal symmetric mass matrix
\begin{equation}
    M^\nu=\begin{pmatrix}
    0_{3\times 3} & \frac{v}{\sqrt{2}} Y_\nu \\
    \frac{v}{\sqrt{2}} Y^T_\nu & M_R 
    \end{pmatrix},
\end{equation}
which may be diagonalized via a unitary transformation
\begin{equation}
    U^TM^\nu U=\tilde{m}^\nu.
\end{equation}
Here $\tilde{m}$ is a diagonal matrix. For convenience, we decompose the neutrino mixing matrix as 
\begin{equation}
    U=\begin{pmatrix}
    U^\dagger_L & U^T_R
    \end{pmatrix},
\end{equation}
where $U_L$ is $3\times 4$ and $U_R$ is $1 \times 4$ matrix. The matrices $U_L$ and $U_R$ also satisfy a few constraints coming from unitarity and the form of the neutrino mass matrix
\begin{equation}
    \begin{matrix}
        U^{}_L U^\dagger_L=\mathbb{1}_{3\times3}, \qquad & U^\star_R U^T_R=1,\qquad & U^\dagger_L U^{}_L+U^T_R U^\star_R=\mathbb{1}_{4\times4},\\
        & &\\
        U^\star_R U^{\dagger}_L=0, & U^{}_L U^T_R=0, & U^\star_L \tilde{m} U^\dagger_L=0.
    \end{matrix}
\end{equation} 
There is another important relation due to the form of the mass matrix $M^\nu$. Since $M^\nu$ is a matrix of rank two, there are only two non-zero eigenvalues, such that $\tilde{m}=\mathrm{diag}(0,0,m_3,m_4)$. Having this it is not hard to find from 
\begin{equation}
    M^\nu U=U^\star \tilde{m}
\end{equation}
that $U_R$ is of the form
\begin{equation}
    U_R=\begin{pmatrix}
    0 & 0 & u_{R3} & u_{R4}
    \end{pmatrix}.
\end{equation}
The two zeroes prove to be useful, for example,
\begin{equation}\label{eq:massless_uni}
    (U^\dagger_L U^{}_L)_{ji}+\cancelto{0}{(U^T_R U^\star_R)_{ji}}\hspace{0.4cm}=\delta_{ji}
\end{equation}
if $j$ and/or $i$ correspond to a massless neutrino (i.e, $i$ or $j$ equal to 1 or 2). 

In addition, at tree-level the matrix $U^\dagger_L$ is not fully determined and the first two rows of $U^\dagger_L$ are constrained only by unitarity. Of course, the first two rows correspond to the two massless neutrinos, which are indistinguishable at tree level. We shall use this remaining freedom at 1-loop level.

Beyond tree level we must add appropriate counterterms as advertised in previous sections. However, in the Grimus-Neufeld model we have massless neutrinos at tree level and it is not straightforward to pick out the mass structures as required by our definitions. To make the situation clearer let us schematically single out a few cases corresponding to $2\times2$ blocks in the 1-loop mass matrix
\begin{equation}
    \tilde{m}_{ji}+\delta m_{ji}\sim
        \left(\begin{array}{c|c}
            \delta m,m_{j,i}=0 & h,\delta Z,\delta m,m_{i}\neq0\\
            \hline 
            h,\delta Z,\delta m,m_{j}\neq0 & h,\delta Z,\delta m,m_{j,i}\neq0
        \end{array}\right).
\end{equation}
Here $h$ is a shorthand notation for the 1-loop rotation matrices $ih_{L,R}$ and is not to be confused with the Higgs, also $\delta m$ and $\delta Z$ serve as shorthand notations for left- and right-handed mass and anti-hermitian part of field counterterms. In the bottom-right block, where $j,i>2$ and the corresponding neutrinos are massive, our definitions for mass and field counterterms work out of the box. In this block it is also possible to perform the 1-loop rotation. 

The off-diagonal $2\times2$ blocks, where $j<3$ and $i>2$ or vice versa, correspond to the cases where either the incoming (indexed by $i$) or outgoing (indexed by $j$) neutrino is massless, in this case there is no way to rely on the $m^2_i-m^2_j$ mass structure as needed in the definition of anti-hermitian part of field counterterms. However, for the off-diagonal blocks we compute in the case where both neutrinos are massive and afterwards simply take the massless limit for one of the neutrinos (incoming or outgoing). In this way, we do not introduce gauge-dependence in the mass counterterm or divergences in the anti-hermitian part of the field renormalization. Hence, the counterterms still satisfy all the properties we found or required in previous sections. However, by taking the massless limit the expressions are no longer symmetric and, for example, the anti-hermiticity of $\delta Z^A_{L,R}$ is not obvious without a computation.

In the top-left block, where $j,i<3$, just as for the off-diagonal blocks we can take the massless limit of the case where both particles are massive. However, taking this limit is especially easy as we do not have to compute self-energies in the massive case and take the limit of the result, but we can take the limit directly in the definition, so for this block we have
\begin{equation}
    \begin{split}
        \delta m^L_{ji}=&\frac{1}{2}\left(\Sigma^{sL}_{ji}(0)+\Sigma^{sR\dagger}_{ji}(0)\right)=\Sigma^{sL}_{ji}(0)
    \end{split}
\end{equation}
and
\begin{equation}
    \begin{split}
        \delta m^R_{ji}=&\frac{1}{2}\left(\Sigma^{sR}_{ji} (0)+\Sigma^{sL\dagger}_{ji}(0)\right)=\Sigma^{sR}_{ji} (0).
    \end{split}
\end{equation}
This block is special in the sense that there are no field renormalization contributions and it is impossible to rotate away the off-diagonal contributions of $\delta m$ via a 1-loop rotation since $m_1=m_2=0$. Nonetheless, this block is gauge-independent and also finite as all of the divergences come with mass structures as can be seen in eq.~(\ref{eq:explicit_div1}) or eq.~(\ref{eq:explicit_div2}). In addition, this finite and gauge-independent block may be diagonalized by choosing appropriate $U_L$ components that were undefined at tree-level. Having all this, we can regard this block as the radiative mass instead of a counterterm (there is nothing to "counter" in the massless block in the bare Lagrangian), this has been noted in~\cite{Grimus1989} as well. In addition, this block, as well as the off-diagonal ones, benefit from the form of $U_R$ and the resulting expressions are fairly tame.

Before moving to examples, we note that the case of neutrinos in the Grimus-Neufeld model provides plenty of scenarios. We shall begin with examples in the quark sector and then move to the neutrino sector.

\subsection{Quarks}
Let us start with the anti-hermitian part of field renormalization. In the case of quarks, the expressions are quite tame and it is even possible to express the anti-hermitian parts in terms of Passarino-Veltman functions. To better illustrate our scheme, let us write down the r.h.s. of eq.~(\ref{eq:dZL_mass}) for up quarks but only take the Standard Model contributions, i.e. no coupling to the second Higgs doublet
\begin{equation}
    \begin{split}
        (\mathrm{r.h.s.})_\mathrm{SM}=&-\frac{V^{}_{jk}V^\star_{ik}}{2^D\pi^{D-2}v^2}\Bigg[B_0((m^u_i)^2,(m^d_k)^2,m^2_W)\Big[(m^d_k)^2((D-3)m^2_W-2(m^u_i)^2)+(m^d_k)^4\\
        &\hspace{6cm}+((m^u_i)^2-m^2_W)((D-2)m^2_W+(m^u_i)^2)\Big]\\
        &+A_0((m^d_k)^2)\left[(m^d_k)^2+(D-2)m^2_W+(m^u_j)^2\right]\\
        &-A_0(m^2_W)\left[(m^d_k)^2+(D-2)m^2_W-(m^u_i)^2\right]\\
        &+A_0(\xi_W m^2_W)((m^u_i)^2+(m^u_j)^2)\\
        &-((m^u_i)^2-(m^u_j)^2)B_0((m^u_i)^2,(m^d_k)^2, \xi_W m^2_W)\left[(m^d_k)^2-(m^u_i)^2+\xi_W m^2_W\right]\Bigg]\\
        &+H.C.
    \end{split}
\end{equation}
Here summation over $k$ is implied and $V$ is the CKM matrix. Note that here $i\neq j$ and we dropped terms with $\delta_{ij}$. Now, to pick out the left-handed anti-hermitian part of the field renormalization for up quarks $\delta Z^{A,u}_{Lji}$ we should pick out terms with ${(m^u_i)^2-(m^u_j)^2}$ mass structures, which is straightforward in this case. However, we note that the line before the last one contains gauge dependence with the mass structure ${(m^u_i)^2+(m^u_j)^2}$, which is not allowed by Nielsen identities as we have shown in Section~\ref{sec:nielsen}. This is easily resolved since only the CKM matrices contain summation over $k$, hence, the problematic term vanishes due to unitarity of the CKM matrix and because $i\neq j$ in the above expression. The only terms contributing to $\delta Z^{A,u}_{Lji}$ are in the last line, in turn we arrive at the following expression
\begin{equation}\label{eq:quark_dZL}
    \begin{split}
        \delta Z^{A,u}_{Lji}=&-\frac{V^{}_{jk}V^\star_{ik}}{2^D\pi^{D-2}v^2}\left[(m^d_k)^2-(m^u_i)^2+\xi_W m^2_W\right]B_0((m^u_i)^2,(m^d_k)^2, \xi_W m^2_W)\\
        &+\frac{V^{}_{jk}V^\star_{ik}}{2^D\pi^{D-2}v^2}\left[(m^d_k)^2-(m^u_j)^2+\xi_W m^2_W\right]B_0((m^u_j)^2,(m^d_k)^2, \xi_W m^2_W)^\star
    \end{split}
\end{equation}
An analogous expression for the down quarks can be acquired by $u \leftrightarrow d$ and complex conjugation of $V$. Note that the second Passarino-Veltman function is complex conjugated because of our approach, this has no effect if the corresponding function is real. Using eq.~(\ref{eq:B_div}) and unitarity of the CKM matrix it is easy to check that the above expression is finite. The right-handed anti-hermitian part of the field renormalization is even more finite as it is equal to 0 for quarks. Even though we picked terms from the SM expression, the anti-hermitian part does not change when the full Grimus-Neufeld model is considered. 

The left-handed mass counterterm for the up-type quarks $\delta m^{u,L}_{ji}$ (in the full Grimus-Neufeld model) is expressible in terms of Passarino-Veltman functions, but is rather unwieldy and is given in Appendix~\ref{seca:full_exp}.

 To complete the quark example, we want to arrive at the divergent parts of the CKM matrix counterterms and compare with the SM results in~\cite{Denner1990_CKMreno, Kniehl2006}. With that in mind, let us first list only the divergent part of $\delta m^{L,u}$ and $\delta m^{R,u}$ in case of the SM (i.e. dropping couplings to the second Higgs doublet)
 
\begin{equation}
\begin{split}
    [\delta m^{L,u}_{ji}]_\mathrm{div.,SM}&=-\frac{1}{\epsilon^{}_{UV}}\frac{3m^u_j (V(m^d)^2V^\dagger)_{ji}}{32\pi^2 v^2},\\
    [\delta m^{R,u}_{ji}]_\mathrm{div.,SM}&=-\frac{1}{\epsilon^{}_{UV}}\frac{3m^u_i (V(m^d)^2V^\dagger)_{ji}}{32\pi^2 v^2}.
\end{split}
\end{equation}
Again, to get the down-type quark result simply exchange $u$ with $d$ and hermitian conjugate\footnote{Previously we had complex conjugation simply due to the notation in which we wrote down our results. Complex conjugation is for indexed notation and hermitian conjugation is for matrix notation.} $V$. 
We can now use the off-diagonal mass counterterms in eq.~(\ref{eq:ihL}) and write the corresponding divergent parts of $ih_L$'s for up and down quarks
\begin{equation}\label{eq:UV_updown_hl}
\begin{split}
    [ih^u_{Lji}]_\mathrm{div.,SM}=&-\frac{3(V(m^d)^2V^\dagger)_{ji}}{32\pi^2v^2\epsilon^{}_{UV}}\cdot\frac{(m^u_i)^2+(m^u_j)^2 }{(m^u_i)^2-(m^u_j)^2},\\
    [ih^d_{Lji}]_\mathrm{div.,SM}=&-\frac{3(V^\dagger(m^u)^2V)_{ji}}{32\pi^2v^2\epsilon^{}_{UV}}\cdot\frac{(m^d_i)^2+(m^d_j)^2 }{(m^d_i)^2-(m^d_j)^2}.
\end{split}
\end{equation}
While not needed for the counterterm of the CKM matrix, it is interesting to write down $ih_R$, for example, for up quarks we have
\begin{equation}
    [ih^u_{Rji}]_\mathrm{div.,SM}=-\frac{3(V(m^d)^2V^\dagger)_{ji}}{32\pi^2v^2\epsilon^{}_{UV}}\cdot\frac{2m^u_im^u_j}{(m^u_i)^2-(m^u_j)^2}.
\end{equation}
We see that there are different mass structures ${(m^u_i)^2+(m^u_j)^2}$ and $2m^u_i m^u_j$ for the left- and right-handed parts. These are also the same structures we noticed when discussing the UV divergences in previous sections.

Now we can write down the divergent part of  the CKM matrix counterterm

\begin{equation}
    \begin{split}
        [\delta V_{ji}]_\mathrm{div.,SM}&=-\sum_{n \neq j}ih^u_{Ljn} V^{}_{ni}+ \sum_{k \neq i}V_{jk} ih^d_{Lki}\\
        &=\frac{3}{32\pi^2v^2\epsilon^{}_{UV}}\Bigg[\sum_{n \neq j}(V(m^d)^2V^\dagger)^{}_{jn}\cdot\frac{(m^u_n)^2+(m^u_j)^2}{(m^u_n)^2-(m^u_j)^2}V^{}_{ni}\\
        &\qquad \qquad \qquad-\sum_{k \neq i}V^{}_{jk}(V^\dagger(m^u)^2V)^{}_{ki}\cdot\frac{(m^d_i)^2+(m^d_k)^2}{(m^d_i)^2-(m^d_k)^2} \Bigg].    
    \end{split}
\end{equation}

Comparing with the divergent parts found in~\cite{Denner1990_CKMreno, Kniehl2006}, we see that the results match up to factors of 2 due to regularization. 

\subsection{Neutrinos}\label{sec:big_neutrinos}

Now we move on to examples with neutrinos where we have plenty of scenarios as noted when introducing the Grimus-Neufeld model. We shall begin with the easiest case where both the incoming and outgoing particles are massless, then consider the fully massive case, and finally in the massive case we will take one of the particles to be massless to get to the partially massive case (i.e., one of the off-diagonal blocks). In addition to the neutrino mass counterterms, we will also need the charged lepton mass counterterms, as it will be needed for the 1-loop rotation and the renormalization of neutrino mixing.

\subsubsection{Fully massless case ($i,j<3$)}

In the fully massless case we only need to evaluate one scalar self-energy function to get the off-diagonal contributions to the mass (for example, $\Sigma^{sL}(0)$). For the neutrino mass contributions $\delta m^{L,R,\nu}_{ji}$ we easily get
\begin{equation}
\begin{split}
    \delta m^{L,\nu}_{ji}&=\frac{1}{2^{D+1}\pi^{D-2}}(U^T_LG^{}_\nu U^\star_R)^{}_{ja}m^\nu_a\Big[s^2_\alpha B_0(0, m^2_h,(m^\nu_a)^2)+c^2_\alpha B_0(0, m^2_H,(m^\nu_a)^2)\\
    &\hspace{5cm}-B_0(0, m^2_A,(m^\nu_a)^2)\Big](U^\dagger_RG^{T}_\nu U^{}_L)^{}_{ai},\\
    \delta m^{R,\nu}_{ji}&=\frac{1}{2^{D+1}\pi^{D-2}}(U^\dagger_LG^{\star}_\nu U^{}_R)^{}_{ja}m^\nu_a\Big[s^2_\alpha B_0(0, m^2_h,(m^\nu_a)^2)+c^2_\alpha B_0(0, m^2_H,(m^\nu_a)^2)\\
    &\hspace{5cm}-B_0(0, m^2_A,(m^\nu_a)^2)\Big](U^T_RG^\dagger_\nu U^\star_L)^{}_{ai}.
\end{split}
\end{equation}
Here summation over $a$ is implied, and we used a different index than $k$ to emphasize that summation runs over all four neutrinos instead of the 3 generations. We got $\delta m^{R,\nu}_{ji}$ via hermitian conjugation of $\delta m^{L,\nu}_{ji}$ and by noticing that the Passarino-Veltman functions are real in the massless case. In evaluating the scalar self-energy function we used unitarity relations of $U_L$ and $U_R$ matrices in the massless case. The above contributions are for $i\neq j$, however, the real part can also be used for $i=j$ as mentioned in Section~\ref{sec:mas_field_def}. In other words, we regard the real part of above contributions for the whole massless block instead of its off-diagonal components only.

It is easy to note that the UV divergent parts cancel since the divergences of $B_0$ functions do not depend on the arguments. Even more easily we see the explicit gauge-independence. As noted in the model introduction, we think of these contributions as a radiative mass. In addition, this contribution can be diagonalized by noticing the remaining freedom in the first two rows of $U^\dagger_L$. To make these contributions diagonal we can choose one of the two rows of $U^T_L$ to be orthogonal to $G^{}_\nu$~\cite{Grimus1989}. Alternatively, diagonalization can be made more simple by defining a new Yukawa coupling to the second Higgs doublet (analogous to the one in~\cite{Dudenas2019})
\begin{equation}\label{eq:gprime}
    U^T_LG^{}_\nu=G^{\prime}_\nu=\begin{pmatrix}
    0 \\
    g^{\nu\prime}_2 \\
    g^{\nu\prime}_3 \\
    g^{\nu\prime}_4
    \end{pmatrix}.
\end{equation}
With this coupling it is evident that at 1-loop the $2\times 2$ massless block is diagonal and that one of the massless tree-level neutrinos acquires a radiative mass, while the other neutrino remains massless. The remaining massless neutrino may acquire a mass beyond 1-loop~\cite{Grimus1989}. In other words, at 1-loop we have
\begin{equation}
    \begin{split}
        m_1&=\mathrm{Re[\delta m^{L,\nu}_{11}}] =\mathrm{Re[\delta m^{R,\nu}_{11}}]=0, \\
        m_2&=\mathrm{Re[\delta m^{L,\nu}_{22}}] =\mathrm{Re[\delta m^{R,\nu}_{22}}]=\mathrm{Re}\left[\tilde{C}(g^{\nu\prime}_2)^2\right].
    \end{split}
\end{equation}
Here we used the dimensionality of $U^{}_R$ and $G^{}_\nu$ in the Grimus-Neufeld model to define the constant $\tilde{C}$
\begin{equation}
\begin{split}
    \tilde{C}=&\frac{1}{2^{D+1}\pi^{D-2}}U^\star_{Ra} U^\dagger_{Ra} m^\nu_a\left[s^2_\alpha B_0(0, m^2_h,(m^\nu_a)^2)+c^2_\alpha B_0(0, m^2_H,(m^\nu_a)^2)-B_0(0, m^2_A,(m^\nu_a)^2)\right]
\end{split}
\end{equation}
where summation over $a$ is again implied and we wrote $U^{}_{R1a}$ as $U^{}_{Ra}$. Analogous results for this model were also found in~\cite{Grimus1989,Dudenas2019}.

As already mentioned, in this fully massless block there is no way to define the 1-loop rotation, but, on the other hand, we do not need it at all as the block can be made diagonal using $U_L$, so we have
\begin{align}
    ih^\nu_{L12}=ih^\nu_{L21}=0.
\end{align}

\subsubsection{Fully massive case ($i,j>2$)}

We next move to the fully massive case, which is more cumbersome as the expressions are larger and we also need to evaluate the Passarino-Veltman functions to see the needed mass structures. We will fully write out the anti-hermitian part of the left-handed field renormalization and only UV divergent parts of the left-handed off-diagonal mass counterterms. 

As in the quark case, to find the left-handed anti-hermitian part of the field renormalization we pick out terms containing ${(m^\nu_i)^2-(m^\nu_j)^2}$ from the evaluated expression in eq.~(\ref{eq:dZL_def}). However, upon evaluation, one may find (logarithmic) terms with mass structures such as 
\begin{equation}
    m^\nu_j\left[3(m^\nu_i)^2+(m^\nu_j)^2\right].
\end{equation}
Terms like these are ambiguous as they can be rewritten in multiple ways in terms of the mass structures we saw already: $m_i,m_j,2m_i m_j, {m^2_i\pm m^2_j}$. For example, the above might be written as
\begin{equation*}
 m^\nu_j\left(2\left[(m^\nu_i)^2+(m^\nu_j)^2\right]+\left[(m^\nu_i)^2-(m^\nu_j)^2\right]\right)
\end{equation*}
or as
\begin{equation*}
 m^\nu_i\left[2m^\nu_i m^\nu_j\right]+m^\nu_j\left[(m^\nu_i)^2+(m^\nu_j)^2\right].
\end{equation*}
In other words, ambiguous terms like these do not allow for a clear separation of the ${m^2_i-m^2_j}$ mass structure, however, Nielsen identities guarantee that no such ambiguous mass structures appear in gauge-dependent terms. In addition, these ambiguous terms cannot be distributed at will, since the massless limit of renormalization constants must exist. For the neutrinos in the Grimus-Neufeld model we were unable to get such a limit when the ambiguous terms were included in field renormalization. Practically this means that gauge-independent (and UV-finite) terms must immediately manifest the ${m_i^2-m_j^2}$ mass structure without any rearrangements if they are to be included into field renormalization, otherwise these terms are to be included in mass renormalization. This respects our definitions and actually makes picking of terms easier. For the current case of the Grimus-Neufeld model we were unable to find such immediate manifestations of the ${m_i^2-m_j^2}$ mass structure for gauge-independent terms. Having this in mind, we write down the off-diagonal left-handed anti-hermitian part of the field renormalization for neutrinos $\delta Z^{A,\nu}_{Lji}$ in the fully massive case

\begin{align}\label{eq:neutrino_dZL}
    \nonumber
   \delta Z^{A,\nu}_{Lji}=&\widetilde{Re}\Big\{U^{}_{Lki}U^\star_{Lkj}\left[\frac{(m^\nu_i)^2-(m^l_k)^2-m^2_W\xi_W}{16\pi^2v^2}\right]\Lambda((m^\nu_i)^2, m^l_k, \sqrt{\xi_W}m_W)\\
    \nonumber
    &+U^{}_{Lki}U^\star_{Lkj}\left[\frac{(m^l_k)^4-((m^\nu_i)^2-m^2_W\xi^2_W)^2}{32\pi^2v^2(m^\nu_i)^2}\right]\log\left(\frac{(m^l_k)^2}{m^2_W\xi_W}\right)\\
    \nonumber
    &+(U^\dagger_LU^{}_L)_{ji}\frac{(m^\nu_i)^2}{16\pi^2v^2}\log\left(\frac{\mu^2}{m^2_W\xi_W}\right)\\
    \nonumber
    &+(U^\dagger_LU^{}_L)_{ja}(U^T_LU^{\star}_L)_{ai}m^\nu_a\left[\frac{(m^\nu_a)^2-(m^\nu_i)^2-m^2_Z\xi_Z}{32\pi^2m^\nu_i}\right]\Lambda((m^\nu_i)^2, m^\nu_a, \sqrt{\xi_Z}m_Z)\\
    \nonumber
    &+(U^\dagger_LU^{}_L)_{ja}(U^\dagger_LU^{}_L)_{ai}\left[\frac{(m^\nu_i)^2-(m^\nu_a)^2-m^2_Z\xi_Z}{32\pi^2v^2}\right]\Lambda((m^\nu_i)^2, m^\nu_a, \sqrt{\xi_Z}m_Z)\\
    \nonumber
    &+(U^\dagger_LU^{}_L)_{ja}(U^T_LU^{\star}_L)_{ai}m^\nu_a\left[\frac{(m^\nu_i)^4-((m^\nu_a)^2-m^2_Z\xi_Z)^2}{64\pi^2v^2(m^\nu_i)^3}\right]\log\left(\frac{(m^\nu_a)^2}{m^2_Z\xi_Z}\right)\\
    \nonumber
    &+(U^\dagger_LU^{}_L)_{ja}(U^\dagger_LU^{}_L)_{ai}\left[\frac{(m^\nu_a)^4-((m^\nu_i)^2-m^2_Z\xi_Z)^2}{64\pi^2v^2(m^\nu_i)^2}\right]\log\left(\frac{(m^\nu_a)^2}{m^2_Z\xi_Z}\right)\\
    %\nonumber
    &+\left[\frac{(U^\dagger_LU^{}_L)_{ji}(m^\nu_i)^2}{32\pi^2v^2}+\frac{(U^\dagger_LU^{}_L(m^\nu)^3U^T_L U^\star_L)_{ji}}{32\pi^2v^2}\right]\log\left(\frac{\mu^2}{m^2_Z\xi_Z}\right)-H.C.\Big\}
\end{align}

Here $i\neq j$, summation over $a$ runs from 1 to 4 and summation over $k$ from 1 to 3, and $\Lambda(m^{2}_1, m^{}_2, m^{}_3)$ is the Disc function as used in \texttt{PackageX} and \texttt{FeynCalc}. It is interesting to note that $\delta Z^{A,\nu}_{L}$ contains only Disc functions and logarithms, since other finite parts behave like UV divergent terms in Section~\ref{sec:UV_div} for which ${m_i^2-m_j^2}$ is non-existent.

To get $\delta Z^{A,\nu}_{Rji}$ one can complex conjugate couplings of the above expression of $\delta Z^{A,\nu}_{Lji}$, but not Disc functions. This partial complex conjugation is due to absorptive parts violating the Majorana condition as discussed in Section~\ref{sec:field_reno}.

Having the field counterterms we can now write down the mass counterterms. However, the expressions are cumbersome enough that we only write down the UV divergent parts in the full Grimus-Neufeld model:

\begin{align}\label{eq:massive_dm}
    \nonumber
    [\delta m^{L,\nu}_{ji}]_\mathrm{div.}&=(U^\dagger_LU^{}_L)_{ji}\frac{m^\nu_j}{4\pi^2v^2\epsilon^{}_{UV}}\mathrm{Tr}\left\{(m^\nu)^4U^\dagger_LU^{}_L+(m^d)^4+(m^u)^4+(m^l)^4\right\}\left[\frac{c^2_\alpha}{m^2_h}+\frac{s^2_\alpha}{m^2_H}\right]\\
    \nonumber
% lambda 3
    &+(U^\dagger_LU^{}_L)_{ji}\frac{s^2_\alpha\lambda_3 m^\nu_j}{64\pi^2v^2m^2_H\epsilon^{}_{UV}}\left[-6c^2_{\alpha}m^2_H+\left(m^2_h(3c_{2\alpha}+1)+2m^2_A+4m^2_{H^\pm}\right)\right]\\
    \nonumber
    &-(U^\dagger_LU^{}_L)_{ji}\frac{c^2_\alpha\lambda_3 m^\nu_j}{64\pi^2v^2m^2_h\epsilon^{}_{UV}}\left[m^2_H(3c_{2\alpha}-1)+2m^2_A+4m^2_{H^\pm}\right]\\
    \nonumber
% lambda 7
    &-(U^\dagger_LU^{}_L)_{ji}\frac{s_{2\alpha}\lambda_7 m^\nu_j}{64\pi^2v^2m^2_H\epsilon^{}_{UV}}\left[3c_{2\alpha}m^2_H+\left(3m^2_h s^2_\alpha+m^2_A+2m^2_{H^\pm}\right)\right]\\
    \nonumber
    &+(U^\dagger_LU^{}_L)_{ji}\frac{s_{2\alpha}\lambda_7 m^\nu_j}{64\pi^2v^2m^2_h\epsilon^{}_{UV}}\left[3m^2_H c^2_\alpha+m^2_A+2m^2_{H^\pm}\right]\\
    \nonumber
% Higgs
    &-(U^\dagger_LU^{}_L)_{ji}\frac{s^2_{\alpha}m^\nu_j}{64\pi^2v^2m^2_H\epsilon^{}_{UV}}\\
    \nonumber
    &\qquad\times\big[4m^4_hs^2_\alpha+2m^2_h m^2_{H^\pm}(3c_{2\alpha}+1)+2m^4_H(c_{2\alpha}+4)-4m^2_A(m^2_{H^\pm}-m^2_A)\big]\\
    \nonumber
    &+(U^\dagger_LU^{}_L)_{ji}\frac{c^2_\alpha m^\nu_j}{64\pi^2v^2m^2_h\epsilon^{}_{UV}}\\
    \nonumber
    &\qquad\times\big[-4m^4_H c^2_\alpha+2m^2_Hm^2_{H^\pm}(3c_{2\alpha}-1)+2m^4_h(c_{2\alpha}-4)+4m^2_A(m^2_{H^\pm}-m^2_A)\big]\\
    \nonumber
%  what's left
    &+(U^\dagger_LU^{}_L)_{ji}\frac{m^\nu_j}{64\pi^2v^2\epsilon^{}_{UV}}\left[4(m^\nu_i)^2+6m^2_{H^\pm}s^2_{2\alpha}\right]+\frac{m^\nu_j}{32\pi^2v^2\epsilon^{}_{UV}}(U^\dagger_LU^{}_L(m^\nu)^2U^\dagger_LU^{}_L)_{ji}\\
    \nonumber
    &-(U^\dagger_LU^{}_L)_{ji}\frac{m^\nu_j}{\epsilon^{}_{UV}}\left[\frac{6m^4_W+3m^4_Z}{16\pi^2v^2}\right]\left[\frac{c^2_\alpha}{m^2_h}+\frac{s^2_\alpha}{m^2_H}\right]-\frac{3m^\nu_j}{32\pi^2v^2\epsilon^{}_{UV}}(U^\dagger_L(m^l)^2U^{}_L)_{ji}\\
    \nonumber
% tadpoles mixing, sine alpha
    &+(U^\dagger_LU^{}_L)_{ji}\frac{s_{2\alpha}m^\nu_j}{16\sqrt{2}\pi^2v\epsilon^{}_{UV}}\left[\frac{1}{m^2_H}-\frac{1}{m^2_h}\right]\\
    \nonumber
    &\qquad\times\mathrm{Tr}\Big\{\frac{(m^\nu)^3}{2}(U^\dagger_L G^\star_\nu U^{}_R)+\frac{(m^\nu)^3}{2}(U^T_R G^\dagger_\nu U^\star_L)\\
    \nonumber
    &\qquad \qquad+G_d(m^d)^3+G_u(m^u)^3+G_l(m^l)^3+C.C.\Big\}\\
    \nonumber
    &+(U^T_LG^{}_\nu U^\star_R)_{ji}\frac{s_{2\alpha}}{8\sqrt{2}\pi^2v\epsilon^{}_{UV}}\left[\frac{1}{m^2_H}-\frac{1}{m^2_h}\right]\mathrm{Tr}\left\{(m^\nu)^4(U^\dagger_LU^{}_L)+(m^d)^4+(m^u)^4+(m^l)^4\right\}\\
% tadpoles, mixing, 3rd power +mA
    \nonumber
    &+\frac{(U^T_LG^{}_\nu U^\star_R)_{ji}}{16\pi^2\epsilon^{}_{UV}}\left(\frac{s^2_\alpha}{m^2_h}+\frac{c^2_\alpha}{m^2_H}+\frac{1}{m^2_A}\right)\\
    \nonumber
    &\qquad\times\mathrm{Tr}\Big\{\frac{(m^\nu)^3}{2}(U^\dagger_LG^\star_\nu U^{}_R)+\frac{(m^\nu)^3}{2}(U^T_RG^\dagger_\nu U^\star_L)+G^{}_d(m^d)^3+G^{\star}_u(m^u)^3+G^{}_l(m^l)^3\Big\}\\
% tadpoles, mixing, 3rd power -mA
    \nonumber
    &+\frac{(U^T_LG^{}_\nu U^\star_R)_{ji}}{16\pi^2\epsilon^{}_{UV}}\left(\frac{s^2_\alpha}{m^2_h}+\frac{c^2_\alpha}{m^2_H}-\frac{1}{m^2_A}\right)\\
    \nonumber
    &\qquad\times\mathrm{Tr}\Big\{\frac{(m^\nu)^3}{2}(U^T_LG^{}_\nu U^\star_R)+\frac{(m^\nu)^3}{2}(U^\dagger_RG^{T}_\nu U^{}_L)+G^{\star}_d(m^d)^3+G^{}_u(m^u)^3+G^{\star}_l(m^l)^3\Big\}\\
% BSM, lambda3
    \nonumber
    &+(U^T_LG^{}_\nu U^\star_R)_{ji}\frac{\lambda_3v s_{2\alpha} }{128\sqrt{2}\pi^2\epsilon^{}_{UV}}\left[\frac{1}{m^2_H}-\frac{1}{m^2_h}\right]\\
    \nonumber
    &\qquad \times\left[3c_{2\alpha}(m^2_h-m^2_H)+m^2_h+m^2_H-2m^2_A-4m^2_{H^\pm}\right]\\
% BSM, lmabda 7
    \nonumber
    &+(U^T_LG^{}_\nu U^\star_R)_{ji}\frac{\lambda_7 v (c_{2\alpha} (m^2_h-m^2_H)+m^2_h+m^2_H)}{128\sqrt{2}\pi^2m^2_hm^2_H\epsilon^{}_{UV}}\\
    \nonumber
    &\qquad\times\left[3c_{2\alpha}(m^2_h-m^2_H)-3m^2_h-3m^2_H-2m^2_A-4m^2_{H^\pm}\right]\\
    \nonumber
    &+(U^T_LG^{}_\nu U^\star_R)_{ji}\frac{s_{2\alpha}}{64\sqrt{2}\pi^2 v\epsilon^{}_{UV}}\left[\frac{1}{m^2_H}-\frac{1}{m^2_h}\right]\\
    \nonumber
    &\qquad\times\left[-m^4_h+c_{2\alpha}(m^2_h-m^2_H)(m^2_h+m^2_H-3m^2_{H^\pm})-6(2m^4_W+m^4_Z)\right]\\
    \nonumber  
    &+(U^T_LG^{}_\nu U^\star_R)_{ji}\frac{s_{2\alpha}}{64\sqrt{2}\pi^2 v\epsilon^{}_{UV}}\left[\frac{1}{m^2_H}-\frac{1}{m^2_h}\right]\\
    \nonumber
    &\qquad\times
    \left[m^2_h(4m^2_h-m^2_{H^\pm})-m^4_H-m^2_Hm^2_{H^\pm}-2m^4_{H^\pm}+2m^2_Am^2_{H^\pm}\right]\\
% other things
    \nonumber
    &-\frac{(U^T_LG^T_l m^l G^{}_\nu U^\star_R)_{ji}}{16\pi^2\epsilon^{}_{UV}}
    +\frac{m^\nu_j(U^\dagger_LG^\dagger_lG^{}_lU^{}_L)_{ji}}{64\pi^2\epsilon^{}_{UV}}\\
    &+\frac{m^\nu_j(U^\dagger_LG^\star_\nu G^T_\nu U^{}_L)_{ji}}{64\pi^2\epsilon^{}_{UV}}
    +\frac{m^\nu_j(U^T_RG^\dagger_\nu G^{}_\nu U^\star_R)_{ji}}{32\pi^2\epsilon^{}_{UV}}+Trans.
\end{align}

Here $i\neq j$, $\mathrm{Tr\{\dots\}}$ is a trace over family indices, $Trans.$ stands for transposition, and $C.C.$ stand for complex conjugation. The divergent parts of $\delta m^{L\nu}$ are symmetric, however, symmetry is spoiled in the full $\delta m^{L,\nu}$ counterterm due to absorptive parts as discussed in Sec.~\ref{sec:field_reno}. Even though symmetry does not hold in full, the hermiticity relation $\delta m^{L,\nu}=(\delta m^{R,\nu})^\dagger$ does hold by construction.

We may compare the SM part (no second Higgs doublet, but with the Majorana neutrino) of our $[\delta m^{L,\nu}_{ji}]_\mathrm{div.}$ with the $(\delta m^{\nu(-)}_\mathrm{div})_{ab}$ found in eq.~(31) of~\cite{Almasy2009}. Upon comparison, there is a mismatch due to our inclusion of \textit{all tadpole diagrams}. The difference of the two counterterms in the SM (we are being carefree with $ab$ indices here) is

\begin{equation}
    \begin{split}
    [\delta m^{L,\nu}_{ji}]_\mathrm{div.,SM}-(\delta m^{\nu(-)}_\mathrm{div})_{ab}&=(U^\dagger_LU^{}_L)_{ij}\frac{m^\nu_i}{\epsilon^{}_{UV}}\mathrm{Tr}\left\{\frac{(m^\nu)^4U^\dagger_LU^{}_L+(m^d)^4+(m^u)^4+(m^l)^4}{4\pi^2m^2_hv^2}\right\}\\
    &-(U^\dagger_LU^{}_L)_{ij}\frac{m^\nu_i}{\epsilon^{}_{UV}}\left[\frac{6m^4_W+3m^4_Z}{16\pi^2m^2_hv^2}\right]+Trans.
\end{split}
\end{equation}

It is important to note that there was no such difference in the quark case due to diagonal quark interactions with the Higgs $h$ and we consider only off-diagonal terms.

In the fully massive case, the 1-loop rotation works without any changes, but $ih^\nu_L$ contains both $2m^\nu_im^\nu_j$ and $(m^\nu_i)^2+(m^\nu_j)^2$ mass structures. This is because neutrinos are Majorana fermions up to absorptive parts. These structures can be noticed by considering how the last term in eq.~(\ref{eq:massive_dm}) and its transpose appear in $ih^\nu_L$.

\subsubsection{Partially massless case ($i>2,j<3$ or $j>2,i<3$)}

As advertised, in the partially massless case we simply take the appropriate massless limit of the fully massive case. Of course, we have to remember eq.~(\ref{eq:massless_uni}), which takes care of many terms in the partially massless case. For clarity, further in this section let us have the index $i$ correspond to a massless incoming particle, i.e. $i=1$ or $i=2$, and $j>2$. 

Let us first consider the off-diagonal mass counterterm $[\delta m^{L,\nu}_{ji}]_\mathrm{div.}$ in this partially massless case. This case is fairly simple as no terms contain terms with negative powers of $m^\nu_i$, so terms either vanish trivially because of $m^\nu_i\rightarrow0$ or due to properties of $U_L$ and $U_R$. In eq.~(\ref{eq:massive_dm}), terms containing $(U^\dagger_LU^{}_L)_{ij}=\delta_{ij}$ vanish since $i \neq j$. Other remaining terms containing $(U^T_LG^{}_\nu U^\star_R)_{ji}$ or  $(U^T_RG^\dagger_\nu G^{}_\nu U^\star_R)_{ji}$ also vanish since $U^\star_{R11}=U^\star_{R12}=0$. In the end, only the second term on line 9, and second and third terms on the last line remain from the visible part. 

The transposed part of eq.~(\ref{eq:massive_dm}) indicated by $Trans.$ also simplifies in the partially massless case. Swapping indices is not hard, so we will describe the transposed part in terms of the visible untransposed part. First 10 lines and the last three terms vanish because of $m^\nu_i\rightarrow0$. All other remaining terms in the transposed part have $(U^T_L G^{}_\nu U^\star_R)_{ij}$, which could be written as $(G^\prime_\nu U^\star_R)_{ij}$. If $G^{\nu\prime}$ gets the index $i$ the appropriate terms are zero for $i=1$ and non-zero for $i=2$, see eq.~(\ref{eq:gprime}), hence, we will additionally indicate these terms with $\delta_{2i}$. In the end, the UV divergent part of the mass counterterm may be written as
\begin{align}
   \nonumber
    [\delta m^{L,\nu}_{ji}]_\mathrm{div.,\cancel{m^\nu_i}}&=\delta_{2i}(G^\prime_\nu U^\star_R)_{ij}\frac{s_{2\alpha}}{8\sqrt{2}\pi^2v\epsilon^{}_{UV}}\left[\frac{1}{m^2_H}-\frac{1}{m^2_h}\right]\\
    \nonumber
    &\times\mathrm{Tr}\left\{(m^\nu)^4(U^\dagger_LU^{}_L)+(m^d)^4+(m^u)^4+(m^l)^4\right\}\\
% tadpoles, mixing, +mA
    \nonumber
    &+\delta_{2i}\frac{(G^\prime_\nu U^\star_R)_{ij}}{16\pi^2\epsilon^{}_{UV}}\left(\frac{s^2_\alpha}{m^2_h}+\frac{c^2_\alpha}{m^2_H}+\frac{1}{m^2_A}\right)\\
    \nonumber&\qquad\times\mathrm{Tr}\Big\{\frac{(m^\nu)^3}{2}(U^\dagger_LG^\star_\nu U^{}_R)+\frac{(m^\nu)^3}{2}(U^T_RG^\dagger_\nu U^\star_L)+G^{}_d(m^d)^3+G^{\star}_u(m^u)^3+G^{}_l(m^l)^3\Big\}\\
% tadpoles, mixing, -mA
    \nonumber
    &+\delta_{2i}\frac{(G^\prime_\nu U^\star_R)_{ij}}{16\pi^2\epsilon^{}_{UV}}\left(\frac{s^2_\alpha}{m^2_h}+\frac{c^2_\alpha}{m^2_H}-\frac{1}{m^2_A}\right)\\
    \nonumber
    &\qquad\times\mathrm{Tr}\Big\{\frac{(m^\nu)^3}{2}(U^T_LG^{}_\nu U^\star_R)+\frac{(m^\nu)^3}{2}(U^\dagger_RG^{T}_\nu U^{}_L)+G^{\star}_d(m^d)^3+G^{}_u(m^u)^3+G^{\star}_l(m^l)^3\Big\}\\
% non trace stuff, lambda 3
    \nonumber
    &+\delta_{2i}(G^\prime_\nu U^\star_R)_{ij}\frac{\lambda_3v s_{2\alpha} (m^2_h-m^2_H)}{128\sqrt{2}\pi^2m^2_hm^2_H\epsilon^{}_{UV}}\left[3c_{2\alpha}(m^2_h-m^2_H)+m^2_h+m^2_H-2m^2_A-4m^2_{H^\pm}\right]\\
% non trace stuff, lambda 7
    \nonumber
    &+\delta_{2i}(G^\prime_\nu U^\star_R)_{ij}\frac{\lambda_7 v (c_{2\alpha} (m^2_h-m^2_H)+m^2_h+m^2_H)}{128\sqrt{2}\pi^2m^2_hm^2_H\epsilon^{}_{UV}}\\
    \nonumber
    &\qquad\times\left[3c_{2\alpha}(m^2_h-m^2_H)-3m^2_h-3m^2_H-2m^2_A-4m^2_{H^\pm}\right]\\
% non trace stuff, non-lambda, mix
    \nonumber
    &+\delta_{2i}(G^\prime_\nu U^\star_R)_{ij}\frac{s_{2\alpha}(m^2_h-m^2_H)}{64\sqrt{2}\pi^2m^2_hm^2_H v\epsilon^{}_{UV}}\\
    \nonumber
    &\qquad\times\left[-m^4_h+c_{2\alpha}(m^2_h-m^2_H)(m^2_h+m^2_H-3m^2_{H^\pm})-6(2m^4_W+m^4_Z)\right]\\
    \nonumber  
    &+\delta_{2i}(G^\prime_\nu U^\star_R)_{ij}\frac{s_{2\alpha}(m^2_h-m^2_H)}{64\sqrt{2}\pi^2m^2_hm^2_H v\epsilon^{}_{UV}}\\
    \nonumber
    &\qquad\times\left[m^2_h(4m^2_h-m^2_{H^\pm})-m^4_H-m^2_Hm^2_{H^\pm}-2m^4_{H^\pm}+2m^2_Am^2_{H^\pm}\right]\\
% BSM, no mix, no tadpole
    \nonumber
    &-\frac{(U^T_LG^T_l m^l G^{}_\nu U^\star_R)_{ij}}{16\pi^2\epsilon^{}_{UV}}-\frac{3m^\nu_j(U^\dagger_L(m^l)^2U^{}_L)_{ji}}{32\pi^2v^2\epsilon^{}_{UV}}\\
    &+\frac{m^\nu_j(U^\dagger_LG^\dagger_lG^{}_lU^{}_L)_{ji}}{64\pi^2\epsilon^{}_{UV}}+\frac{m^\nu_j(U^\dagger_LG^\star_\nu G^T_\nu U^{}_L)_{ji}}{64\pi^2\epsilon^{}_{UV}}
\end{align}
Here we explicitly see the result of taking one of the masses to 0. As noted before, in the partially massless case the expressions lack symmetry, i.e. one o the masses are missing, and one should be careful in, for example, getting $[\delta m^R_{ji}]_{\cancel{m^\nu_i}}$ from $[\delta m^L_{ji}]_{\cancel{m^\nu_i}}$. In addition, it is also not too hard to see that if we took the second mass to zero, there would be no more UV divergent terms and we would arrive at the fully massless limit. 

For completeness, we also take the mass $m^\nu_i$ to zero for the anti-hermitian part of field renormalization in eq.~(\ref{eq:neutrino_dZL}). However, this part is slightly more cumbersome as we first have to take the small $m^\nu_i$ limit for Disc functions for important cancellations to take place. The small $m$ limit of a Disc function is the following
\begin{equation}
    \Lambda(m^2,b,c)=\left[\frac{b^2-c^2}{2 m^2}-\frac{b^2+c^2}{2(b^2-c^2)}\right]\log\left(\frac{b^2}{c^2}\right)-1.
\end{equation}

On the other hand, a lot of terms vanish before we have to deal with Disc functions. Let us notice that in eq.~(\ref{eq:neutrino_dZL}) a lot of terms contain $(U^\dagger_LU^{}_L)_{ja}(U^T_LU^{\star}_L)_{ai}$, which for $m^\nu_i=0$ simplifies tremendously

\begin{equation}
\begin{split}
    (U^\dagger_LU^{}_L)_{ja}(U^T_LU^{\star}_L)_{ai}&=(U^\dagger_LU^{}_L)_{ja}\delta_{ai}\\
    &=(U^\dagger_LU^{}_L)_{ji}=\delta_{ji}=0.
\end{split}
\end{equation}
Here summation over $a$ is implied and $i \neq j$, also notice that this result holds even if there are other diagonal matrices with the index $a$. Having this we now notice that only terms in first two lines of eq.~(\ref{eq:neutrino_dZL}) along with their $H.C.$ counterparts remain. By first taking the small $m^\nu_i$ limit and then the massless limit we get
\begin{align}
    \nonumber
    \delta Z^{A,\nu}_{Lji,m^\nu_i\approx 0}=&\widetilde{Re}\Big\{U^{}_{Lki}U^\star_{Lkj}\left[\frac{(m^\nu_i)^2-(m^l_k)^2-m^2_W\xi^{}_W}{16\pi^2v^2}\right]\\
    \nonumber
    &\qquad\times\left[\left[\frac{(m^l_k)^2-m^2_W\xi^{}_W}{2 (m^\nu_i)^2}-\frac{(m^l_k)^2+m^2_W\xi^{}_W}{2((m^l_k)^2-m^2_W\xi^{}_W)}\right]\log\left(\frac{(m^l_k)^2}{m^2_W\xi^{}_W}\right)-1\right]\\
    %\nonumber
    &+U^{}_{Lki}U^\star_{Lkj}\left[\frac{(m^l_k)^4-((m^\nu_i)^2-m^2_W\xi_W)^2}{32\pi^2v^2(m^\nu_i)^2}\right]\log\left(\frac{(m^l_k)^2}{m^2_W\xi^{}_W}\right)-H.C.^\star\Big\},\\
    \nonumber
   \delta Z^{A,\nu}_{Lji,\cancel{m^\nu_i}}=&\widetilde{Re}\Big\{\frac{U^{}_{Lki}U^\star_{Lkj}}{16\pi^2v^2}(m^l_k)^2\left[\frac{(m^l_k)^2+m^2_W\xi_W}{(m^l_k)^2-m^2_W\xi^{}_W}\right]\log\left(\frac{(m^l_k)^2}{m^2_W\xi^{}_W}\right)+\frac{(U^\dagger_L(m^l)^2U^{}_L)_{ji}}{16\pi^2v^2}\\
    \nonumber\label{eq:mi0}
    &-U^{}_{Lki}U^\star_{Lkj}\left[\frac{(m^\nu_j)^2-(m^l_k)^2-m^2_W\xi^{}_W}{16\pi^2v^2}\right]\Lambda((m^\nu_j)^2, m^l_k, \sqrt{\xi^{}_W}m_W)^\star\\
    %\nonumber
    &-U^{}_{Lki}U^\star_{Lkj}\left[\frac{(m^l_k)^4-((m^\nu_j)^2-m^2_W\xi_W)^2}{32\pi^2v^2(m^\nu_j)^2}\right]\log\left(\frac{(m^l_k)^2}{m^2_W\xi^{}_W}\right)\Big\}.
\end{align}
Here $H.C.^\star$ means that we take the hermitian conjugate part from eq.~(\ref{eq:neutrino_dZL}) and do not expand the Disc function as it depends on $m^\nu_j$ and not $m^\nu_i$. In eq.~(\ref{eq:mi0}) we took $m^\nu_i$ to zero and also used eq.~(\ref{eq:massless_uni}), the last two lines are contributions from $H.C.^\star$. In addition, if we also went to the fully massless case by taking $m^\nu_j$ to zero, we would explicitly get $\delta Z^{A,\nu}_{Lji,\cancel{m^\nu}}=0$ as is not hard to check.

In this block the 1-loop rotation is also a simple limit of $ih_{L,Rji}$ and is found trivially by simply setting appropriate masses to zero.

Having considered explicit examples in the fully massless, fully massive, and partially massless cases we see that our scheme is valid in all needed scenarios.

\subsubsection{Charged leptons}

Before moving to the renormalization of the neutrino mixing matrix we must also discuss the renormalization of charged leptons - this is similar to the case of CKM matrix where both up- and down-type quarks are needed.

The renormalization of charged leptons is very similar to that of quarks in the sense, that we are able to express all the counterterms in terms of Passarino-Veltman functions. Let's us write down the anti-hermitian part of left-handed field renormalization for charged leptons $\delta Z^{A,l}_{Lji}$
\begin{equation}
\begin{split}
    \delta Z^{A,l}_{Lji}=&-\frac{U^{}_{Lja}U^\star_{Lia}}{2^{D}\pi^{D-2}v^2}\left[(m^\nu_a)^2-(m^l_i)^2+m^2_W\xi_W\right]B_0((m^l_i)^2,(m^\nu_a)^2,m^2_W\xi_W)\\
    &+\frac{U^{}_{Lja}U^\star_{Lia}}{2^{D}\pi^{D-2}v^2}\left[(m^\nu_a)^2-(m^l_j)^2+m^2_W\xi_W\right]B_0((m^l_j)^2,(m^\nu_a)^2,m^2_W\xi_W)^\star.
\end{split}    
\end{equation}
Note that here $i,j=1,2,3$ and $i \neq j$. The result is similar to the one in the case of quarks in eq.~(\ref{eq:quark_dZL}) and it is also easy to check that $\delta Z^{A,l}_{Lji}$ is finite. In addition, $\delta Z^{A,l}_{Rji}=0$ as was the case for quarks due to the chiral $W$ interaction. 

As in the quark case, the full charged lepton mass counterterm is given in Appendix~\ref{seca:full_exp} and here we only write down the divergent part $[\delta m^{L,l}_{ji}]_\mathrm{div.}$
\begin{align}
    \nonumber
    [\delta m^{L,l}_{ji}]_\mathrm{div.}=&
    -\frac{3m^l_j(U^{}_L(m^\nu)^2U^\dagger_L)_{ji}}{32\pi^2v^2\epsilon^{}_{UV}}
    -\frac{(G^{}_lU^{}_Lm^\nu U^\dagger_RG^T_\nu)_{ji}}{16\pi^2\epsilon^{}_{UV}}\\
    \nonumber
    &+\frac{m^l_j(G^\star_\nu G^T_\nu)_{ji}}{64\pi^2\epsilon^{}_{UV}}
    +\frac{m^l_i(G^{}_lG^\dagger_{l})_{ji}}{32\pi^2\epsilon^{}_{UV}}
    +\frac{m^l_j(G^\dagger_lG^{}_{l})_{ji}}{64\pi^2\epsilon^{}_{UV}}\\
% tadpoles, sine
    \nonumber
    &+\frac{s_{2\alpha}(G_l)_{ji}(m^2_h-m^2_H)}{8\sqrt{2}\pi^2 m^2_hm^2_Hv\epsilon^{}_{UV}}\left[(m^\nu)^4(U^\dagger_LU^{}_L)+(m^d)^4+(m^u)^4)+(m^l)^4\right]\\
% tadpoles, other mix
    \nonumber
    &+\frac{(G^{}_l)_{ji}}{16\pi^2\epsilon^{}_{UV}}\left(\frac{s^2_{\alpha}}{m^2_h}+\frac{c^2_{\alpha}}{m^2_H}+\frac{1}{m^2_A}\right)\mathrm{Tr}\Big\{(m^d)^3G^\star_d+(m^u)^3G^{}_u+(m^l)^3G^\star_l\\
    \nonumber
    &\hspace{5.5cm}+\frac{(m^\nu)^3}{2}(U^T_LG^{}_\nu U^\star_R+U^\dagger_RG^{T}_\nu U^{}_L)\Big\} \\
    \nonumber
    &+\frac{(G^{}_l)_{ji}}{16\pi^2\epsilon^{}_{UV}}\left(\frac{s^2_{\alpha}}{m^2_h}+\frac{c^2_{\alpha}}{m^2_H}-\frac{1}{m^2_A}\right)\mathrm{Tr}\Big\{(m^d)^3G^{}_d+(m^u)^3G^{\star}_u+(m^l)^3G^{}_l\\
    \nonumber
    &\hspace{5.5cm}+\frac{(m^\nu)^3}{2}(U^\dagger_LG^\star_\nu U^{}_R+U^T_L G^{}_\nu U^\star_R)\Big\}\\
% non-tadpole, Higgs potential, l7
    \nonumber
    &-\frac{(G_l)_{ji}\lambda_7 v}{128\sqrt{2}\pi^2m^2_hm^2_H\epsilon^{}_{UV}}\Big(c_{2\alpha}(m^2_h-m^2_H)+m^2_h+m^2_H\Big)\\
    \nonumber
    &\hspace{4cm}\times\Big(3c_{2\alpha}(m^2_h-m^2_H)-3m^2_h-3m^2_H-2m^3_A-4m^2_{H^\pm}\Big)\\
% non-tadpole, Higgs potential, l3
    \nonumber
    &+\frac{s_{2\alpha}(G_l)_{ji}\lambda_3 v}{128\sqrt{2}\pi^{2}\epsilon^{}_{UV}}\left[\frac{1}{m^2_H}-\frac{1}{m^2_h}\right]\left(3c_{2\alpha}(m^2_h-m^2_H)+m^2_h+m^2_H-2m^2_A-4m^2_{H^\pm}\right)\\
% other contributions...
    \nonumber
    &+\frac{s_{2\alpha}(G_l)_{ji}}{32\sqrt{2}\pi^2v\epsilon^{}_{UV}}\left[\frac{1}{m^2_h}-\frac{1}{m^2_H}\right]\Big[6m^4_W+3m^4_Z-c_{2\alpha}(m^2_h-m^2_H)(m^2_h+m^2_H-3m^2_A)\\
    \nonumber
    &\hspace{2cm}+m^4_h+m^4_H+2m^4_A-m^2_h(4m^2_H-m^2_{H^\pm})+m^2_H m^2_{H^\pm}-2m^2_A m^2_{H^\pm}\Big)\Big]
\end{align}
The case is again similar to that of quarks, only the first term is not due to an interaction with the second Higgs doublet and mimics the case of CKM matrix in the SM, which is also in agreement with the results in~\cite{Almasy2009}.

\subsubsection{1-loop rotation}

In the case of quarks, the $W$ vertex contains the CKM matrix arising from diagonalization of up- and down-type quark mass matrices. In the case of neutrinos, the $W$ vertex in principle contains the mixing matrix arising from diagonalization of charged lepton and neutrino mass matrices. However, we chose a basis where only the matrix $U_L$ appears in the $Wl\nu$ vertex. It is possible to arrive at this basis by diagonalizing the charged lepton fields and using the same rotation on the left-handed  neutrinos, after which the diagonalization of the neutrino mass matrix follows. In other words, we are concerned with the counterterm of the matrix $U^{}_L$ and not the the full neutrino mixing matrix, which also contains $U^{}_R$. This is also due to the fact that we did not introduce additional charged leptons or changed the SM $Wl\nu$ vertex in any way other than via diagonalization of the neutrino mass matrix.

Analogously to quarks, we may get the $U^{}_L$ counterterm $\delta U^{}_L$ via a 1-loop rotation that diagonalizes the masses of neutrinos and charged leptons at 1-loop. As we have already discussed all the needed mass counterterms and all of the needed expressions are in the text, we only write down the definition of the counterterm $\delta U^{}_L$, which is analogous to the one for the $B$ matrix in~\cite{Almasy2009}, we have
\begin{equation}
    \delta U^{}_{Lji}=-\sum_{k\neq j}^{3}ih^l_{Ljk}U^{}_{Lki}+\sum_{a\neq i}^{4}U^{}_{Lja}ih^\nu_{Lai}.
\end{equation}
Note that here the indices $k$ and $a$ have different bounds and the matrices $ih^l_L$ and $ih^\nu_L$ are of different dimensions, which is to be expected since $U^{}_L$ is a $3\times 4$ matrix. 

As already discussed in the text the UV divergent SM parts of $\delta U^{}_{Lji}$ do match the ones in~\cite{Almasy2009} up to some tadpole contributions, which were not considered by the authors of~\cite{Almasy2009}. We do not expect to get the same finite parts due to different approaches towards absorptive parts.

Finally, we briefly note that for large Majorana mass $M^{}_R$ the last column of $U^{}_L$ is approximately zero, then the remaining $3\times 3$ block corresponds to the usual Pontecorvo-Maki-Nakagawa-Sakata (PMNS) matrix~\cite{Pontecorvo1957, Maki1962}.

\section{Conclusions}\label{sec:big_conclusions}
In this paper we described a novel way of defining off-diagonal mass and anti-hermitian parts of field counterterms for fermions, which relies on various mass structures and imposing the no-mixing condition on incoming particles. It is important to note, that in our definitions we do not rely on dropping absorptive parts, in addition the hermiticity of the Lagrangian is also preserved. However, Majorana fermions are not fully compatible with Hermiticity of the Lagrangian and the no-mixing conditions and one may have to make an exception by dropping the absorptive parts, see the note on Majorana fermions in Section~\ref{sec:field_reno}. Using the 1-loop logic the scheme is extended to arbitrary orders in Appendix~\ref{sec:beyond_1loop}, although without examples in the Grimus-Neufeld model and without absorptive parts.

By using Nielsen identities and properties of Passarino-Veltman functions we also found the anti-hermitian part of field counterterms to be always finite and to contain all possible gauge-dependence. On the other hand, mass counterterms are UV divergent and contain no gauge-dependence. Considering this and the CKM matrix renormalization approach in~\cite{Denner1990_CKMreno}, where no off-diagonal mass counterterms are present, we conclude two things: 1) one needs to have off-diagonal mass counterterms to avoid propagation of UV divergences from the mass term to other terms in the Lagrangian, 2) there is no need to renormalize the SM CKM mixing matrix on the basis of UV divergences. In the section on 1-loop rotations we also extended the latter conclusion to the general case --- for full consistency mixing matrices should not be renormalized in any basis, although, that does not mean that the usual approach where mixing matrices are renormalized cannot be useful.

Finally, by using the Grimus-Neufeld model for examples we computed the off-diagonal counterterms for quarks, charged leptons, and Majorana neutrinos. The Grimus-Neufeld model also allowed us to test our scheme for massive and massless fermions as well as in the case of radiative mass generation. By performing the 1-loop rotation (regarding external legs) we were also able to compare with the results in~\cite{Denner1990_CKMreno, Kniehl2006, Almasy2009} - we reproduced the UV divergent terms (up to additional tadpole contributions in the case of neutrinos), while we did not compare the finite parts due to different approaches towards absorptive parts. In addition, these computations provide new results in the Grimus-Neufeld model.

\begin{acknowledgments}
The author would like to thank his supervisor T.~Gajdosik and also V.~Dūdėnas for reading of the manuscript as well as for helpful comments and discussions.

This work was in part funded by the Vilnius University Experimental Nuclear and Particle Physics Center.

\end{acknowledgments}

\appendix

\section{Extending the scheme beyond 1-loop}\label{sec:beyond_1loop}
In this appendix we extend our 1-loop approach to higher orders. Beyond 1-loop there are more contributions that we need to keep track of and there are a few new objects that we define below. However, the logic to all orders remains the same --- we look at the gauge dependence of the field renormalization and see whether the ${m_i^2-m_j^2}$ structure in the gauge-dependent part cancels. If it does cancel, it means that we can define the anti-hermitian part of field renormalization as the coefficient of ${m_i^2-m_j^2}$ and solve for the mass counterterm as we did in our 1-loop discussion. After introducing the setup we re-derive tree and 1-loop results as well as derive the 2-loop results. Having the hang of the approach we extend the scheme to arbitrary orders. 

\subsection{The setup}

First off, to avoid pesky numerical factors, the fields are renormalized without square roots on field renormalization constants
\begin{equation}
    \psi_0\rightarrow Z \psi
\end{equation}
with all the decompositions remaining the same as in eq.~(\ref{eq:reno_conditions}). 

Secondly, in this appendix we do not decompose the self-energy according to eq.~(\ref{eq:SE_deco}) to make the write up more compact. In addition, we introduce the bare self-energy computed with bare parameters $\tilde{\Sigma}^0(\cancel{p})$, the bare self-energy computed with renormalized parameters $\Sigma^0(\cancel{p})$ (no counterterms included) and the renormalized self-energy $\Sigma^R(\cancel{p})$. The two bare self-energies are topologically/pictorially the same and the difference comes from the parameters used to compute the diagrams. Counterterm insertions, at least pictorially, appear only in the renormalized self-energy and not in the bare ones. 

We also relate all the self-energies. In momentum space this is done by looking at the kinetic term in the bare Lagrangian and then renormalizing the fields:
\begin{equation}
    \bar{\psi}_0 \tilde{\Sigma}^0(\cancel{p})\psi_0=\bar{\psi} \gamma^0Z^\dagger\gamma^0 \tilde{\Sigma}^0(\cancel{p})Z\psi=\bar{\psi}\Sigma^R(\cancel{p})\psi,
\end{equation}
which brings us to the following relation
\begin{equation}\label{eq:barebare_SE}
    \tilde{\Sigma}^0(\cancel{p})=\gamma^0(Z^\dagger)^{-1}\gamma^0\Sigma^R(\cancel{p})Z^{-1}.
\end{equation}

To relate the remaining bare self-energy we expand the bare parameters $p^0_k$ in $\tilde{\Sigma}^0(\cancel{p})$ around their renormalized parts $p_k$. Since we need to take Taylor series expansions for every parameter it is convenient to define the series operator
\begin{equation}
    \textbf{T}_{p^0}=\sum_{n=0}^\infty\frac{(\delta p)^n}{n!}\frac{\partial^n}{\partial (p^0)^n}
\end{equation}
with $(\delta p)^n$ being the $n$th power of the parameter counterterm. Using the series operator and the fact that
\begin{equation}
   \left.\frac{\partial \tilde{\Sigma}^0(\cancel{p})}{\partial p^0_k}\right|_{p^0_k=p_k}=\frac{\partial \Sigma^0(\cancel{p})}{\partial p_k}
\end{equation}
we can relate the two bare self-energies
\begin{equation}\label{eq:barebare_relation}
    \tilde{\Sigma}^{0}(\cancel{p})=\left(\prod_k\textbf{T}_{p^0_k}\right)\tilde{\Sigma}^{0}(\cancel{p})\Bigg|_{p_k^0=p_k}=\left(\prod_k\textbf{T}_{p_k}\right)\Sigma^{0}(\cancel{p}).
\end{equation}
The restriction to renormalized parameters is to be understood as acting on (bare) parameter derivatives of $\tilde{\Sigma}^{0}$, the operators $\textbf{T}_{p^0}$ and $\textbf{T}_{p}$ are understood to only differ in their derivatives, the product $\prod_k$ goes over all parameters of the theory. Thus all three self-energies are compactly related as follows (making the flavour indices explicit).
\begin{equation}\label{eq:SE_relation}
    \tilde{\Sigma}^0_{ji}(\cancel{p})=\left(\prod_k\textbf{T}_{p_k}\right)\Sigma^0_{ji}(\cancel{p})=\gamma^0(Z^\dagger)^{-1}_{jl}\gamma^0\Sigma^R_{lr}(\cancel{p})Z^{-1}_{ri}.
\end{equation}
Here $j,i,l$ and $r$ are flavor indices, $l$ and $r$ are summed over. It is important to note that the derivative terms correspond to 1PI diagrams with counterterm insertions. 

Third, $n$-loop order means certain powers of renormalized couplings that come out of $n$-loop diagrams, hence, anything that has such powers of renormalized couplings we call of $n$th order. For example, the field counterterm can be expanded order by order
\begin{equation}
    Z=1+\delta Z=1+\delta Z^{(1)}+\delta Z^{(2)}+\dots.
\end{equation}
Here the order is indicated with superscripts in parentheses. In a similar manner, the series operator can also be expanded due to the powers of counterterms appearing in its definition, for example, the second order of the series operator is 
\begin{equation}
    \textbf{T}^{(2)}_{p^0}= \delta p^{(2)}\frac{\partial}{\partial p^0}+\frac{1}{2}(2 \delta p^{(1)} \delta p^{(1)})\frac{\partial^2}{\partial (p^0)^2}.
\end{equation}
Here the second term came form the square of the counterterm. 

Since all the self-energies can be written in terms of renormalized couplings, they can also be expanded in these orders
\begin{equation}
    \Sigma(\cancel{p})=\Sigma^{(0)}(\cancel{p})+\Sigma^{(1)}(\cancel{p})+\Sigma^{(2)}(\cancel{p})+\dots.
\end{equation}
It is worth noting that the superscript for $\Sigma^{0(n)}(\cancel{p})$ also indicates the number of loops, while for $\tilde{\Sigma}^{0(n)}(\cancel{p})$ and $\Sigma^{R(n)}(\cancel{p})$ it indicates the order only. 

Fourth, having the relations between self-energies we discuss the renormalization conditions. Since we go beyond 1-loop we must also discuss renormalization conditions for diagonal counterterms as they appear in the above relation even for off-diagonal cases. Considering the diagonal components of self-energies is made harder by the absorptive parts already at 1-loop. Due to hermiticity of the Lagrangian the diagonal mass counterterms must be real, which means that the mass counterterm cannot account for the absorptive parts. To also include these absorptive parts in the diagonal contributions it is needed to have a consistent definition of the decay width, which we successfully avoided in the preceding discussion. Hence, for consistency and simpler discussion \textit{we drop all the absorptive parts}.

Without the absorptive parts we can write down the standard On-Shell renormalization conditions~\cite{Aoki1982}:
\begin{enumerate}
    \item No mixing for external legs ($i\neq j$)
    \begin{equation}
        \Sigma^{R(n>0)}_{ji}(\cancel{p})u_i=0.
    \end{equation}
    \item The pole mass $m^P$ is equal to the renormalized mass $m_i$
    \begin{equation}
        m^P_{i}=m_i \rightarrow  \Sigma^{R(n>0)}_{ii}(\cancel{p})u_i=0.
    \end{equation}
    \item Residue at the pole is unity
\begin{equation}
\begin{split}
    &\lim_{\cancel{p}\rightarrow m_i}\frac{1}{\cancel{p}-m_i}\Sigma^{R}_{ii}(\cancel{p})u_i=u_i \\
    \rightarrow&\lim_{\cancel{p}\rightarrow m_i}\frac{1}{\cancel{p}-m_i}\Sigma^{R(n>0)}_{ii}(\cancel{p})u_i=\frac{d}{d\cancel{p}}\Sigma^{R(n>0)}_{ii}(\cancel{p})u_i=0.
\end{split}
\end{equation}
\end{enumerate}

Conveniently, the second condition simply extends the first condition to all $i$ and $j$, while the third condition is used to fix the diagonal component of the field renormalization.

Finally, we will take gauge derivatives of the various self-energies and use Nielsen identities. For the bare self-energies the form of the Nielsen identity is the same as in eq.~(\ref{eq:NI}), however, for $\tilde{\Sigma}^0$ the derivative must be taken w.r.t. the bare gauge parameter $\xi^0$ and for $\Sigma^0$ w.r.t. the renormalized one $\xi$. For the $\tilde{\Sigma}^0$ case we simply add tildes on $\Lambda$'s to indicate that they are computed in terms of bare parameters.

The gauge derivative of the renormalized self-energy is taken w.r.t. the renormalized gauge parameter, however, now there can be modifications of the Nielsen identity as in eq.~(6) in~\cite{Gambino2000}. There are four renormalization-related modifications: due to physical parameters, due to $\Lambda$'s in eq.~(\ref{eq:NI}), due to tadpoles, and due to gauge parameters. However, most of these modifications are irrelevant to us. 

The physical parameters contribute to modifications if their counterterms are gauge-dependent --- we choose them to be gauge-independent. 

As $\Lambda$'s must also be renormalized a modification can come if the self-energies and $\Lambda$'s are renormalized with different schemes (e.g., On-Shell vs. $\overline{\mathrm{MS}}$). On the one hand this simply induces a shift in the renormalized $\Lambda$'s, on the other hand we will not be directly using the Nielsen identity on the renormalized self-energy and so renormalized $\Lambda$'s are not going to appear. 

As mentioned in Section~\ref{sec:SE_decomposition} we do not renormalize the tadpoles, which is equivalent to the FJ scheme, so there is no modification due to tadpole renormalization. Although, using the language of the FJ scheme, the bare fields do not develop the vacuum expectation values due to additional and arbitrary shifts of the bare vacuum expectation value~\cite{Denner2016} --- this allows to have a well defined generating functional~\cite{Gambino2000} and references therein.

The only relevant modification is due to gauge-dependence of the counterterms of gauge parameters. This modification first contributes at second order and comes from taking the (renormalized) gauge derivative of the series operator $\prod_k\textbf{T}_{p_k}$. This modification contributes to the field renormalization.

Now we can re-derive our results (\textit{without} absorptive parts) as well as expand them beyond 1-loop order.

\subsection{Zeroth order}

Taking the 0th order terms in eq.~(\ref{eq:SE_relation}) we simply get the following
\begin{equation}
    \tilde{\Sigma}^{0(0)}_{ji}(\cancel{p})=\Sigma^{0(0)}_{ji}(\cancel{p})=\Sigma^{R(0)}_{ji}(\cancel{p})=(\cancel{p}-m_i)\delta_{ji}
\end{equation}
and there is no need to do anything else.

\subsection{First order}
Now we take the first order terms in eq.~(\ref{eq:SE_relation}) and get the following
\begin{equation}\label{eq:1st_order}
\begin{split}
    \tilde{\Sigma}^{0(1)}_{ji}(\cancel{p})=\Sigma^{0(1)}_{ji}(\cancel{p})+\sum_k\frac{\partial \Sigma^{0(0)}_{ji}(\cancel{p})}{\partial p_k}\delta p^{(1)}_k&=\Sigma^{R(1)}_{ji}(\cancel{p})-\gamma^0\delta Z^{\dagger(1)}_{jk}\gamma^0\Sigma^{R(0)}_{ki}(\cancel{p})-\Sigma^{R(0)}_{jk}(\cancel{p})\delta Z^{(1)}_{ki} \\
    &=\Sigma^{R(1)}_{ji}(\cancel{p})-\gamma^0\delta Z^{\dagger(1)}_{ji}\gamma^0(\cancel{p}-m_i)-(\cancel{p}-m_j)\delta Z^{(1)}_{ji}
\end{split}
\end{equation}
Let us take care of the derivative term. The only parameter entering the tree level self-energy is the mass, so this term gives us the mass counterterm and nothing else. The mass is fundamentally a matrix, so one must take parameter derivatives with respect to every element. In the above we have the derivative w.r.t. the renormalized mass which is diagonalized, unlike the bare mass. To correctly account for this diagonalization we take the derivative w.r.t. to $m_l\delta_{kl}$, where the Kronecker delta acts as a symbolic place holder for off-diagonal indices, we get
\begin{equation}\label{eq:mass_derivative}
    \sum_{k,l}\frac{\partial (\cancel{p}-m_i)\delta_{ji}}{\partial (m_l \delta_{kl})}\delta m^{(1)}_{kl}=-\sum_{k,l}\delta_{kj}\delta_{li}\delta m^{(1)}_{kl}=-\delta m^{(1)}_{ji}.
\end{equation}
One arrives at the same result by taking the derivative of $\tilde{\Sigma}^{0(0)}(\cancel{p})$ w.r.t. the bare mass parameter
\begin{equation}
\begin{split}
	\sum_{k,l}\frac{\partial (\cancel{p}\delta_{ji}-m^0_{ji})}{\partial m^0_{kl}}\Bigg|_{m^0=m}\delta m^{(1)}_{kl}&=-\sum_{k,l}\delta_{kj}\delta_{li}\delta m^{(1)}_{kl}\\
    &=-\delta m^{(1)}_{ji}.
\end{split}
\end{equation}
With this result we can already recognize the 1-loop renormalized self-energy we had in previous sections. 

Now we can impose the renormalization condition to eq.~(\ref{eq:1st_order})
\begin{equation}\label{eq:1st_order_Z}
    \left[\Sigma^{0(1)}_{ji}(\cancel{p})-\delta m^{(1)}_{ji}\right]u_i=-\left[(\cancel{p}-m_j)Z^{(1)}_{ji}\right]u_i
\end{equation}
Once decomposed in terms of scalar functions, the above equation gives the same relation between mass and field counterterms as in eqs.~(\ref{eq:dZL},~\ref{eq:dZR}). 

To get the gauge dependence we use the Nielsen identity and arrive at
\begin{equation}\label{eq:1st_order_Zgauge}
        \partial_\xi\delta Z^{(1)}_{ji}u_i=-\frac{(\cancel{p}+m_j)}{(m_i^2-m_j^2)}\left[(\cancel{p}-m_j)\bar{\Lambda}^{(1)}_{ji}(\cancel{p})\right]u_i=-\bar{\Lambda}^{(1)}_{ji}(\cancel{p})u_i.
\end{equation}
Here we already used our 1-loop result that the mass counterterm is gauge independent. The above can be decomposed to give the results in eq.~(\ref{eq:gauge_dep}) without mass counterterms. Obviously, this again shows that the ${m_i^2-m_j^2}$ mass structure cancels for the gauge dependent part and so the discussion of previous sections applies: taking the coefficient of ${m_i^2-m_j^2}$ is a sensible definition for (the anti-hermitian part of) off-diagonal components of the field counterterms. 

It is very important to note that this result also holds for diagonal parts. To see this we impose the third renormalization condition in eq.~(\ref{eq:1st_order}) and get
%\begin{equation}
%\begin{split}
%    \lim_{\cancel{p}\rightarrow m_i}\frac{1}{\cancel{p}-m_i}\left[\Sigma^{0(1)}_{ii}(\cancel{p})-\delta m^{(1)}_{ii}\right]u_i&=\lim_{\cancel{p}\rightarrow m_i}\frac{1}{\cancel{p}-m_i}\left[\Sigma^{R(1)}_{ii}(\cancel{p})-\gamma^0\delta Z^{\dagger(1)}_{ii}\gamma^0(\cancel{p}-m_i)-(\cancel{p}-m_i)\delta Z^{(1)}_{ii}\right]u_i\\
%   &=-\left[\delta Z^{\dagger(1)}_{ii}+\delta Z^{(1)}_{ii}\right]u_i=-2\delta Z^{H(1)}_{ii}u_i.
%\end{split}
%\end{equation}
\begin{equation}
\begin{split}
    \lim_{\cancel{p}\rightarrow m_i}\frac{1}{\cancel{p}-m_i}\left[\Sigma^{R(1)}_{ii}(\cancel{p})-\gamma^0\delta Z^{\dagger(1)}_{ii}\gamma^0(\cancel{p}-m_i)-(\cancel{p}-m_i)\delta Z^{(1)}_{ii}\right]u_i&=-\left[\delta Z^{\dagger(1)}_{ii}+\delta Z^{(1)}_{ii}\right]u_i\\
    &=-2\delta Z^{H(1)}_{ii}u_i.
\end{split}
\end{equation}
Here we used a trick from the Dirac algebra playbook when taking the limit in the $\gamma^0\delta Z^\dagger\gamma^0$ term. In this term both the numerator and the denominator vanish at $\cancel{p}=m_i$, so one must take the derivative $\frac{d}{d\cancel{p}}$, which has the Dirac properties of $\cancel{p}$. Commuting the derivative through $\gamma^0\delta Z^\dagger\gamma^0$ gets rid of the surrounding gamma matrices.

Now we take the gauge derivative of $\delta Z^{H(1)}_{ii}$ and arrive at
\begin{equation}\label{eq:diagonal_gauge}
\begin{split}
    \partial_\xi\delta Z^{H(1)}_{ii}u_i&=-\frac{1}{2}\lim_{\cancel{p}\rightarrow m_i}\frac{1}{\cancel{p}-m_i}\left[\Lambda^{(1)}_{ii}(\cancel{p})(\cancel{p}-m_i)+(\cancel{p}-m_i)\bar{\Lambda}^{(1)}_{ii}(\cancel{p})\right]u_i\\
    &=-\frac{1}{2}\lim_{\cancel{p}\rightarrow m_i}\frac{1}{\cancel{p}-m_i}\left[\Lambda^{(1)}_{ii}(\cancel{p})(\cancel{p}-m_i)\right]u_i-\frac{1}{2}\bar{\Lambda}^{(1)}_{ii}(\cancel{p})u_i
    %&=-\frac{1}{2}\left[\gamma^0\Lambda^{(1)}_{ii}(\cancel{p}^\dagger)\gamma^0+\bar{\Lambda}^{(1)}_{ii}(\cancel{p})\right]u_i=-\frac{1}{2}\left[\left(\bar{\Lambda}^{(1)}(\cancel{p}^\dagger)\right)^\dagger_{ii}+\bar{\Lambda}^{(1)}_{ii}(\cancel{p})\right]u_i=-\bar{\Lambda}^{H(1)}_{ii}(\cancel{p})u_i\\
    %&=-\bar{\Lambda}^{(1)}_{ii}(\cancel{p})u_i.
\end{split}
\end{equation}

In the second term the problematic $(\cancel{p}-m_i)$ factor cancels and there is no need to take the limit. On the other hand, in the first term the cancellation is not trivial and it is needed to use L'Hopital's rule. The important thing to note is that taking the limit does not require to commute $\cancel{p}$ through the projectors $P_{L,R}$ to act on the spinor. Let us decompose the $\Lambda(\cancel{p})$, take only the term containing $\cancel{p}P_L$, and apply the L'Hopital's rule
\begin{equation}
    \lim_{\cancel{p}\rightarrow m_i}\frac{d}{d\cancel{p}}\left[\Lambda^{L(1)}_{ii}(p^2)\cancel{p}P_L(\cancel{p}-m_i)\right]=\lim_{\cancel{p}\rightarrow m_i}\Lambda^{L(1)}_{ii}(p^2)\cancel{p}P_R\frac{d}{d\cancel{p}}(\cancel{p}-m_i)=\Lambda^{L(1)}_{ii}(m_i^2)m_iP_R.
\end{equation}
Terms with the derivative not acting on $\cancel{p}-m_i$ vanished. In a similar manner, the derivative exchanges the projectors $P_L\leftrightarrow P_R$ in all terms in the decomposition. Since we have dropped the absorptive parts, pseudo-hermiticity gives $\gamma^0\Lambda(\cancel{p})\gamma^0=\left(\bar{\Lambda}(\cancel{p})\right)^{\dagger}$, where the hermitian conjugation is acting on flavour and Dirac structures. This gives relations between scalar functions, namely $\Lambda^{L,R}=\bar{\Lambda}^{L,R\dagger}$ and $\Lambda^{sL,sR}=\bar{\Lambda}^{sR,sL\dagger}$. Then, the full term with the limit gives the same as
%\begin{equation}
%    \lim_{\cancel{p}\rightarrow m_i}\frac{d}{d\cancel{p}}\left[\Lambda^{(1)}_{ii}(\cancel{p})(\cancel{p}-m_i)\right]\rightarrow\left[\bar{\Lambda}^{L(1)\dagger}_{ii}(\cancel{p})\cancel{p} P_L+\bar{\Lambda}^{R(1)\dagger}_{ii}(\cancel{p})\cancel{p} P_R+\bar{\Lambda}^{sL(1)\dagger}_{ii}(\cancel{p}) P_L+\bar{\Lambda}^{sR(1)\dagger}_{ii}(\cancel{p})P_R\right]u_i\equiv\bar{\Lambda}^{(1)\dagger}_{ii}(\cancel{p})u_i.
%\end{equation}
\begin{equation}
    \begin{split}
        \lim_{\cancel{p}\rightarrow m_i}\frac{d}{d\cancel{p}}\left[\Lambda^{(1)}_{ii}(\cancel{p})(\cancel{p}-m_i)\right]\rightarrow &\Big[\bar{\Lambda}^{L(1)\dagger}_{ii}(\cancel{p})\cancel{p} P_L+\bar{\Lambda}^{R(1)\dagger}_{ii}(\cancel{p})\cancel{p}P_R\\  &+\bar{\Lambda}^{sL(1)\dagger}_{ii}(\cancel{p}) P_L+\bar{\Lambda}^{sR(1)\dagger}_{ii}(\cancel{p})P_R\Big]u_i\\
        \equiv&\bar{\Lambda}^{(1)\dagger}_{ii}(\cancel{p})u_i
    \end{split}
\end{equation}

Plugging the above into eq.~(\ref{eq:diagonal_gauge}) we get
\begin{equation}
\begin{split}
    \partial_\xi\delta Z^{H(1)}_{ii}u_i&=-\frac{1}{2}\left[\bar{\Lambda}^{(1)\dagger}_{ii}(\cancel{p})+\bar{\Lambda}^{(1)}_{ii}(\cancel{p})\right]u_i\\
    &\equiv -\bar{\Lambda}^{H(1)}_{ii}(\cancel{p})u_i=-\bar{\Lambda}^{(1)}_{ii}(\cancel{p})u_i,
\end{split}
\end{equation}
where $\bar{\Lambda}^{H(1)}$ is defined to have for its scalar functions the hermitian parts of the scalar functions of $\bar{\Lambda}^{(1)}$. Dropping the absorptive parts means that the diagonal components of self-energies are real and so are the diagonal components of $\Lambda$'s, this gave the final equality. Since one cannot fix the anti-hermitian part of the field renormalization for the diagonal components we simply set it to zero at 1-loop as is usually done. Eventually, ${\delta Z^{H(1)}_{ii}=\delta Z^{(1)}_{ii}}$ and the gauge dependence matches the one found for off-diagonal components. That the gauge dependence is the same for diagonal and off-diagonal components has been explicitly checked at 1-loop in~\cite{Espriu2002} (although in the $Z$ and $\bar{Z}$ approach with absorptive parts included). Here we find the same result in a more general setting. This is an essential piece of information when going beyond 1-loop, since gauge-derivatives of diagonal parts also appear in expressions for off-diagonal components.

\subsection{Second order}
Now we take the second order terms in eq.~(\ref{eq:SE_relation}) and get the following
\begin{equation}\label{eq:2nd_order}
\begin{split}
    \tilde{\Sigma}^{0(2)}_{ji}(\cancel{p})=&\Sigma^{0(2)}_{ji}(\cancel{p})-\delta m^{(2)}_{ji}+\sum_k\frac{\partial \Sigma^{0(1)}_{ji}(\cancel{p})}{\partial p_k}(\delta p^{(1)}_k)\\
    =&\Sigma^{R(2)}_{ji}(\cancel{p})+\gamma^0\delta Z^{\dagger(1)}_{jk}\gamma^0\Sigma^{R(0)}_{kl}(\cancel{p})Z^{(1)}_{li}-\gamma^0\delta Z^{\dagger(1)}_{jk}\gamma^0\Sigma^{R(1)}_{ki}(\cancel{p})-\Sigma^{R(1)}_{jk}(\cancel{p})Z^{(1)}_{ki}\\
    &+\gamma^0\left((\delta Z^{\dagger(1)})^2_{jk}-\delta Z^{\dagger(2)}_{jk}\right)\gamma^0\Sigma^{R(0)}_{ki}(\cancel{p})+\Sigma^{R(0)}_{jk}(\cancel{p})\left((\delta Z^{(1)})^2_{ki}-\delta Z^{(2)}_{ki}\right).
\end{split}
\end{equation}

Upon using the first two renormalization conditions and inserting the definitions of renormalized self-energies in terms of bare ones the above equation simplifies considerably and we get

\begin{equation}\label{eq:2nd_order_simple}
    \begin{split}
        (m^2_i-m^2_j)\delta Z^{(2)}_{ji}u_i=&-(\cancel{p}+m_j)\Bigg[\Sigma^{0(2)}_{ji}(\cancel{p})-\delta m^{(2)}_{ji}+\sum_k\frac{\partial \Sigma^{0(1)}_{ji}(\cancel{p})}{\partial p_k}\delta p^{(1)}_k\\
        &\qquad \qquad \quad+\Sigma^{0(1)}_{jk}(\cancel{p})\delta Z^{(1)}_{ki}-\delta m^{(1)}_{jk}\delta Z^{(1)}_{ki}\Bigg]u_i.
    \end{split}
\end{equation}

Just as for the previous order we may use the Nielsen identity, but first let us separately take the gauge derivative on the term containing parameter derivatives. Here it is also important to separate a few cases: the mass parameter, other physical parameters, and non-physical parameters (e.g. gauge-fixing parameters) as they all have slightly different behaviour.

We take the counterterms of physical parameters to be gauge-independent so that the gauge derivative effectively only acts on the self-energy. If these physical parameters are not the mass parameter we get
\begin{equation}
    \sum_k\frac{\partial}{\partial p_k}\partial_\xi \Sigma^{0(1)}_{ji}(\cancel{p}) \delta p^{(1)}_k=\sum_k\delta p^{(1)}_k\left[\frac{\partial}{\partial p_k}\left(\Lambda^{(1)}_{ji}(\cancel{p})\right)(\cancel{p}-m_i)+(\cancel{p}-m_j)\frac{\partial}{\partial p_k}\left(\bar{\Lambda}^{(1)}_{ji}(\cancel{p})\right)\right].
\end{equation}
On the other hand, if the physical parameter is the mass parameter, there is an additional piece coming from the parameter derivative acting on ${(\cancel{p}-m)}$. Performing the derivative as in eq.~(\ref{eq:mass_derivative}) the additional piece is
\begin{equation}\label{eq:mass_piece}
    -\Lambda^{(1)}_{jk}(\cancel{p})\delta m^{(1)}_{ki}-\delta m^{(1)}_{jk}\bar{\Lambda}^{(1)}_{ki}(\cancel{p}).
\end{equation}
With this piece, some cancellations take place in eq.~(\ref{eq:2nd_order_simple}) once the gauge derivative is taken.

If the parameters are the gauge parameters their parameter derivatives already give the Nielsen identity when acting on self-energies, in addition, the counterterms can now depend on gauge parameters. Indicating $\Lambda$'s with the superscript $l$ to mean that they come from the Nielsen identity for the gauge parameter $\xi_l$ we get
\begin{equation}
\begin{split}
    \partial_\xi \sum_l\frac{\partial}{\partial \xi_l}\Sigma^{0(1)}_{ji}(\cancel{p}) \delta \xi^{(1)}_l
    =&\sum_l\delta \xi^{(1)}_l\frac{\partial}{\partial \xi_l}\left[\Lambda^{(1)}_{ji}(\cancel{p})(\cancel{p}-m_i)+(\cancel{p}-m_j)\bar{\Lambda}^{(1)}_{ji}(\cancel{p})\right]\\
    &+\sum_l\partial_\xi(\delta \xi^{(1)}_l)\left[\Lambda^{l(1)}_{ji}(\cancel{p})(\cancel{p}-m_i)+(\cancel{p}-m_j)\bar{\Lambda}^{l(1)}_{ji}(\cancel{p})\right].
\end{split}
\end{equation}
Here the second line is due to gauge-dependence of gauge parameter counterterms --- this is the modification of the renormalized Nielsen identity. The important thing to note is that from the parameter derivative terms we get the additional piece in eq.~(\ref{eq:mass_piece}), the modification part in the above, and that the terms with ${\cancel{p}-m_i}$ on the right will vanish due to the spinor $u_i$. 

Finally, we take the gauge derivative of eq.~(\ref{eq:2nd_order_simple}) assuming that the mass counterterm is gauge-independent, using the known gauge dependence of field renormalization for all $i$ and $j$, and using eq.~(\ref{eq:1st_order_Z}) to get a term with $\delta m^{(1)}$ we arrive at
\begin{equation}\label{eq:2loop_gauge}
    \begin{split}
        (m_i^2-m_j^2)\partial_\xi\delta Z^{(2)}_{ji}u_i=-(m_i^2-m_j^2)&\Bigg[\bar{\Lambda}^{(2)}_{ji}(\cancel{p})+\bar{\Lambda}^{(1)}_{jk}(\cancel{p})\delta Z^{(1)}_{ki}
    +\sum_k \delta p^{(1)}_k\frac{\partial}{\partial p_k}\left(\bar{\Lambda}^{(1)}_{ji}(\cancel{p})\right)\\
    &+\sum_l \partial_\xi\left(\delta \xi^{(1)}_l\right)\bar{\Lambda}^{l(1)}_{ji}(\cancel{p})\Bigg]u_i,    
    \end{split}
\end{equation}
which again shows that the $m_i^2-m_j^2$ factor cancels for the gauge-dependent part. Most importantly, this tells us that the definition we have made for the field and mass counterterms at 1-loop can very well be extended to 2-loop order. In complete analogy, the anti-hermitian part of field renormalization is defined by taking the coefficient of ${m_i^2-m_j^2}$ in eq.~(\ref{eq:2nd_order_simple}), afterwards one can simply solve for the mass counterterms. The mass counterterm is then gauge-independent by definition. Comparing with the first order discussion, now there is a contribution to the field renormalization associated with the modification of the Nielsen identity.

To fully complete the discussion we should also determine the gauge-dependence of diagonal components, however, this notation already seems plentiful and so we deal with the diagonal components in the next section where we consider arbitrary orders. 

\subsection{Arbitrary order}

In this section we extend the scheme to arbitrary orders. The discussion comes in two main parts: w.r.t. the bare gauge parameter and then the renormalized one. Deriving results with the bare gauge parameter helps avoid modifications of the Nielsen identity and makes the discussion slightly more compact. Afterwards we will simply relate the bare and renormalized gauge-derivatives. For convenience, we summarize definitions at the end.

\subsubsection{Bare gauge parameter}
In this subsection the discussion is w.r.t. the bare gauge parameter $\xi^0$ and the bare self-energy in terms of bare parameters $\tilde{\Sigma}^{0}(\cancel{p})$. 

To begin, we rewrite eq.~(\ref{eq:barebare_SE}) and immediately apply the renormalization condition
\begin{equation}\label{eq:bareSE_order1}
    \tilde{\Sigma}^{0}_{jk}(\cancel{p})Z_{ki}u_i=\gamma^0(Z^\dagger)^{-1}_{jk}\gamma^0\Sigma^{R}_{ki}(\cancel{p})u_i=0.
\end{equation}
At arbitrary order $n$ we assume that all the counterterms have been defined up to order $n-1$. Having this in mind we can rewrite the above at order $n$ 
\begin{equation}\label{eq:bareSE_order2}
\begin{split}
    (\cancel{p}-m_j)\delta Z^{(n)}_{ji}u_i=-\left[\tilde{\Sigma}^{0(n)}_{ji}(\cancel{p})+\sum_{l=1}^{n-1}\tilde{\Sigma}^{0(n-l)}_{jk}(\cancel{p})\delta Z^{(l)}_{ki}\right]u_i.
\end{split}
\end{equation}
One can check that this reproduces the tree, 1-loop, and 2-loop level results we had previously by using eq.~(\ref{eq:SE_relation}).

Before looking at the gauge dependence we should also fix the diagonal components at arbitrary order. To do so we rewrite eq.~(\ref{eq:barebare_SE}) for the diagonal part
\begin{equation}\label{eq:diagonal_order}
    \gamma^0 Z^\dagger_{ir}\gamma^0\tilde{\Sigma}^{0}_{rk}(\cancel{p})Z_{ki}=\Sigma^R_{ii}(\cancel{p})
\end{equation}
and use the third renormalization condition
\begin{equation}
\lim_{\cancel{p}\rightarrow m_i}\frac{1}{\cancel{p}-m_i}\left[\gamma^0 Z^\dagger_{ir}\gamma^0\tilde{\Sigma}^{0}_{rk}(\cancel{p})Z_{ki}\right]u_i=\frac{d}{d\cancel{p}}\Sigma^R_{ii}(\cancel{p})u_i=0.
\end{equation}
Expanding the above at arbitrary order $n$ and taking trivial limits, one arrives at
\begin{equation}\label{eq:daigonal_order_full}
\begin{split}
    \delta Z^{H(n)}_{ii}u_i=-\frac{1}{2}\lim_{\cancel{p}\rightarrow m_i}\frac{1}{\cancel{p}-m_i}&\left[\tilde{\Sigma}^{0(n)}_{ii}(\cancel{p})
%    +(\cancel{p}-m_i)\delta Z^{(n)}_{ii}
%    +\gamma^0\delta Z^{\dagger(n)}_{ii}\gamma^0(\cancel{p}-m_i)
    +\sum_{l=1}^{n-1}\tilde{\Sigma}^{0(n-l)}_{ik}(\cancel{p})\delta Z^{(l)}_{ki}
    +\sum_{l=1}^{n-1}\gamma^0 \delta Z^{\dagger(l)}_{ik}\gamma^0\tilde{\Sigma}^{0(n-l)}_{ki}(\cancel{p})\right.\\
    &\left.
%    +\sum_{l=1}^{n-1}\gamma^0 \delta Z^{\dagger(n-l)}_{ik}\gamma^0(\cancel{p}-m_k)\delta Z^{(l)}_{ki}
    +\sum_{l=1}^{n-1}\sum_{f=1}^{n-l}\gamma^0 \delta Z^{\dagger(l)}_{ir}\gamma^0\tilde{\Sigma}^{0(n-l-f)}_{rk}(\cancel{p})\delta Z^{(f)}_{ki}\right]u_i.
\end{split}
\end{equation}

If it was not for the limit, one could take eq.~(\ref{eq:bareSE_order2}) to replace the ${(n-l)}$ order self-energy in the last term on the first line to cancel the sum on the second 
line, but this is not the case. Since the off-diagonal components seem a lot simpler, we first deal with their gauge dependence. However, for the moment we assume that the diagonal components result in the same gauge dependence as off-diagonal ones.

We remind that we can use the Nielsen identities also for the self-energies with bare parameters, the only difference is that the $\Lambda$'s are now also in terms of bare parameters, hence, the tildes. Taking the bare gauge derivative of eq.~(\ref{eq:bareSE_order2}), multiplying by ${\cancel{p}+m_j}$ and also dropping the $\cancel{p}$ arguments we get

\begin{equation}
\begin{split}
    (m^2_i-m^2_j)\partial_{\xi^0} \delta Z^{(n)}_{ji}u_i=-(\cancel{p}+m_j)&\Bigg[(\cancel{p}-m_j)\tilde{\bar{\Lambda}}^{(n)}_{ji}+\sum^{n-1}_{l=1}\left[\tilde{\Lambda}^{(l)}_{jk}\tilde{\Sigma}^{0(n-l)}_{ki}+\tilde{\Sigma}^{0(n-l)}_{jk}\tilde{\bar{\Lambda}}^{(l)}_{ki}\right]\\
    &+\sum_{l=1}^{n-1}\tilde{\Sigma}^{0(n-l)}_{jk}\partial_{\xi^0} \delta Z^{(l)}_{ki}\\
    &+\sum_{l=1}^{n-1}\sum^{n-l}_{f=1}\left[\tilde{\Lambda}^{(f)}_{jr}\tilde{\Sigma}^{0(n-l-f)}_{rk}+\tilde{\Sigma}^{0(n-l-f)}_{jr}\tilde{\bar{\Lambda}}^{(f)}_{rk}\right]\delta Z^{(l)}_{ki}\Bigg]u_i.
\end{split}
\end{equation}

Here we separated the $\Lambda^{(n)}$ contribution from the Nielsen identity of the $n$th order self-energy. As can be seen in eq.~(\ref{eq:bareSE_order1}) and eq.~(\ref{eq:bareSE_order2}) we can rewrite the bare self-energy $\tilde{\Sigma}^{0(n-l)}$ in terms of lower order bare self-energies and field counterterms --- this gives the cancellation with the second term on the second line. The simplified expression is then

\begin{equation}
    \begin{split}
        (m^2_i-m^2_j)\partial_{\xi^0} \delta Z^{(n)}_{ji}u_i=-(\cancel{p}+m_j)&\Bigg[(\cancel{p}-m_j)\tilde{\bar{\Lambda}}^{(n)}_{ji}+\sum^{n-1}_{l=1}\tilde{\Sigma}^{0(n-l)}_{jk}\left(\tilde{\bar{\Lambda}}^{(l)}_{ki}+\partial_{\xi^0} \delta Z^{(l)}_{ki}\right)\\
        &+\sum_{l=1}^{n-1}\sum^{n-l}_{f=1}\tilde{\Sigma}^{0(n-l-f)}_{jr}\tilde{\bar{\Lambda}}^{(f)}_{rk}\delta Z^{(l)}_{ki}\Bigg]u_i.
    \end{split}
\end{equation}

Here  one can already get the 1-loop and 2-loop gauge-dependence (up to modification terms) we have found in previous sections and subsections by taking $n=1$ and $n=2$, but the ${m^2_i-m^2_j}$ factor is not obvious for higher orders. One can notice that in the first sum the self-energy has the order between $1$ and $n-2$, since for $n-1$ (i.e. $l=1$) the gauge dependence in the parentheses cancels. In the second sum the order of the self-energy is between $0$ and $n-2$. We separate the $0$th order term from the second sum and combine the two remaining sums since the orders of self-energies match. We get the following

\begin{equation}
\begin{split}
    (m^2_i-m^2_j)\partial_{\xi^0} \delta Z^{(n)}_{ji}u_i
    =-(\cancel{p}+m_j)&\Bigg[(\cancel{p}-m_j)\left[\tilde{\bar{\Lambda}}^{(n)}_{ji}
    +\sum^{n-1}_{l=1}\tilde{\bar{\Lambda}}^{(n-l)}_{jk}\delta Z^{(l)}_{ki}\right]\\
    &+\sum^{n-1}_{l=2}\tilde{\Sigma}^{0(n-l)}_{jk}\left(\tilde{\bar{\Lambda}}^{(l)}_{ki}+\sum^{l-1}_{f=1}\tilde{\bar{\Lambda}}^{(l-f)}_{kr}\delta Z^{(f)}_{ri}+\partial_{\xi^0} \delta Z^{(l)}_{ki}\right)\Bigg]u_i.
\end{split}
\end{equation}

Here, on the second line the sum over $l$ starts at 2 so that the self-energy has the correct range of orders. Now, the combined sum on the second line first appears for $n=3$, but vanishes by inserting the gauge dependence of $\partial_{\xi^0} \delta Z^{(2)}$. In this way we see that when inserting the explicit result for $\partial_{\xi^0} \delta Z^{(l)}$ the combined sum always vanishes, hence, at arbitrary order we get the following result
\begin{equation}\label{eq:all_order_gauge}
    \partial_{\xi^0} \delta Z^{(n)}_{ji}u_i
    =-\left[\tilde{\bar{\Lambda}}^{(n)}_{ji}
    +\sum^{n-1}_{l=1}\tilde{\bar{\Lambda}}^{(n-l)}_{jk}\delta Z^{(l)}_{ki}\right]u_i
\end{equation}
showing that for off-diagonal field counterterms the ${m_i^2-m_j^2}$ mass structure cancels in the gauge-dependent part at all orders. So far we have assumed that the diagonal field counterterms have the same gauge dependence as off-diagonal ones at all orders --- now we set out to prove this.

Consider the gauge derivative of eq.~(\ref{eq:daigonal_order_full})

\begin{align}
    \nonumber
    \partial_{\xi^0} \delta Z^{H(n)}_{ii}u_i=&-\frac{1}{2}\lim_{\cancel{p}\rightarrow m_i}\frac{1}{\cancel{p}-m_i}\Bigg[\tilde{\Lambda}^{(n)}_{ii}(\cancel{p}-m_i)+(\cancel{p}-m_i)\tilde{\bar{\Lambda}}^{(n)}_{ii}+\sum^{n-1}_{l=1}\left[\underbrace{\tilde{\Lambda}^{(l)}_{ik}\tilde{\Sigma}^{0(n-l)}_{ki}}_{a}+\underbrace{\tilde{\Sigma}^{0(n-l)}_{ik}\tilde{\bar{\Lambda}}^{(l)}_{ki}}_{a^\prime}\right]\\
%second line
    \nonumber
    &+\sum_{l=1}^{n-1}\underbrace{\tilde{\Sigma}^{0(n-l)}_{ik}\partial_{\xi^0} \delta Z^{(l)}_{ki}}_{c^\prime}+\sum_{l=1}^{n-1}\sum^{n-l}_{f=1}\left[\underbrace{\tilde{\Lambda}^{(f)}_{ir}\tilde{\Sigma}^{0(n-l-f)}_{rk}}_{a}+\underbrace{\tilde{\Sigma}^{0(n-l-f)}_{ir}\tilde{\bar{\Lambda}}^{(f)}_{rk}}_{a^\prime}\right]\delta Z^{(l)}_{ki}\\
%third line
    \nonumber
    &+\sum_{l=1}^{n-1}\underbrace{\gamma^0 \partial_{\xi^0}\delta Z^{\dagger(l)}_{ik}\gamma^0\tilde{\Sigma}^{0(n-l)}_{ki}}_{c}
    +\sum_{l=1}^{n-1}\sum^{n-l}_{f=1}\gamma^0\delta Z^{\dagger(l)}_{ik}\gamma^0\left[\underbrace{\tilde{\Lambda}^{(f)}_{kr}\tilde{\Sigma}^{0(n-l-f)}_{ri}}_{b}+\underbrace{\tilde{\Sigma}^{0(n-l-f)}_{kr}\tilde{\bar{\Lambda}}^{(f)}_{ri}}_{b^\prime}\right]\\
%fifth line
    \nonumber
    &+\sum_{l=1}^{n-1}\sum_{f=1}^{n-l}\underbrace{\gamma^0 \partial_{\xi^0}\delta Z^{\dagger(l)}_{ir}\gamma^0\tilde{\Sigma}^{0(n-l-f)}_{rk}\delta Z^{(f)}_{ki}}_{c}
    +\sum_{l=1}^{n-1}\sum_{f=1}^{n-l}\underbrace{\gamma^0 \delta Z^{\dagger(l)}_{ir}\gamma^0\tilde{\Sigma}^{0(n-l-f)}_{rk}\partial_{\xi^0}\delta Z^{(f)}_{ki}}_{c^\prime}\\
%sixth line
    &+\left.\sum_{l=1}^{n-2}\sum_{f=1}^{n-l-1}\sum_{q=1}^{n-l-f}\gamma^0 \delta Z^{\dagger(l)}_{ir}\gamma^0\left[\underbrace{\tilde{\Lambda}^{(q)}_{rt}\tilde{\Sigma}^{0(n-l-f-q)}_{tk}}_{b}+\underbrace{\tilde{\Sigma}^{0(n-l-f-q)}_{rt}\tilde{\bar{\Lambda}}^{(q)}_{tk}}_{b^\prime}\right]\delta Z^{(f)}_{ki}\right]u_i.   
\end{align}
%\end{split}
%\end{equation}
Here, the underbraces label the terms for later rearrangements. Although crowded the above is simple to achieve by applying the Nielsen identities. On the last line the sum over $l$ goes up to $n-2$ since the gauge derivative acting on the 0th order self-energy gives 0. Due to the limit we cannot use eq.~(\ref{eq:bareSE_order2}) to lower orders of self-energies and get cancellations, hence, we simply rearrange to better see logical structures. From the $a^\prime$ term on the second line and the $b$ term on the third line we separate the 0th order self-energies, while we combine the remaining terms with the same labels, the result is as follows:

\begin{equation}
\begin{split}\label{eq:factors}
    \partial_{\xi^0} \delta Z^{H(n)}_{ii}u_i=&-\frac{1}{2}\lim_{\cancel{p}\rightarrow m_i}\frac{1}{\cancel{p}-m_i}\Bigg[
    (\cancel{p}-m_i)\sum^{n-1}_{l=1}\tilde{\bar{\Lambda}}^{(n-l)}_{ik}\delta Z^{(l)}_{ki}
    +\sum^{n-1}_{l=1}\gamma^0\delta Z^{\dagger(l)}_{ik}\gamma^0\tilde{\Lambda}^{(n-l)}_{ki}(\cancel{p}-m_i)\\
%second line
    &+\sum^{n-1}_{l=1}\left(\underbrace{\tilde{\Lambda}^{(l)}_{ik}}_{a}+\sum^{l-1}_{f=1}\underbrace{\gamma^0\delta Z^{\dagger(f)}_{it}\gamma^0\tilde{\Lambda}^{(l-f)}_{tk}}_{b}+\underbrace{\gamma^0\partial_{\xi^0} \delta Z^{\dagger(l)}_{ik}\gamma^0}_{c}\right)\\
    &\qquad \qquad \times\left(\tilde{\Sigma}^{0(n-l)}_{ki}+\sum^{n-l}_{q=1}\tilde{\Sigma}^{0(n-l-q)}_{kr}\delta Z^{(q)}_{ri}\right)\\
% third line
    &+\sum^{n-1}_{l=1}\left(\tilde{\Sigma}^{0(n-l)}_{ik}+\sum^{n-l}_{q=1}\gamma^0 \delta Z^{\dagger(q)}_{it}\gamma^0\tilde{\Sigma}^{0(n-l-q)}_{tk}\right)\\
    &\qquad \qquad \times \left(\underbrace{\tilde{\bar{\Lambda}}^{(l)}_{ki}}_{a^\prime}+\sum^{l-1}_{f=1}\underbrace{\tilde{\bar{\Lambda}}^{(l-f)}_{kr}\delta Z^{(f)}_{ri}}_{b^\prime}+\underbrace{\partial_{\xi^0} \delta Z^{(l)}_{ki}}_{c^\prime}\right)\Bigg]u_i-\tilde{\bar{\Lambda}}^{H(n)}_{ii}u_i.   
\end{split}
\end{equation}
Here we already used our off-diagonal case knowledge to separate $\tilde{\bar{\Lambda}}^{H(n)}$ and also kept the underbraces to indicate the origin of the terms. On lines 2-4 there are two huge terms that each have two clear factors emphasized by the $\times$ symbol. The factors containing $\Lambda$'s would vanish when acted on by the spinor and when assuming that the gauge-dependence of the field counterterms is the same for all $i$ and $j$. The factors without $\Lambda$'s represent the renormalization condition and would also vanish when acted on by the spinors. Of course, the present limit does not allow to simply take these terms to 0. To get rid of the limit we notice that the left factors would vanish when acted on by the $\bar{u}_i$ spinor\footnote{We can use this because the absorptive parts are dropped and the renormalization conditions work for outgoing particles as well.} on the left and so they must be proportional to ${\cancel{p}-m_i}$. For example, taking the self-energy factor on the fourth line and eq.~(\ref{eq:SE_relation}) we can expand around the $\bar{u}_i\Sigma^{R}_{ik}=0$ renormalization condition
\begin{equation}
    \tilde{\Sigma}^{0(n-l)}_{ik}+\sum^{n-l}_{q=1}\gamma^0 \delta Z^{\dagger(q)}_{it}\gamma^0\tilde{\Sigma}^{0(n-l-q)}_{tk}=(\Sigma^RZ^{-1})^{(n-l)}_{ik}=\sum_{g=1}(\cancel{p}-m_i)^{g} (A_g)^{(n-l)}_{ik}. 
\end{equation}
Here $A_g$ is some coefficient of order $(n-l)$ coming from $\cancel{p}$ derivatives of $\Sigma^RZ^{-1}$. The $A_g$'s also have Dirac structure, but the important part is that the ${\cancel{p}-m_i}$ factor is on the left side and further determination is not needed. Analogously, we can take the factor on the second line in eq.\ref{eq:factors} containing  $\Lambda$'s, which vanishes at $\cancel{p}=m_i$, and expand it around $m_i$ with some coefficients $B_g$. Putting all of this in we get
\begin{equation}
\begin{split}
    \partial_{\xi^0} \delta Z^{H(n)}_{ii}u_i=-\frac{1}{2}&\lim_{\cancel{p}\rightarrow m_i}\frac{1}{\cancel{p}-m_i}\Bigg[
    (\cancel{p}-m_i)\sum^{n-1}_{l=1}\tilde{\bar{\Lambda}}^{(n-l)}_{ik}\delta Z^{(l)}_{ki}
    +\sum^{n-1}_{l=1}\gamma^0\delta Z^{\dagger(l)}_{ik}\gamma^0\tilde{\Lambda}^{(n-l)}_{ki}(\cancel{p}-m_i)\\
%second line
    &+\sum^{n-1}_{l=1}\left(\sum_{g=1}(\cancel{p}-m_i)^{g}(B_g)^{(l)}_{ik}\right)\left(\tilde{\Sigma}^{0(n-l)}_{ki}+\sum^{n-l}_{q=1}\tilde{\Sigma}^{0(n-l-q)}_{kr}\delta Z^{(q)}_{ri}\right)\\
% third line
    &+\sum^{n-1}_{l=1}\left(\sum_{g=1}(\cancel{p}-m_i)^g (A_g)^{(n-l)}_{ik}\right)\left(\tilde{\bar{\Lambda}}^{(l)}_{ki}+\sum^{l-1}_{f=1}\tilde{\bar{\Lambda}}^{(l-f)}_{kr}\delta Z^{(f)}_{ri}+\partial_{\xi^0} \delta Z^{(l)}_{ki}\right)\Bigg]u_i\\
    &-\tilde{\bar{\Lambda}}^{H(n)}_{ii}u_i.   
\end{split}
\end{equation}
Conveniently, the $\frac{1}{\cancel{p}-m_i}$ simply lowers the order of ${(\cancel{p}-m_i)^g}$ by one and there is no more limit. After the limit is gone we can freely act with the spinor $u_i$ on the right side of the terms containing $B$'s and $A$'s. The term containing $B$'s vanishes due to the renormalization condition in eq.~(\ref{eq:bareSE_order1}), the term containing $A$'s vanishes as in the off-diagonal case as long as the gauge dependence of the field renormalization is the same for all $i$ and $j$.
Eventually, we are left with a fairly simple answer
\begin{equation}
\begin{split}
     \partial_{\xi^0} \delta Z^{H(n)}_{ii}u_i&=-\tilde{\bar{\Lambda}}^{H(n)}_{ii}u_i
     -\frac{1}{2}\lim_{\cancel{p}\rightarrow m_i}\frac{1}{\cancel{p}-m_i}\Bigg[
     (\cancel{p}-m_i)\sum^{n-1}_{l=1}\tilde{\bar{\Lambda}}^{(n-l)}_{ik}\delta Z^{(l)}_{ki}\\
     &\hspace{5cm}+\sum^{n-1}_{l=1}\gamma^0\delta Z^{\dagger(l)}_{ik}\gamma^0\tilde{\Lambda}^{(n-l)}_{ki}(\cancel{p}-m_i)
     \Bigg]u_i\\
     &=-\left[\tilde{\bar{\Lambda}}^{(n)}_{ii}+\sum^{n-1}_{l=1}\left(\tilde{\bar{\Lambda}}^{(n-l)}_{ik}\delta Z^{(l)}_{ki}+\tilde{\bar{\Lambda}}^{\dagger(n-l)}_{ki}\delta Z^{\dagger(l)}_{ik}\right)\right]u_i.
\end{split}
\end{equation}
Here we used that $\Lambda^{H}_{ii}=\Lambda_{ii}$, in the second equality for compact notation we are forced to change the order of Dirac structures in the $\delta Z^\dagger$ term. However, this is still the hermitian part of eq.~(\ref{eq:all_order_gauge}) for $i=j$. If we want this to hold, we can no longer set the diagonal anti-hermitian part of the field renormalization to zero beyond 1-loop, since otherwise the sums do not fully cancel. Again, since there is no other way to determine the anti-hermitian part for the diagonal field counterterm we can choose it to have the same gauge dependence as the anti-hermitian part of eq.~(\ref{eq:all_order_gauge}). This is most easily achieved if the diagonal component is determined from the off-diagonal one by simply setting $i=j$. This is allowed since the off-diagonal anti-hermitian part is the coefficient of ${m_i^2-m_j^2}$ and no singular behaviour occurs for $i=j$. In addition, at 1-loop this choice gives the usual 0 since only absorptive parts could contribute to the diagonal anti-hermitian part --- we have dropped these parts in this appendix. With this choice we see that the gauge dependence for field counterterms can be described with the same eq.~(\ref{eq:all_order_gauge}) for all $i$ and $j$ to all orders in general without any explicit computation of $\Lambda$'s, which is an unexpected and non-trivial result. This result can also be written in a very compact form
\begin{equation}\label{eq:all_order_gague_nice}
    \partial_{\xi^0}Zu_i=-\tilde{\bar{\Lambda}}Zu_i.
\end{equation}
This is analogous to the result achieved for the LSZ factor in eq.(48) of~\cite{Gambino2000}, however, our approach allows to define field and mass counterterms order by order. Importantly, having off-diagonal mass counterterms does not change gauge-dependence properties, but rather allows to get rid of singular behavior and the need to renormalize mixing matrices, hence, not all contributions are to be included into field renormalization or LSZ factors.

\subsubsection{Renormalized gauge parameter}
So far we have found the gauge-dependence of the field renormalization in terms of the bare gauge parameter. This avoided the discussion of the terms modifying the Nielsen identity. In this subsection we relate the bare and renormalized gauge derivatives, which is made easier by the results of the previous subsection. 

We begin by taking the gauge derivative of eq.~(\ref{eq:bareSE_order1}) w.r.t. the renormalized gauge parameter $\xi$ and explicitly expand the bare self-energy $\tilde{\Sigma}^0$ in terms of renormalized parameters
\begin{equation}\label{eq:NI_modification}
    \partial_\xi\left(\prod_k\textbf{T}_{p^0_k}\tilde{\Sigma}^0\big|_{p^0=p} Z\right)u_i=\partial_\xi\left(\prod_k\textbf{T}_{p^0_k}\right)\tilde{\Sigma}^0\big|_{p^0=p} Zu_i
    +\left(\prod_k\textbf{T}_{p_k}\partial_\xi\Sigma^0\right) Zu_i
    +\tilde{\Sigma}^0\partial_\xi Zu_i=0.
\end{equation}
It should be understood that the series operator acts only up to the restriction $\big|_{p^0=p}$. The first term is the gauge derivative of the counterterms appearing in the series operators, in the second term we used eq.~(\ref{eq:barebare_relation}) before taking the gauge derivative, the third term simply contains the gauge derivative of the field renormalization and we un-expanded the self-energy. 

Let us consider the term containing the gauge derivative of the self-energy as we can immediately use the Nielsen identity. After using the identity we can conveniently get rid of the series operator by simply writing everything in terms of the bare parameters:
\begin{equation}\label{eq:gauge_result1}
    \left(\prod_k\textbf{T}_{p_k}\partial_\xi\Sigma^0\right) Zu_i=\left(\prod_k\textbf{T}_{p_k}\left(\Lambda\Sigma^0+\Sigma^0\bar{\Lambda}\right)\right) Zu_i=\left(\tilde{\Lambda}\tilde{\Sigma}^0+\tilde{\Sigma}^0\tilde{\bar{\Lambda}}\right)Zu_i=-\tilde{\Sigma}^0\partial_{\xi^0}Zu_i.
\end{equation}
To get the final equality we used eq.~(\ref{eq:all_order_gague_nice}) and eq.~(\ref{eq:bareSE_order1}) which showed that $\tilde{\Lambda}\tilde{\Sigma}^0Zu_i$ vanishes due to the renormalization condition.

Now we take care of the remaining term with the gauge derivative of counterterms. First, we use  $Z=1+ \delta Z$  and the fact the spinor $u_i$ can go through the series operators --- this allows to use eq.~(\ref{eq:bareSE_order2}), we get
\begin{equation}\label{eq:operator_gauge}
\begin{split}
    \partial_\xi\left(\prod_k\textbf{T}_{p^0_k}\right)\tilde{\Sigma}^0\big|_{p^0=p} Zu_i&=    \partial_\xi\left(\prod_k\textbf{T}_{p^0_k}\right)\tilde{\Sigma}^0u_i\big|_{p^0=p}
    +\partial_\xi\left(\prod_k\textbf{T}_{p^0_k}\right)\tilde{\Sigma}^0\big|_{p^0=p}\delta Z u_i\\
    &=-\partial_\xi\left(\prod_k\textbf{T}_{p^0_k}\right)\tilde{\Sigma}^0\delta Z\big|_{p^0=p}u_i
    +\partial_\xi\left(\prod_k\textbf{T}_{p^0_k}\right)\tilde{\Sigma}^0\big|_{p^0=p} \delta Z u_i.
\end{split}
\end{equation}
Now we have the gauge derivative of the series operator acting on $\tilde{\Sigma}^0\delta Z$. Conveniently, the series expansion remains the same if one expands the counterterm and the self-energy separately. The gauge derivative then acts on two series operators, for which we simply have the Leibniz rule, eventually we get
\begin{equation}
\begin{split}
    -\partial_\xi\left(\prod_k\textbf{T}_{p^0_k}\right)\tilde{\Sigma}^0\delta Z\big|_{p^0=p}u_i&=
    -\partial_\xi\left(\prod_k\textbf{T}_{p^0_k}\right)\tilde{\Sigma}^0\big|_{p^0=p}\prod_k\textbf{T}_{p^0_k}\delta Z\big|_{p^0=p}u_i\\
    &\hspace{0.5cm}-\prod_k\textbf{T}_{p^0_k}\tilde{\Sigma}^0\big|_{p^0=p}\partial_\xi\left(\prod_k\textbf{T}_{p^0_k}\right)\delta Z\big|_{p^0=p}u_i\\
    &=-\partial_\xi\left(\prod_k\textbf{T}_{p^0_k}\right)\tilde{\Sigma}^0\big|_{p^0=p}\delta Zu_i
    -\tilde{\Sigma}^0\partial_\xi\left(\prod_k\textbf{T}_{p^0_k}\right)\delta Z\big|_{p^0=p}u_i.
\end{split}
\end{equation}
To get the final equality we removed the series expansions and left the quantities written in terms of bare parameters where possible. Putting this back in eq.~(\ref{eq:operator_gauge}) there are cancellations and only one term remains
\begin{equation}\label{eq:gauge_result2}
    \partial_\xi\left(\prod_k\textbf{T}_{p^0_k}\right)\tilde{\Sigma}^0\big|_{p^0=p} Zu_i=-\tilde{\Sigma}^0\partial_\xi\left(\prod_k\textbf{T}_{p^0_k}\right)\delta Z\big|_{p^0=p}u_i=-\tilde{\Sigma}^0\partial_\xi\left(\prod_k\textbf{T}_{p^0_k}\right)Z\big|_{p^0=p}u_i.
\end{equation}
In the final equality we simply added 1 to the field counterterm since the gauge derivative of the series operator has only derivative terms acting on the field counterterm.

Finally, we collect our results in eq.~(\ref{eq:gauge_result1}) and eq.~(\ref{eq:gauge_result2}), put them in eq.~(\ref{eq:NI_modification})
\begin{equation}
  \tilde{\Sigma}^0\left[-\partial_\xi\left(\prod_k\textbf{T}_{p^0_k}\right)Z\big|_{p^0=p}- \partial_{\xi^0}Z+\partial_\xi Z\right]u_i=0
\end{equation}
Dropping the self-energy and rearranging we get
\begin{equation}
    \partial_\xi Z u_i=\left[\partial_{\xi^0}Z+\partial_\xi\left(\prod_k\textbf{T}_{p^0_k}\right)Z\big|_{p^0=p}\right]u_i.
\end{equation}
Again, there are no singularities coming from the ${m_i^2-m_j^2}$ mass structure, meaning that our definitions of mass and field counterterms remain valid with the renormalized gauge parameter. As a quick check, at second order the above correctly reproduces eq.~(\ref{eq:2loop_gauge}) with the expected term related to the modification of the Nielsen identity. If one chose to renormalize the physical parameters in a gauge-dependent way, the term with the series operator would also reproduce the needed modification of the Nielsen identity. 

\subsubsection{Summary of the counterterms}
Now we can go back to eq.~(\ref{eq:bareSE_order2}) and eq.~(\ref{eq:daigonal_order_full}) and extract (or simply copy for convenience) the $n$th order definitions of mass and field counterterms. 
First, to separate the mass counterterm at order $n$ one must expand the bare self-energy $\tilde{\Sigma}^{0(n)}$ and separate the term where $\delta m^{(n)}$ appears, thus we define $\hat{\Sigma}^{0(n)}$ by
\begin{equation}
    \tilde{\Sigma}^{0(n)}_{ji}(\cancel{p})=\hat{\Sigma}^{0(n)}_{ji}(\cancel{p})-\delta m^{(n)}_{ji}.
\end{equation}
Then, leaving the algebra to the reader and reminding that \textit{the absorptive parts are dropped}, the hermitian and anti-hermitian parts of the off-diagonal field counterterms\footnote{Remember that in the appendix there are no square roots and the field renormalization is defined as $\psi_0=Z\psi$.} are as follows\footnote{Here $((m_i^2-m_j^2)A+B)\bigg|_{m_i^2-m_j^2}=A$}
\begin{equation}
\begin{split}
    \delta Z^{H(n)}_{ji}u_i=-\frac{1}{2(m_i^2-m_j^2)}\left[(\cancel{p}+m_j)\left(\hat{\Sigma}^{0(n)}_{ji}(\cancel{p})+\sum_{l=1}^{n-1}\tilde{\Sigma}^{0(n-l)}_{jk}(\cancel{p})\delta Z^{(l)}_{ki}\right)
    - H.C.\right]u_i,
\end{split}
\end{equation}
\begin{equation}
    \delta Z^{A(n)}_{ji}u_i\equiv-\frac{1}{2}\left[(\cancel{p}+m_j)\left(\hat{\Sigma}^{0(n)}_{ji}(\cancel{p})+\sum_{l=1}^{n-1}\tilde{\Sigma}^{0(n-l)}_{jk}(\cancel{p})\delta Z^{(l)}_{ki}\right)
    + H.C.\right]u_i\Bigg|_{m_i^2-m_j^2},
\end{equation}
In the above $H.C.$ means Hermitian conjugation, although, we note that Hermitian conjugation makes more sense (is easier to apply) after using $u_i$ on $\cancel{p}$. As in the 1-loop definitions, for the anti-hermitian part we simply take the coefficient of $m_i^2-m_j^2$.
The hermitian part of the diagonal field counterterm is 
\begin{equation}
\begin{split}
    \delta Z^{H(n)}_{ii}u_i=-\frac{1}{2}\lim_{\cancel{p}\rightarrow m_i}\frac{1}{\cancel{p}-m_i}&\left[\tilde{\Sigma}^{0(n)}_{ii}(\cancel{p})
    +\sum_{l=1}^{n-1}\tilde{\Sigma}^{0(n-l)}_{ik}(\cancel{p})\delta Z^{(l)}_{ki}
    +\sum_{l=1}^{n-1}\gamma^0 \delta Z^{\dagger(l)}_{ik}\gamma^0\tilde{\Sigma}^{0(n-l)}_{ki}(\cancel{p})\right.\\
    &\left.
    +\sum_{l=1}^{n-1}\sum_{f=1}^{n-l}\gamma^0 \delta Z^{\dagger(l)}_{ir}\gamma^0\tilde{\Sigma}^{0(n-l-f)}_{rk}(\cancel{p})\delta Z^{(f)}_{ki}\right]u_i.
\end{split}
\end{equation}
and the anti-hermitian part is determined from the off-diagonal one
\begin{equation}
    \delta Z^{A(n)}_{ii}\equiv\left.\delta Z^{A(n)}_{ji}\right|_{i=j}.
\end{equation}

To reiterate, with these definitions the gauge-dependence of the field renormalization is described in the same way for all $i$ and $j$. In terms of the bare gauge parameter we have 
\begin{equation}
    \partial_{\xi^0} Z_{ji}u_i
    =-\left(\tilde{\bar{\Lambda}}(\cancel{p})Z\right)_{ji}u_i
\end{equation}
and in terms of the renormalized one
\begin{equation}
    \partial_\xi Z_{ji} u_i=\left[\partial_{\xi^0} Z_{ji}+\partial_\xi\left(\prod_k\textbf{T}_{p^0_k}\right)Z_{ji}\big|_{p^0=p}\right]u_i
\end{equation}
both of which hold order by order.

Finally, the mass counterterm for all $i$ and $j$ is
\begin{equation}
\begin{split}
    \delta m^{(n)}_{ji}u_i=\left[\hat{\Sigma}^{0(n)}_{ji}(\cancel{p})+\sum_{l=1}^{n-1}\tilde{\Sigma}^{0(n-l)}_{jk}(\cancel{p})\delta Z^{(l)}_{ki}+(\cancel{p}-m_j)\delta Z^{(n)}_{ji}\right]u_i.
\end{split}
\end{equation}
It can also be checked that the mass counterterm is gauge-independent by definition. For the diagonal component one can see that the $n$th order field counterterms do not contribute by multiplying both sides by ${\cancel{p}+m_i}$. Finally, we remind that the tildes above self-energies and $\Lambda$'s mean that the computation is done with bare parameters. To get these relations for renormalized parameters one needs to expand the bare parameters around their renormalized values, which gives contributions of the counterterms (up to order $n-1$) from other parameters in the theory.

We have shown that our 1-loop logic and results with the absorptive parts dropped can be extended to all orders. In addition, other 1-loop benefits are preserved to all orders: no singular behaviour in the degenerate mass limit, process independence, no need for counterterms of mixing matrices, basis rotations and renormalization commute, etc.
Although, the one thing we cannot show is the UV finiteness of the anti-hermitian part of the field renormalization --- such finiteness is not obvious beyond 1-loop and may be an accident.

\section{Full mass counterterms}\label{seca:full_exp}
In this appendix we simply provide the off-diagonal mass counterterms that were too big for the main body of the paper.
\subsection{Up-type quarks}
\begin{align}
% sm contribution
    \nonumber
    \delta m^{u,L}_{ji}=&\frac{V^{}_{jk}V^\star_{ik}}{2^D\pi^{D-2} v^2m^u_j}B_0((m^u_j)^2,(m^d_k)^2,m^2_W)^\star\\
    \nonumber
    &\qquad\times\Big[(m^d_k)^2((D-3)m^2_W-2(m^u_j)^2)+((m^u_j)^2-m^2_W)((D-2)m^2_W+(m^u_j)^2)\Big]\\
% sm higgs, sine
    \nonumber
    &+\frac{s^2_\alpha(G^{}_u)_{jk}}{2^D\pi^{D-2}}B_0((m^u_i)^2, m^2_h, (m^u_k)^2)\left[\frac{(G_u)^\star_{ik}((m^u_k)^2+(m^u_i)^2-m^2_h)}{8m^u_i}+\frac{m^u_k (G_u)_{ki}}{4}\right]\\
    \nonumber
    &+\frac{s^2_\alpha(G^{}_u)_{ki}}{2^D\pi^{D-2}}B_0((m^u_j)^2, m^2_h, (m^u_k)^2)^\star\left[\frac{(G_u)^\star_{kj}((m^u_k)^2+(m^u_j)^2-m^2_h)}{8m^u_j}+\frac{m^u_k (G_u)_{jk}}{4}\right]\\
% sm higgs, sine2 j   
    \nonumber
    &+\left[\frac{s_{2\alpha}m^u_j}{2^D\pi^{D-2}m^u_i}B_0((m^u_i)^2, m^2_h, (m^u_j)^2)+H.C.\right]\left[\frac{(G^{}_u)^\star_{ij}(m^2_h-(m^u_i)^2-(m^u_j)^2)}{8\sqrt{2}v}\right]\\
    \nonumber
    &+\left[\frac{s_{2\alpha}m^u_j}{2^D\pi^{D-2}m^u_i}B_0((m^u_i)^2, m^2_h, (m^u_j)^2)+H.C.\right]\left[\frac{m^u_i m^u_j(G_u)_{ji}}{4\sqrt{2}v}\right]\\
% sm higgs, sine2 i    
    \nonumber
    &+\frac{s_{2\alpha}(G^{}_u)_{ji}}{2^D\pi^{D-2}}\left[B_0((m^u_i)^2, m^2_h, (m^u_i)^2)\left[\frac{m^2_h-4(m^u_i)^2}{8\sqrt{2}v}\right]+H.C.\right]\\
% heavy Higgs, cos
    \nonumber
    &+\frac{c^2_\alpha(G^{}_u)_{jk}}{2^D\pi^{D-2}}B_0((m^u_i)^2, m^2_H, (m^u_k)^2)\left[\frac{(G_u)^\star_{ik}((m^u_k)^2+(m^u_i)^2-m^2_H)}{8m^u_i}+\frac{m^u_k (G_u)_{ki}}{4}\right]\\
    \nonumber
    &+\frac{c^2_\alpha(G^{}_u)_{ki}}{2^D\pi^{D-2}}B_0((m^u_j)^2, m^2_H, (m^u_k)^2)^\star\left[\frac{(G_u)^\star_{kj}((m^u_k)^2+(m^u_j)^2-m^2_H)}{8m^u_j}+\frac{m^u_k (G_u)_{jk}}{4}\right]\\
% heavy higgs, sin2 j
    \nonumber
    &+\left[\frac{s_{2\alpha}m^u_j}{2^D\pi^{D-2}m^u_i}B_0((m^u_i)^2, m^2_H, (m^u_j)^2)+H.C.\right]\left[\frac{(G^{}_u)^\star_{ij}((m^u_i)^2+(m^u_j)^2)-m^2_H}{8\sqrt{2}v}\right]\\
    \nonumber
    &+\left[\frac{s_{2\alpha}m^u_j}{2^D\pi^{D-2}m^u_i}B_0((m^u_i)^2, m^2_H, (m^u_j)^2)+H.C.\right]\left[\frac{m^u_i m^u_j(G_u)_{ji}}{4\sqrt{2}v}\right]\\
% heavy higgs, sin2 i
    \nonumber
    &+\frac{s_{2\alpha}(G^{}_u)_{ji}}{2^D\pi^{D-2}}\left[B_0((m^u_i)^2, m^2_H, (m^u_i)^2)\left[\frac{4(m^u_i)^2-m^2_H}{8\sqrt{2}v}\right]+H.C.\right]\\
% pseudo higgs
    \nonumber
    &+\frac{(G^{}_u)_{jk}}{2^D\pi^{D-2}}B_0((m^u_i)^2,m^2_A,(m^u_k)^2)\left[\frac{(G_u)^\star_{ik}((m^u_k)^2+(m^u_i)^2-m^2_A)}{8m^u_i}-\frac{m^u_k (G_u)_{ki}}{4}\right]\\
    \nonumber
    &+\frac{(G^{}_u)_{ki}}{2^D\pi^{D-2}}B_0((m^u_j)^2,m^2_A,(m^u_k)^2)^\star\left[\frac{(G_u)^\star_{kj}((m^u_k)^2+(m^u_j)^2-m^2_A)}{8m^u_j}-\frac{m^u_k (G_u)_{jk}}{4}\right]\\
% charged higgs
    \nonumber
    &+\frac{1}{2^D\pi^{D-2}}B_0((m^u_i)^2,m^2_{H^\pm},(m^u_k)^2)\left[(G_uV)_{jk}(V^\dagger G^\dagger_u)_{ki}\frac{(m^d_k)^2+(m^u_i)^2-m^2_{H^\pm}}{4m^u_i}\right]\\
    \nonumber
    &+\frac{1}{2^D\pi^{D-2}}B_0((m^u_i)^2,m^2_{H^\pm},(m^u_k)^2)\left[-\frac{(G_uV)_{jk}m^d_k(G_dV^\dagger)_{ki}}{2}\right]\\
    \nonumber
    &+\frac{1}{2^D\pi^{D-2}}B_0((m^u_j)^2,m^2_{H^\pm},(m^u_k)^2)^\star\left[(VG^\dagger_d)_{ik}(G_dV^\dagger)_{kj}\frac{(m^d_k)^2+(m^u_j)^2-m^2_{H^\pm}}{4m^u_j}\right]\\
    \nonumber
    &+\frac{1}{2^D\pi^{D-2}}B_0((m^u_j)^2,m^2_{H^\pm},(m^u_k)^2)^\star\left[-\frac{(G_uV)_{ki}m^d_k(G_dV^\dagger)_{kj}}{2}\right]\\
% SM A0 part, no tad stuff
    \nonumber
    &+\frac{A_0(m^2_W)}{2^D\pi^{D-2}}\frac{(V(m^d)^2V^\dagger)_{ji}}{2m^u_j}-\frac{A_0((m^d_k)^2)V_{jk}V^\star_{ik}}{2^D\pi^{D-2}}\left[\frac{(m^d_k)^2+(D-2)m^2_W+(m^u_j)^2}{2v^2m^u_j}\right]\\
% BSM A0 part, lambda 7
    \nonumber
    &-\frac{\lambda_7 v (G_u)_{ji}}{2^{D+1}\sqrt{2}\pi^{D-2}}\left[2A_0(m^2_{H^\pm})+A_0(m^2_A)+3s^2_\alpha A_0(m^2_h)+3c^2_\alpha A_0(m^2_H))\right]\left[\frac{s^2_\alpha}{m^2_h}+\frac{c^2_\alpha}{m^2_H}\right]\\
% BSM A0 part, lambda 3
    \nonumber
    &+\frac{s_{2\alpha}\lambda_3 v (G_u)_{ji}}{2^{D+2}\sqrt{2}\pi^{D-2}}\left[A_0(m^2_A)+2A_0(m^2_{H^{pm}})\right]\left[\frac{1}{m^2_h}-\frac{1}{m^2_H}\right]\\
    \nonumber
    &+\frac{s_{2\alpha}\lambda_3 v (G_u)_{ji}}{2^{D+2}\sqrt{2}\pi^{D-2}}\left[A_0(m^2_h)\left[\frac{3s^2_\alpha}{m^2_h}+\frac{3c_{2\alpha}+1}{2m^2_H}\right]+A_0(m^2_H)\left[\frac{3c_{2\alpha}-1}{2m^2_h}-\frac{3c^2_{\alpha}}{m^2_H}\right]\right]\\
% BSM A0 part, no lambda
    \nonumber
    &-\frac{A_0((m^d_k)^2)}{2^D\pi^{D-2}}\left[\frac{m^u_i(VG^\dagger_d)_{ik}(G_dV^\dagger)_{kj}+m^u_j(G_uV)_{jk}(V^\dagger G^\dagger_u)_{ki}}{4m^u_im^u_j}\right]\\
    \nonumber
    &-\frac{A_0((m^u_k)^2)}{2^D\pi^{D-2}}\left[\frac{m^u_i (G_u)_{ki}(G_u)^\star_{kj}+m^u_j(G_u)_{jk}(G_u)^\star_{ik}}{4m^u_im^u_j}\right]\\
% BSM A0 part, no lambda, no mix
    \nonumber
    &+\frac{s_{2\alpha}(G_u)_{ji}}{2^{D+1}\sqrt{2}\pi^{D-2}v^2}\left[\frac{1}{m^2_h}-\frac{1}{m^2_H}\right]\\
    \nonumber
    &\qquad\times\Big[2(D-1)m^2_WA_0(m^2_W)(D-1)m^2_ZA_0(m^2_Z)+(m^2_A-m^2_{H^\pm})A_0(m^2_A)\Big]\\
% BSM A0 part, no lambda, mix, SM higgs
    \nonumber
    &-\frac{s_{2\alpha}}{2^{D}\sqrt{2}\pi^{D-2}v}A_0(m^2_h)\left[\frac{((m^u_i)^2+(m^u_j)^2)(G_u)^\star_{ij}}{8m^u_im^u_j}+\frac{(c_{2\alpha}-4)(G_u)_{ji}}{4}\right]\\
    \nonumber
    &-\frac{s_{2\alpha}(G_u)_{ji}}{2^{D}\sqrt{2}\pi^{D-2}v}A_0(m^2_h)\left[\frac{3s^2_\alpha m^2_{H^\pm}}{2m^2_h}+\frac{2s^2_\alpha m^2_h-(3c_{2\alpha}+1)m^2_{H^\pm}}{4m^2_H}\right]\\
% BSM A0 part, no lambda, mix, heavy higgs
    \nonumber
    &+\frac{s_{2\alpha}}{2^{D}\sqrt{2}\pi^{D-2}v}A_0(m^2_H)\left[\frac{((m^u_i)^2+(m^u_j)^2)(G_u)^\star_{ij}}{8m^u_im^u_j}-\frac{(c_{2\alpha}+4)(G_u)_{ji}}{4}\right]\\
    \nonumber
    &+\frac{s_{2\alpha}(G_u)_{ji}}{2^{D}\sqrt{2}\pi^{D-2}v}A_0(m^2_H)\left[\frac{3c^2_\alpha m^2_{H^\pm}}{2m^2_H}+\frac{2c^2_\alpha m^2_h-(3c_{2\alpha}-1)m^2_{H^\pm}}{4m^2_h}\right]\\
% BSM A0, no lambda, pseudo higgs 
    \nonumber
    &+\frac{A_0(m^2_A)}{2^D\pi^{D-2}}\left[\frac{(G^{}_u G^\dagger_u)_{ji}}{8m^u_i}+\frac{(G^{\dagger}_u G^{}_u)_{ji}}{8m^u_j}+\frac{s_{2\alpha}(G_u)_{ji}(m^2_A-m^2_{H^\pm})}{2\sqrt{2}v}\left[\frac{1}{m^2_h}-\frac{1}{m^2_H}\right]\right]\\
% BSM A0, no lambda, charged
    \nonumber
    &+\frac{A_0(m^2_{H^\pm})}{2^D\pi^{D-2}}\left[\frac{(G^{}_u G^\dagger_u)_{ji}}{4m^l_i}+\frac{(VG^\dagger_d G^{}_d V^\dagger)_{ji}}{4m^l_j}\right]\\
% tadpole part, 4th power
    \nonumber
    &+\frac{s_{2\alpha}(G_u)_{ji}\sqrt{2}}{2^{D}\pi^{D-2}v}\left[\frac{1}{m^2_H}-\frac{1}{m^2_h}\right]\mathrm{Tr}\left\{A_0((m^\nu)^2)(m^\nu)^2(U^\dagger_L U^{}_L)\right.\\
    \nonumber
    &\hspace{2.5cm}\left.+A_0((m^d)^2)(m^d)^2+A_0((m^u)^2)(m^u)^2+A_0((m^l)^2)(m^l)^2\right\}\\
% tadpole part, +m_A
    \nonumber
    &+\frac{(G_u)_{ji}}{2^{D+1}\pi^{D-2}}\left[\frac{s^2_\alpha}{m^2_h}+\frac{c^2_\alpha}{m^2_H}+\frac{1}{m^2_A}\right]\mathrm{Tr}\Big\{A_0((m^\nu)^2)\left(m^\nu U^\dagger_L G^{\star}_\nu U^{}_R+m^\nu U^T_R G^\dagger_\nu U^\star_L\right)\Big\}\\
    \nonumber
    &+\frac{(G_u)_{ji}}{2^D\pi^{D-2}}\left[\frac{s^2_\alpha}{m^2_h}+\frac{c^2_\alpha}{m^2_H}+\frac{1}{m^2_A}\right]\\
    \nonumber
    &\qquad\times\mathrm{Tr}\Big\{A_0((m^d)^2)m^dG_d+A_0((m^u)^2)m^uG^\star_uA_0((m^l)^2)m^lG_l\Big\}\\
% tadpole part, -m_A    
    \nonumber
    &+\frac{(G_u)_{ji}}{2^{D+1}\pi^{D-2}}\left[\frac{s^2_\alpha}{m^2_h}+\frac{c^2_\alpha}{m^2_H}-\frac{1}{m^2_A}\right]\mathrm{Tr}\left\{A_0((m^\nu)^2)\left(m^\nu U^\dagger_RG^{T}_\nu U^{}_L+m^\nu U^T_L G^{}_\nu U^\star_R\right)\right\}\\
    \nonumber
    &+\frac{(G_u)_{ji}}{2^D\pi^{D-2}}\left[\frac{s^2_\alpha}{m^2_h}+\frac{c^2_\alpha}{m^2_H}-\frac{1}{m^2_A}\right]\\
    &\qquad\times\mathrm{Tr}\Big\{A_0((m^d)^2)m^dG^\star_d+A_0((m^u)^2)m^uG_uA_0((m^l)^2)m^lG^\star_l\Big\}
\end{align}
Here $m_H$, $m_A$, and $m_{H^\pm}$ are masses of the additional scalar, pseudoscalar, and charged scalar, respectively, coming from the second Higgs doublet. The SM Higgs contributes only in terms containing sines of $\alpha$ as the interaction of the first Higgs doublet is diagonal and we are computing $i\neq j$ terms. Contributions with traces $\mathrm{Tr\{\dots\}}$ over family indices come from tadpole diagrams with fermion loops. Summation over the index $k$ is implied. The first line and a few contributions with $A_0$ are the SM parts, while the rest are beyond SM contributions. It is easy to see that $\delta m^L$ is explicitly gauge independent as advertised by our scheme. The reader may also get $\delta m^R$ by hermitian conjugation of $\delta m^L$.

\subsection{Charged leptons}
\begin{align}
% SM B0
    \nonumber
    \delta m^{L,l}_{ji}&=\frac{U^{}_{Lja}U^{\star}_{Lia}}{2^{D+1}\pi^{D-2}v^2m^l_j}B_0((m^l_j)^2, (m^\nu_a)^2,m^2_W)^\star\\
    \nonumber
    &\qquad\times\left[(m^\nu_a)^4+(m^\nu_a)^2((D-3)m^2_W-2(m^l_j)^2)+((m^l_j)^2-m^2_W)((D-2)m^2_W+(m^l_j)^2)\right]\\
% SM A0
    \nonumber
    &+\frac{U^{}_{Lja}U^{\star}_{Lia}}{2^{D+1}\pi^{D-2}v^2m^l_j}A_0(m^2_W)\left[(m^\nu_a)^2-(m^l_j)^2+(D-2)m^2_W\right]\\
    \nonumber
    &-\frac{U^{}_{Lja}U^{\star}_{Lia}}{2^{D+1}\pi^{D-2}v^2m^l_j}A_0((m^\nu_a)^2)\left[(m^\nu_a)^2+(m^l_j)^2+(D-2)m^2_W\right]\\
% BSM, sm higgs, 
    \nonumber
    &+\frac{s^2_\alpha(G^{}_l)_{jk}}{2^D\pi^{D-2}}B_0((m^l_i)^2,m^2_h,(m^l_k)^2)\left[\frac{((m^l_k)^2+(m^l_i)^2-m^2_h)(G^{}_l)^\star_{ik}}{8m^l_i}+\frac{m^l_k(G^{}_l)_{ki}}{4}\right]\\
    \nonumber
    &+\frac{s^2_\alpha(G^{}_l)_{ki}}{2^D\pi^{D-2}}B_0((m^l_j)^2,m^2_h,(m^l_k)^2)^\star\left[\frac{((m^l_k)^2+(m^l_j)^2-m^2_h)(G^{}_l)^\star_{kj}}{8m^l_i}+\frac{m^l_k(G^{}_l)_{jk}}{4}\right]\\
% BSM, sm higgs, sin2 j
    \nonumber
    &+\left[\frac{s_{2\alpha}m^l_j}{2^D\pi^{D-2}m^l_i}B_0((m^l_i)^2,m^2_h,(m^l_j)^2)+H.C.\right]\left[\frac{(G_l)^\star_{ij}(m^2_h-(m^l_i)^2-(m^l_j)^2)}{8\sqrt{2}v}\right]\\
    \nonumber
    &+\left[\frac{s_{2\alpha}m^l_j}{2^D\pi^{D-2}m^l_i}B_0((m^l_i)^2,m^2_h,(m^l_j)^2)+H.C.\right]\left[\frac{m^l_i m^l_j(G_l)_{ji}}{4\sqrt{2}v}\right]\\
% BSM, sm higgs, sin2 i
    \nonumber
    &+\frac{s_{2\alpha}(G_l)_{ji}}{2^D\pi^{D-2}}\left[B_0((m^l_i)^2,m^2_h,(m^l_i)^2)\left[\frac{m^2_h-4(m^l_i)^2}{8\sqrt{2}v}\right]+H.C.\right]\\
% BSM, heavy higgs mix
    \nonumber
    &+\frac{c^2_\alpha(G^{}_l)_{jk}}{2^D\pi^{D-2}}B_0((m^l_i)^2,m^2_H,(m^l_k)^2)\left[\frac{((m^l_k)^2+(m^l_i)^2-m^2_H)(G^{}_l)^\star_{ik}}{8m^l_i}+\frac{m^l_k(G^{}_l)_{ki}}{4}\right]\\
    \nonumber
    &+\frac{c^2_\alpha(G^{}_l)_{ki}}{2^D\pi^{D-2}}B_0((m^l_j)^2,m^2_H,(m^l_k)^2)^\star\left[\frac{((m^l_k)^2+(m^l_j)^2-m^2_H)(G^{}_l)^\star_{kj}}{8m^l_i}+\frac{m^l_k(G^{}_l)_{jk}}{4}\right]\\
% BSM, heavy higgs, sin2 j
    \nonumber
    &+\left[\frac{s_{2\alpha}m^l_j}{2^D\pi^{D-2}m^l_i}B_0((m^l_i)^2,m^2_H,(m^l_j)^2)+H.C.\right]\left[\frac{(G_l)^\star_{ij}((m^l_i)^2+(m^l_j)^2-m^2_H)}{8\sqrt{2}v}\right]\\
    \nonumber
    &+\left[\frac{s_{2\alpha}m^l_j}{2^D\pi^{D-2}m^l_i}B_0((m^l_i)^2,m^2_H,(m^l_j)^2)+H.C.\right]\left[-\frac{m^l_i m^l_j(G_l)_{ji}}{4\sqrt{2}v}\right]\\    
% BSM, heavy higgs, sin2 i
    \nonumber
    &+\frac{s_{2\alpha}(G_l)_{ji}}{2^D\pi^{D-2}}\left[B_0((m^l_i)^2,m^2_H,(m^l_i)^2)\left[\frac{4(m^l_i)^2-m^2_H}{8\sqrt{2}v}\right]+H.C.\right]\\
% BSM, pseudo higgs
    \nonumber
    &+\frac{(G^{}_l)_{jk}}{2^D\pi^{D-2}}B_0((m^l_i)^2,m^2_A,(m^l_k)^2)\left[\frac{((m^l_k)^2+(m^l_i)^2-m^2_A)(G^{}_l)^\star_{ik}}{8m^l_i}-\frac{m^l_k(G^{}_l)_{ki}}{4}\right]\\
    \nonumber
    &+\frac{(G^{}_l)_{ki}}{2^D\pi^{D-2}}B_0((m^l_j)^2,m^2_A,(m^l_k)^2)^\star\left[\frac{((m^l_k)^2+(m^l_j)^2-m^2_A)(G^{}_l)^\star_{kj}}{8m^l_i}-\frac{m^l_k(G^{}_l)_{jk}}{4}\right]\\
% BSM, charged higgs
    \nonumber
    &+\frac{(G^{}_lU^{}_L)_{ja}}{2^D\pi^{D-2}}B_0((m^l_i)^2,m^2_{H^\pm},(m^\nu_a)^2)\left[\frac{((m^l_i)^2+(m^\nu_a)^2-m^2_{H^\pm})(G^\star_lU^\star_L)_{ia}}{4m^l_i}-\frac{m^\nu_a(G^{}_\nu U^\star_R)_{ia}}{2}\right]\\
    \nonumber
    &+\frac{(G^{}_\nu U^\star_R)_{ia}}{2^D\pi^{D-2}}B_0((m^l_j)^2,m^2_{H^\pm},(m^\nu_a)^2)^\star\left[\frac{((m^l_k)^2+(m^\nu_a)^2-m^2_{H^\pm})(G^\star_\nu  U^{}_R)_{ja}}{4m^l_j}-\frac{m^\nu_a(G^{}_l U^{}_L)_{ja}}{2}\right]\\
% BSM A0, lambda7
    \nonumber
    &-\frac{\lambda_7 v (G_l)_{ji}}{2^{D+1}\sqrt{2}\pi^{D-2}}\left[2A_0(m^2_{H^\pm})+A_0(m^2_A)+3s^2_\alpha A_0(m^2_h)+3c^2_\alpha A_0(m^2_H))\right]\left[\frac{s^2_\alpha}{m^2_h}+\frac{c^2_\alpha}{m^2_H}\right]\\
% BSM A0, lambda3 
    \nonumber
    &+\frac{s_{2\alpha}\lambda_3 v (G_l)_{ji}}{2^{D+2}\sqrt{2}\pi^{D-2}}\left[A_0(m^2_A)+2A_0(m^2_{H^{\pm}})\right]\left[\frac{1}{m^2_h}-\frac{1}{m^2_H}\right]\\
    \nonumber
    &+\frac{s_{2\alpha}\lambda_3 v (G_l)_{ji}}{2^{D+2}\sqrt{2}\pi^{D-2}}\left[A_0(m^2_h)\left[\frac{3s^2_\alpha}{m^2_h}+\frac{3c_{2\alpha}+1}{2m^2_H}\right]+A_0(m^2_H)\left[\frac{3c_{2\alpha}-1}{2m^2_h}-\frac{3c^2_{\alpha}}{m^2_H}\right]\right]\\
% BSM A0, no lambda, sine sm higgs
    \nonumber
    &+\frac{s_{\alpha} A_0(m^2_h)}{2^D\pi^{D-2}}\left[\frac{s_\alpha(G^{}_lG^\dagger_l)_{ji}}{8m^l_i}+\frac{s_\alpha(G^{\dagger}_lG^{}_l)_{ji}}{8m^l_j}+\frac{c_\alpha((m^l_i)^2+(m^l_j)^2)(G_l)^\star_{ij}}{4\sqrt{2}m^l_i m^l_jv}+\frac{2(c_{2\alpha}-4)(G_l)_{ji}}{2\sqrt{2}v}\right]\\
    \nonumber
    &-\frac{s_{2\alpha}(G_l)_{ji} }{2^{D+2}\sqrt{2}\pi^{D-2}v}A_0(m^2_h)\left[\frac{s^2_\alpha 6m^2_{H^\pm}}{m^2_h}+\frac{s^2_\alpha m^2_h+(3c_{2\alpha}+1)m^2_{H^\pm}}{m^2_H}\right]\\
% BSM A0, no lambda, Heavy higgs
    \nonumber
    &+\frac{c_{\alpha} A_0(m^2_H)}{2^D\pi^{D-2}}\left[\frac{c_\alpha(G^{}_lG^\dagger_l)_{ji}}{8m^l_i}+\frac{c_\alpha(G^{\dagger}_lG^{}_l)_{ji}}{8m^l_j}+\frac{s_\alpha((m^l_i)^2+(m^l_j)^2)(G_l)^\star_{ij}}{4\sqrt{2}m^l_i m^l_jv}-\frac{2(c_{2\alpha}+4)(G_l)_{ji}}{2\sqrt{2}v}\right]\\
    \nonumber
    &-\frac{s_{2\alpha}(G_l)_{ji} }{2^{D+2}\sqrt{2}\pi^{D-2}v}A_0(m^2_H)\left[\frac{c^2_\alpha 6m^2_{H^\pm}}{m^2_H}+\frac{c^2_\alpha m^2_H-(3c_{2\alpha}-1)m^2_{H^\pm}}{m^2_h}\right]\\
% BSM A0, no lambda, pseudo higgs 
    \nonumber
    &+\frac{A_0(m^2_A)}{2^D\pi^{D-2}}\left[\frac{(G^{}_lG^\dagger_l)_{ji}}{8m^l_i}+\frac{(G^{\dagger}_lG^{}_l)_{ji}}{8m^l_j}+\frac{s_{2\alpha}(G_l)_{ji}(m^2_A-m^2_{H^\pm})}{\sqrt{2}v}\left[\frac{1}{m^2_h}-\frac{1}{m^2_H}\right]\right]\\
% BSM A0, no lambda, charged
    \nonumber
    &+\frac{A_0(m^2_{H^\pm})}{2^D\pi^{D-2}}\left[\frac{(G^{}_lG^\dagger_l)_{ji}}{4m^l_i}+\frac{(G^{}_\nu G^\dagger_\nu)_{ij}}{4m^l_j}\right]\\
% BSM A0, no lambda, neutrino
    \nonumber
   & -\frac{1}{2^D\pi^{D-2}}A_0((m^\nu_a)^2)\left[\frac{(G_\nu U^\star_R)_{ia}(G^\star_\nu U^{}_R)_{ja}}{4m^l_j}+[\frac{(G_l U_L)_{ja}(G^\star_l U^\star_l)_{ia}}{4m^l_i}\right]\\
% BSM A0, no lambda, charged lepton
    \nonumber
    &-\frac{A_0((m^\nu_k)^2)}{2^D\pi^{D-2}}\left[\frac{(G_l)^{}_{ki}(G_l)^\star_{kj}}{4m^l_j}+\frac{(G_l)^{}_{jk}(G_l)^\star_{ik}}{4m^l_i}\right]\\
% BSM A0, no lambda, gauge bosons
    \nonumber
    &+\frac{s_{2\alpha}(G_l)_{ji}(D-1)}{2^{D+1}\sqrt{2}\pi^{D-2}}\left[m^2_W 2A_0(m^2_W)+m^2_Z A_0(m^2_Z)\right]\left[\frac{1}{m^2_h}-\frac{1}{m^2_H}\right]\\
% tadpole part, 4th power
    \nonumber
    &+\frac{s_{2\alpha}(G_l)_{ji}\sqrt{2}}{2^{D}\pi^{D-2}v}\left[\frac{1}{m^2_H}-\frac{1}{m^2_h}\right]\mathrm{Tr}\left\{A_0((m^\nu)^2)(m^\nu)^2(U^\dagger_L U^{}_L)\right.\\
    \nonumber
    &\hspace{2.5cm}\left.+A_0((m^d)^2)(m^d)^2+A_0((m^u)^2)(m^u)^2+A_0((m^l)^2)(m^l)^2\right\}\\    
% tadpole part, -m_A
    \nonumber
    &+\frac{(G_l)_{ji}}{2^{D+1}\pi^{D-2}}\left[\frac{s^2_\alpha}{m^2_h}+\frac{c^2_\alpha}{m^2_H}-\frac{1}{m^2_A}\right]\mathrm{Tr}\left\{A_0((m^\nu)^2)\left(m^\nu U^\dagger_L G^{\star}_\nu U^{}_R+\right)m^\nu U^T_R G^\dagger_\nu U^\star_L\right\}\\
    \nonumber
    &+\frac{(G_l)_{ji}}{2^D\pi^{D-2}}\left[\frac{s^2_\alpha}{m^2_h}+\frac{c^2_\alpha}{m^2_H}-\frac{1}{m^2_A}\right]\\
    \nonumber
    &\qquad\times\mathrm{Tr}\Big\{A_0((m^d)^2)m^dG_d+A_0((m^u)^2)m^uG^\star_u+A_0((m^l)^2)m^lG_l\Big\}\\
% tadpole part, +m_A    
    \nonumber
    &+\frac{(G_l)_{ji}}{2^{D+1}\pi^{D-2}}\left[\frac{s^2_\alpha}{m^2_h}+\frac{c^2_\alpha}{m^2_H}+\frac{1}{m^2_A}\right]\mathrm{Tr}\left\{A_0((m^\nu)^2)\left(m^\nu U^\dagger_RG^{T}_\nu U^{}_L+\right)m^\nu U^T_L G^{}_\nu U^\star_R\right\}\\
    \nonumber
    &+\frac{(G_l)_{ji}}{2^D\pi^{D-2}}\left[\frac{s^2_\alpha}{m^2_h}+\frac{c^2_\alpha}{m^2_H}+\frac{1}{m^2_A}\right]\\
    &\qquad\times\mathrm{Tr}\Big\{A_0((m^d)^2)m^dG^\star_d+A_0((m^u)^2)m^uG_u+A_0((m^l)^2)m^lG^\star_l\Big\}\
\end{align}
Here $i \neq j$, summations over indices $k$ and $a$ are implied. The index $k$ sums over the 3 usual generations, while $a$ goes up to 4. As is obvious from the expression, there is no gauge-dependence as should be the case.

\bibliography{maintext_jhep}

\providecommand{\href}[2]{#2}\begingroup\raggedright\begin{thebibliography}{10}

\bibitem{Denner1990_CKMreno}
A.~Denner and T.~Sack, \emph{{Renormalization of the quark mixing matrix}},
  \href{https://doi.org/10.1016/0550-3213(90)90557-T}{\emph{Nuclear Physics B}
  {\bfseries 347} (1990) 203}.

\bibitem{Cabibbo1963}
N.~Cabibbo, \emph{{Unitary Symmetry and Leptonic Decays}},
  \href{https://doi.org/10.1103/PhysRevLett.10.531}{\emph{Physical Review
  Letters} {\bfseries 10} (1963) 531}.

\bibitem{Kobayashi1973}
M.~Kobayashi and T.~Maskawa, \emph{{$CP$-Violation in the Renormalizable Theory
  of Weak Interaction}},
  \href{https://doi.org/10.1143/PTP.49.652}{\emph{Progress of Theoretical
  Physics} {\bfseries 49} (1973) 652}.

\bibitem{Gambino1999}
P.~Gambino, P.~Grassi and F.~Madricardo, \emph{{Fermion mixing renormalization
  and gauge invariance}},
  \href{https://doi.org/10.1016/S0370-2693(99)00321-4}{\emph{Physics Letters B}
  {\bfseries 454} (1999) 98} [\href{https://arxiv.org/abs/9811470}{{\ttfamily
  9811470}}].

\bibitem{Kniehl1996}
B.A.~Kniehl and A.~Pilaftsis, \emph{{Mixing renormalization in Majorana
  neutrino theories}},
  \href{https://doi.org/10.1016/0550-3213(96)00280-5}{\emph{Nuclear Physics B}
  {\bfseries 474} (1996) 286}.

\bibitem{Kniehl2006}
B.A.~Kniehl and A.~Sirlin, \emph{{Simple on-shell renormalization framework for
  the Cabibbo-Kobayashi-Maskawa matrix}},
  \href{https://doi.org/10.1103/PhysRevD.74.116003}{\emph{Physical Review D}
  {\bfseries 74} (2006) 116003}.

\bibitem{Feynman1949}
R.P.~Feynman, \emph{{Space-time approach to quantum electrodynamics}},
  \href{https://doi.org/10.1103/PhysRev.76.769}{\emph{Physical Review}
  {\bfseries 76} (1949) 769}.

\bibitem{Kniehl2009a}
B.A.~Kniehl and A.~Sirlin, \emph{{A Novel Formulation of
  Cabibbo-Kobayashi-Maskawa Matrix Renormalization}},
  \href{https://doi.org/10.1016/j.physletb.2009.02.024}{\emph{Physics Letters,
  Section B: Nuclear, Elementary Particle and High-Energy Physics} {\bfseries
  673} (2008) 208} [\href{https://arxiv.org/abs/0901.0114}{{\ttfamily
  0901.0114}}].

\bibitem{Yamada2001}
Y.~Yamada, \emph{{Gauge dependence of the on-shell renormalized mixing
  matrices}}, \href{https://doi.org/10.1103/PhysRevD.64.036008}{\emph{Physical
  Review D} {\bfseries 64} (2001) 036008}
  [\href{https://arxiv.org/abs/0103046}{{\ttfamily 0103046}}].

\bibitem{Diener2001}
K.-P.~Diener and B.~Kniehl, \emph{{On-mass-shell renormalization of fermion
  mixing matrices}},
  \href{https://doi.org/10.1016/S0550-3213(01)00453-9}{\emph{Nuclear Physics B}
  {\bfseries 617} (2001) 291} [\href{https://arxiv.org/abs/0109110}{{\ttfamily
  0109110}}].

\bibitem{Pilaftsis2002}
A.~Pilaftsis, \emph{{Gauge and scheme dependence of mixing matrix
  renormalization}},
  \href{https://doi.org/10.1103/PhysRevD.65.115013}{\emph{Physical Review D}
  {\bfseries 65} (2002) 115013}
  [\href{https://arxiv.org/abs/0203210}{{\ttfamily 0203210}}].

\bibitem{Denner2004}
A.~Denner, E.~Kraus and M.~Roth, \emph{{Physical renormalization condition for
  the quark-mixing matrix}},
  \href{https://doi.org/10.1103/PhysRevD.70.033002}{\emph{Physical Review D}
  {\bfseries 70} (2004) 033002}
  [\href{https://arxiv.org/abs/0402130}{{\ttfamily 0402130}}].

\bibitem{Zhou2003}
Y.~Zhou, \emph{{Renormalization of the Cabibbo-Kobayashi-Maskawa Matrix in
  Standard Model}},
  \href{https://doi.org/10.1016/j.physletb.2003.10.013}{\emph{Physics Letters
  B} {\bfseries 577} (2003) 67}
  [\href{https://arxiv.org/abs/0304003}{{\ttfamily 0304003}}].

\bibitem{Barroso2000}
A.~Barroso, L.~Br{\"{u}}cher and R.~Santos, \emph{{Renormalization of the
  Cabibbo-Kobayashi-Maskawa matrix}},
  \href{https://doi.org/10.1103/PhysRevD.62.096003}{\emph{Physical Review D}
  {\bfseries 62} (2000) 096003}.

\bibitem{Almasy2009}
A.A.~Almasy, B.A.~Kniehl and A.~Sirlin, \emph{{On-shell renormalization of the
  mixing matrices in Majorana neutrino theories}},
  \href{https://doi.org/10.1016/j.nuclphysb.2009.03.025}{\emph{Nuclear Physics
  B} {\bfseries 818} (2009) 115}
  [\href{https://arxiv.org/abs/0902.3793}{{\ttfamily 0902.3793}}].

\bibitem{Denner2007}
A.~Denner, \emph{{Techniques for the calculation of electroweak radiative
  corrections at the one-loop level and results for W-physics at LEP200}},
  {\emph{Fortschritte der Physik/Progress of Physics} {\bfseries 41} (2007)
  307} [\href{https://arxiv.org/abs/0709.1075}{{\ttfamily 0709.1075}}].

\bibitem{Grimus1989}
W.~Grimus and H.~Neufeld, \emph{{Radiative neutrino masses in an SU(2) × U(1)
  model}}, \href{https://doi.org/10.1016/0550-3213(89)90370-2}{\emph{Nuclear
  Physics B} {\bfseries 325} (1989) 18}.

\bibitem{Espriu2002}
D.~Espriu, J.~Manzano and P.~Talavera, \emph{{Flavour Mixing, Gauge Invariance
  and Wave-function Renormalisation}},
  \href{https://doi.org/10.1103/PhysRevD.66.076002}{\emph{Physical Review D -
  Particles, Fields, Gravitation and Cosmology} {\bfseries 66} (2002) 076002}
  [\href{https://arxiv.org/abs/0204085}{{\ttfamily 0204085}}].

\bibitem{Aoki1982}
K.-i.~Aoki, Z.~Hioki, R.~Kawabe, M.~Konuma and T.~Muta, \emph{{Electroweak
  Theory}}, \href{https://doi.org/10.1143/PTPS.73.1}{\emph{Progress of
  Theoretical Physics Supplement} {\bfseries 73} (1982) 1}.

\bibitem{Espriu2000}
D.~Espriu and J.~Manzano, \emph{{CP Violation and Family Mixing in the
  Effective Electroweak Lagrangian}},
  \href{https://doi.org/10.1103/PhysRevD.63.073008}{\emph{Physical Review D}
  {\bfseries 63} (2000) } [\href{https://arxiv.org/abs/0011036}{{\ttfamily
  0011036}}].

\bibitem{Pal2010}
P.B.~Pal, \emph{{Dirac, Majorana and Weyl fermions}},
  \href{https://doi.org/10.1119/1.3549729}{\emph{American Journal of Physics}
  {\bfseries 79} (2010) 485} [\href{https://arxiv.org/abs/1006.1718}{{\ttfamily
  1006.1718}}].

\bibitem{Denner2020}
A.~Denner and S.~Dittmaier, \emph{{Electroweak radiative corrections for
  collider physics}},
  \href{https://doi.org/10.1016/j.physrep.2020.04.001}{\emph{Physics Reports}
  {\bfseries 864} (2020) 1} [\href{https://arxiv.org/abs/1912.06823}{{\ttfamily
  1912.06823}}].

\bibitem{Fleischer1981}
J.~Fleischer and F.~Jegerlehner, \emph{{Radiative corrections to Higgs-boson
  decays in the Weinberg-Salam model}},
  \href{https://doi.org/10.1103/PhysRevD.23.2001}{\emph{Physical Review D}
  {\bfseries 23} (1981) 2001}.

\bibitem{Denner2016}
A.~Denner, L.~Jenniches, J.N.~Lang and C.~Sturm, \emph{{Gauge-independent
  {$\overline{\mathrm{MS}}$} renormalization in the 2HDM}},
  \href{https://doi.org/10.1007/JHEP09(2016)115}{\emph{Journal of High Energy
  Physics} {\bfseries 2016} (2016) 115}
  [\href{https://arxiv.org/abs/1607.07352}{{\ttfamily 1607.07352}}].

\bibitem{Nielsen1975}
N.~Nielsen, \emph{{On the gauge dependence of spontaneous symmetry breaking in
  gauge theories}},
  \href{https://doi.org/10.1016/0550-3213(75)90301-6}{\emph{Nuclear Physics B}
  {\bfseries 101} (1975) 173}.

\bibitem{Gambino2000}
P.~Gambino and P.A.~Grassi, \emph{{Nielsen identities of the SM and the
  definition of mass}},
  \href{https://doi.org/10.1103/PhysRevD.62.076002}{\emph{Physical Review D}
  {\bfseries 62} (2000) 076002}
  [\href{https://arxiv.org/abs/9907254}{{\ttfamily 9907254}}].

\bibitem{Passarino1979}
G.~Passarino and M.~Veltman, \emph{{One-loop corrections for e+e−
  annihilation into $\mu$+$\mu$− in the Weinberg model}},
  \href{https://doi.org/10.1016/0550-3213(79)90234-7}{\emph{Nuclear Physics B}
  {\bfseries 160} (1979) 151}.

\bibitem{Staub2010}
F.~Staub, \emph{{Automatic Calculation of supersymmetric Renormalization Group
  Equations and Self Energies}},
  \href{https://doi.org/10.1016/j.cpc.2010.11.030}{\emph{Computer Physics
  Communications} {\bfseries 182} (2010) 808}
  [\href{https://arxiv.org/abs/1002.0840}{{\ttfamily 1002.0840}}].

\bibitem{Lehmann1955}
H.~Lehmann, K.~Symanzik and W.~Zimmermann, \emph{{Zur Formulierung
  quantisierter Feldtheorien}},
  \href{https://doi.org/10.1007/BF02731765}{\emph{Il Nuovo Cimento} {\bfseries
  1} (1955) 205}.

\bibitem{Baro2009}
N.~Baro and F.~Boudjema, \emph{{Automatized full one-loop renormalization of
  the MSSM. II. The chargino-neutralino sector, the sfermion sector, and some
  applications}},
  \href{https://doi.org/10.1103/PhysRevD.80.076010}{\emph{Physical Review D}
  {\bfseries 80} (2009) 076010}
  [\href{https://arxiv.org/abs/0906.1665}{{\ttfamily 0906.1665}}].

\bibitem{Fritzsche2002}
T.~Fritzsche and W.~Hollik, \emph{{Complete one-loop corrections to the mass
  spectrum of charginos and neutralinos in the MSSM}},
  \href{https://doi.org/10.1007/s10052-002-0992-0}{\emph{The European Physical
  Journal C} {\bfseries 24} (2002) 619}.

\bibitem{Chatterjee2011}
A.~Chatterjee, M.~Drees, S.~Kulkarni and Q.~Xu, \emph{{On-shell renormalization
  of the chargino and neutralino masses in the MSSM}},
  \href{https://doi.org/10.1103/PhysRevD.85.075013}{\emph{Physical Review D}
  {\bfseries 85} (2012) 075013}
  [\href{https://arxiv.org/abs/1107.5218}{{\ttfamily 1107.5218}}].

\bibitem{Fritzsche2012}
T.~Fritzsche, S.~Heinemeyer, H.~Rzehak and C.~Schappacher, \emph{{Heavy scalar
  top quark decays in the complex MSSM: A full one-loop analysis}},
  \href{https://doi.org/10.1103/PhysRevD.86.035014}{\emph{Physical Review D}
  {\bfseries 86} (2012) 035014}
  [\href{https://arxiv.org/abs/1111.7289}{{\ttfamily 1111.7289}}].

\bibitem{Altenkamp2017}
L.~Altenkamp, S.~Dittmaier and H.~Rzehak, \emph{{Renormalization schemes for
  the Two-Higgs-Doublet Model and applications to h → WW/ZZ → 4 fermions}},
  \href{https://doi.org/10.1007/JHEP09(2017)134}{\emph{Journal of High Energy
  Physics} {\bfseries 2017} (2017) 134}
  [\href{https://arxiv.org/abs/1704.02645}{{\ttfamily 1704.02645}}].

\bibitem{Lee1973}
T.D.~Lee, \emph{{A Theory of Spontaneous $T$ Violation}},
  \href{https://doi.org/10.1103/PhysRevD.8.1226}{\emph{Physical Review D}
  {\bfseries 8} (1973) 1226}.

\bibitem{Degrande2015}
C.~Degrande, \emph{{Automatic evaluation of UV and R2 terms for beyond the
  Standard Model Lagrangians: a proof-of-principle}},
  \href{https://doi.org/10.1016/j.cpc.2015.08.015}{\emph{Computer Physics
  Communications} {\bfseries 197} (2014) 239}
  [\href{https://arxiv.org/abs/1406.3030}{{\ttfamily 1406.3030}}].

\bibitem{Peskin1995}
M.E.~Peskin and D.V.~Schroeder, \emph{{An Introduction to quantum field
  theory}}, Addison-Wesley, Reading, USA (1995).

\bibitem{Alloul2014}
A.~Alloul, N.D.~Christensen, C.~Degrande, C.~Duhr and B.~Fuks, \emph{{FeynRules
  2.0 - A complete toolbox for tree-level phenomenology}},
  \href{https://doi.org/10.1016/j.cpc.2014.04.012}{\emph{Computer Physics
  Communications} {\bfseries 185} (2014) 2250}
  [\href{https://arxiv.org/abs/1310.1921}{{\ttfamily 1310.1921}}].

\bibitem{Hahn2001}
T.~Hahn, \emph{{Generating Feynman Diagrams and Amplitudes with FeynArts 3}},
  \href{https://doi.org/10.1016/S0010-4655(01)00290-9}{\emph{Computer Physics
  Communications} {\bfseries 140} (2000) 418}
  [\href{https://arxiv.org/abs/0012260}{{\ttfamily 0012260}}].

\bibitem{Mertig1991}
R.~Mertig, M.~B{\"{o}}hm and A.~Denner, \emph{{Feyn Calc - Computer-algebraic
  calculation of Feynman amplitudes}},
  \href{https://doi.org/10.1016/0010-4655(91)90130-D}{\emph{Computer Physics
  Communications} {\bfseries 64} (1991) 345}.

\bibitem{Shtabovenko2016}
V.~Shtabovenko, R.~Mertig and F.~Orellana, \emph{{New developments in FeynCalc
  9.0}}, \href{https://doi.org/10.1016/j.cpc.2016.06.008}{\emph{Computer
  Physics Communications} {\bfseries 207} (2016) 432}
  [\href{https://arxiv.org/abs/1601.01167}{{\ttfamily 1601.01167}}].

\bibitem{Shtabovenko2020}
V.~Shtabovenko, R.~Mertig and F.~Orellana, \emph{{FeynCalc 9.3: New features
  and improvements}},
  \href{https://doi.org/10.1016/j.cpc.2020.107478}{\emph{Computer Physics
  Communications} {\bfseries 256} (2020) }
  [\href{https://arxiv.org/abs/2001.04407}{{\ttfamily 2001.04407}}].

\bibitem{Patel2015}
H.H.~Patel, \emph{{Package-X: A Mathematica package for the analytic
  calculation of one-loop integrals}},
  \href{https://doi.org/10.1016/j.cpc.2015.08.017}{\emph{Computer Physics
  Communications} {\bfseries 197} (2015) 276}
  [\href{https://arxiv.org/abs/1503.01469}{{\ttfamily 1503.01469}}].

\bibitem{Dudenas2019}
V.~Dūdėnas, \emph{{Renormalization of neutrino masses in the Grimus-Neufeld
  model}}, Vilniaus universiteto leidykla, Vilnius (2019).

\bibitem{Dudenas2019a}
V.~Dūdėnas, T.~Gajdosik and D.~Jur{\v{c}}iukonis, \emph{{Pole masses of
  neutrinos in the Grimus–Neufeld model}},
  \href{https://doi.org/10.5506/APhysPolB.50.1737}{\emph{Acta Physica Polonica
  B} {\bfseries 50} (2019) 1737}.

\bibitem{Ferreira2009}
P.M.~Ferreira, H.E.~Haber and J.P.~Silva, \emph{{Generalized CP symmetries and
  special regions of parameter space in the two-Higgs-doublet model}},
  \href{https://doi.org/10.1103/PhysRevD.79.116004}{\emph{Physical Review D -
  Particles, Fields, Gravitation and Cosmology} {\bfseries 79} (2009) 116004}
  [\href{https://arxiv.org/abs/0902.1537}{{\ttfamily 0902.1537}}].

\bibitem{Branco2011}
G.C.~Branco, P.M.~Ferreira, L.~Lavoura, M.N.~Rebelo, M.~Sher and J.P.~Silva,
  \emph{{Theory and phenomenology of two-Higgs-doublet models}},
  \href{https://doi.org/10.1016/j.physrep.2012.02.002}{\emph{Physics Reports}
  {\bfseries 516} (2011) 1} [\href{https://arxiv.org/abs/1106.0034}{{\ttfamily
  1106.0034}}].

\bibitem{Pontecorvo1957}
B.~Pontecorvo, \emph{{Inverse beta processes and nonconservation of lepton
  charge}}, {\emph{Zh. Eksp. Teor. Fiz.} {\bfseries 34} (1957) 247}.

\bibitem{Maki1962}
Z.~Maki, M.~Nakagawa and S.~Sakata, \emph{{Remarks on the Unified Model of
  Elementary Particles}},
  \href{https://doi.org/10.1143/PTP.28.870}{\emph{Progress of Theoretical
  Physics} {\bfseries 28} (1962) 870}.

\end{thebibliography}\endgroup
\end{document}